\pdfoutput=1 

\documentclass[traditabstract]{aa} 

\usepackage{natbib}
\usepackage[normalem]{ulem}
\usepackage{xcolor}
\usepackage{graphicx}
\usepackage{url}
\usepackage{subfig}
\usepackage[colorlinks=true, citecolor=blue, linkcolor=blue, breaklinks=true]{hyperref}
\usepackage{txfonts}

\begin{document} 

   \title{From spirals to lenticulars: evidence from the rotation curves and mass models of three early-type galaxies}
   \titlerunning{Rotation curves and mass models of early-type galaxies}

   \author{A. V. Shelest\inst{1,2}\and F. Lelli\inst{3,1}}
   \institute{ESO, Karl-Schwarschild-Strasse 2, 85748 Garching bei M\"{u}nchen, Germany; \email ashelest@g.harvard.edu
   \and Department of Physics, Harvard University, 17 Oxford Street, Cambridge, MA 02138, USA
   \and School of Physics and Astronomy, Cardiff University, Queens Buildings, The Parade, Cardiff, CF24 3AA, UK}

   \date{Received xxxxx; accepted xxxxx}

 
  \abstract{
  Rotation curves have traditionally been difficult to trace for early-type galaxies (ETGs) because they often lack a high-density disk of cold gas as in late-type galaxies (LTGs). We derive rotation curves for three lenticular galaxies from the ATLAS$^{\textrm{3D}}$ survey, combining CO data in the inner parts with deep H{\footnotesize I} data in the outer regions, extending out to 10-20 effective radii. We also use Spitzer photometry at 3.6 $\mu$m to decompose the rotation curves into the contributions of baryons and dark matter (DM). We find that (1) the rotation-curve shapes of these ETGs are similar to those of LTGs of similar mass and surface brightness; (2) the dynamically$-$inferred stellar mass-to-light ratios are small for quiescent ETGs but similar to those of star-forming LTGs; (3) the DM halos follow the same scaling relations with galaxy luminosity as those of LTGs; (4) one galaxy (NGC\,3626) is poorly fitted by cuspy DM profiles, suggesting that DM cores may exist in high-mass galaxies too. Our results indicate that these lenticular galaxies have recently transitioned from LTGs to ETGs without altering their DM halo structure (e.g., via a major merger) and could be faded spirals. We also confirm that ETGs follow the same radial acceleration relation as LTGs, reinforcing the notion that this is a universal law for all galaxy types.}


   \maketitle
%

\section{Introduction}
Extended rotation curves and mass models have been derived for hundreds of late-type galaxies (LTGs: spirals and irregulars), but only for a few early-type galaxies (ETGs: ellipticals and lenticulars) because they often lack high-density H{\footnotesize I} disks \citep[e.g.,][]{SPARC}. Traditionally, dynamical studies of ETGs relied on different kinematic tracers. One approach is to combine stellar kinematics in the inner parts \citep[e.g.,][]{Cappellari16} with discrete tracers like planetary nebulae and globular clusters in the outer regions \citep[e.g.,][]{Forbes}. Alternatively, strong lensing (e.g., \citealp{Gerhard}) and X-ray halos (e.g., \citealp{Buote}) have also been used to trace the mass distribution of ETGs. These methods allowed studying the dark matter (DM) content of ETGs, but a direct comparison with LTGs has been highly non-trivial since different methods require different assumptions (e.g., \citealp{Janz}, \citealp{RAR}). For example, studies of stellar kinematics and discrete tracers rely on assumptions about the orbital anisotropy and the tracer population, studies of X-ray halos hold under the assumption of hydrostatic equilibrium, while strong lensing studies may depend on the lensing model and are generally limited to the inner galactic radii.

On the other hand, cold gas (both atomic and molecular) is a more straightforward tracer of the gravitational potential for a number of reasons: (1) the cold gas forms thin rotating disks in which the pressure support is negligible, so the rotation velocity directly traces the circular velocity of a test particle in the given gravitational potential, (2) non-circular motions are generally small, contrary to the ionized gas traced by H$\alpha$ and [O{\footnotesize III}] lines, and (3) the atomic gas -- if present -- can extend to large radii, well beyond the bright stellar component. However, analyzing gas kinematics of ETGs as it is done for LTGs has long been challenging due to the low surface brightness of the gas emission (e.g., \citealp{Weijmans}, \citealp{Struve}). Deep, interferometric radio surveys have been able to detect and spatially resolve the H{\footnotesize I} emission in ETGs in group/field environments (\citealp{ATLASHI}), showing that the atomic gas often forms regularly rotating disks extending out to large radii. Moreover, deep CO observations revealed that many ETGs host inner disks of molecular gas (\citealp{ATLASCO}, \citealp{Davis13}). These key observational advances allow us to study the cold gas kinematics in ETGs in detail, providing new insights into the DM halos of different galaxy types.

In this work, we derive rotation curves and mass models of three ETGs from the ATLAS$^{\text{3D}}$ survey \citep{ATLAS1}: NGC\,2824, NGC\,3626 and UGC\,6176. These are the only three galaxies in the ATLAS$^{\text{3D}}$ sample that have both an inner CO disk and an outer H{\footnotesize I} disk. We combine CO and H{\footnotesize I} kinematics to determine accurate rotation curves in a uniform way over the entire extent of the gaseous body. Some of the known properties of the galaxies are summarized in Table\,\ref{Tab: Distances}. They are morphologically classified as lenticulars (S0) and identified as ``fast rotators'' based on their observed stellar kinematics \citep{ATLAS1}. In particular, NGC\,3626 is a well studied object since it has a stellar component that counter-rotates with respect to the gaseous disk (\citealp{Ciri}, \citealp{Jore}, \citealp{Haynes}, \citealp{Silchenko}, \citealp{Mazzei}). NGC\,3626 also has a companion galaxy (UGC\,6341) with significant H{\footnotesize I} emission \citep{ATLASHI}, which we do not study here.

\begin{table}  
\caption{The galaxy sample.}
\label{Tab: Distances} 
\centering
\small
	\begin{tabular}{lccc }
		\hline
		\hline
		& NGC\,2824 & NGC\,3626 & UGC\,6176\\
		\hline
		RA (J2000) & 09 19 02.23 & 11 20 03.79 & 11 07 24.66\\
		DEC (J2000) & +26 16 12.00 & +18 21 24.45 & +21 39 25.87\\
		Morph. type & S0 & SA0 & SB0 \\
		$D$ [Mpc] & 39.6 & 22.9 & 39.3 \\
		M$_K$ [mag] & $-22.87$ & $-23.65$ & $-22.56$ \\
		$\log{M_{H{\footnotesize I}}/M_\odot}$ & 8.93 & 9.08 & 9.00 \\
		$\log{M_{H_2}/M_\odot}$ & 8.89 & 8.46 & 8.74 \\
		$\log{L_{[3.6]}/L_\odot}$ & 10.6 & 10.9 & 10.5 \\
		$R_\text{eff}$ [kpc] & 1.09 & 2.43 & 1.18 \\
		$\Sigma_\text{eff}$ [L$_\odot$/pc$^2$]& 5210 & 2018 & 3318 \\
		\hline
	\end{tabular}
\tablefoot{The galaxy coordinates are from \protect\cite{ATLAS1}. The morphological type is from \protect\cite{RC3}. NGC\,3626 hosts a bar and has been reclassified as SAB0/a by \protect\citet{Buta}. The galaxy distance $D$ is from \protect\cite{RAR}. The $K$-band absolute magnitude M$_{\rm K}$ (corrected for foreground galactic extinction), the atomic gas mass $M_{\rm HI}$, and the molecular gas mass $M_{\rm H_2}$ are from \protect\cite{ATLAS1}, \protect\cite{ATLASHI} and \protect\cite{ATLASCO}, respectively, and converted to the adopted distances. The HI mass of NGC\,2824 is estimated in this work using deeper observations. The luminosity at 3.6 $\mu$m, the effective radius $R_\text{eff}$ encompassing half the total luminosity, and the effective surface brightness $\Sigma_\text{eff}$ are from \protect\cite{RAR}.}
\end{table}		

\section{H{\footnotesize I} and CO moment maps}

We analyze public H{\footnotesize I} and CO cubes from the Atlas$^{\textrm{3D}}$ website (\citealp{ATLASHI}, \citealp{ATLASCO}). For NGC\,2824, we use a deeper H{\footnotesize I} cube that was kindly provided to us by Dr. Paolo Serra. The cube properties are summarized in Table \ref{Tab: cube params}. We perform our analysis using the Groningen imaging processing system (GIPSY, \citealp{GIPSY}) and $^{\text{3D}}$Barolo \citep{Barolo}. We create masks that identify regions of gas emissions using the ``smooth and search'' algorithm of $^{\text{3D}}$Barolo with the parameter \textit{threshVelocity} set to 1 and \textit{minPix} set to the beam size of each cube. We apply these masks to the full-resolution cubes to create intensity maps (moment-zero) summing all channels that contain line emission, and velocity maps (moment-one), estimating an intensity-weighted mean velocity. Velocity dispersion maps (moment-two) have little physical meaning since the line broadening is largely driven by beam-smearing effects rather than representing the true, intrinsic velocity dispersion of the gas. The average root-mean-square noise in the cubes is determined using channels that are free of line emission. In the intensity maps, a pseudo-3$\sigma$ contour is estimated using the formulae from \cite{PseudoSigma1} and \cite{Lelli14}, which account for the different number of summed channels at each spatial pixel and for the channel dependencies.

\begin{figure*}[!]
	\centering
	\subfloat{%
		\includegraphics[height=0.77\textheight]{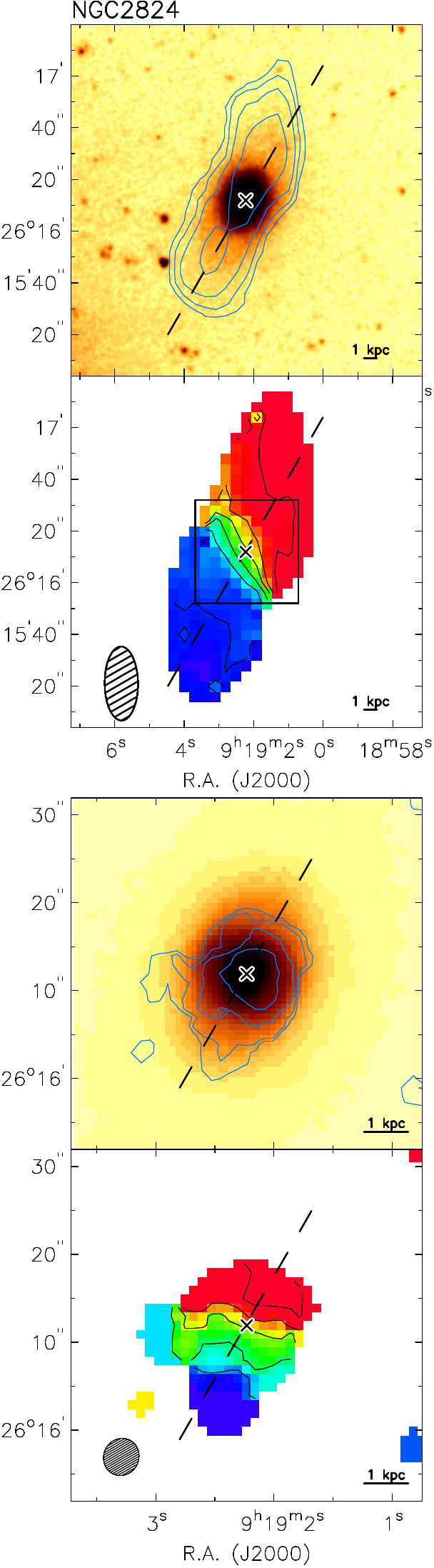}}
	\hspace{3mm}
	\subfloat{\includegraphics[height=0.77\textheight]{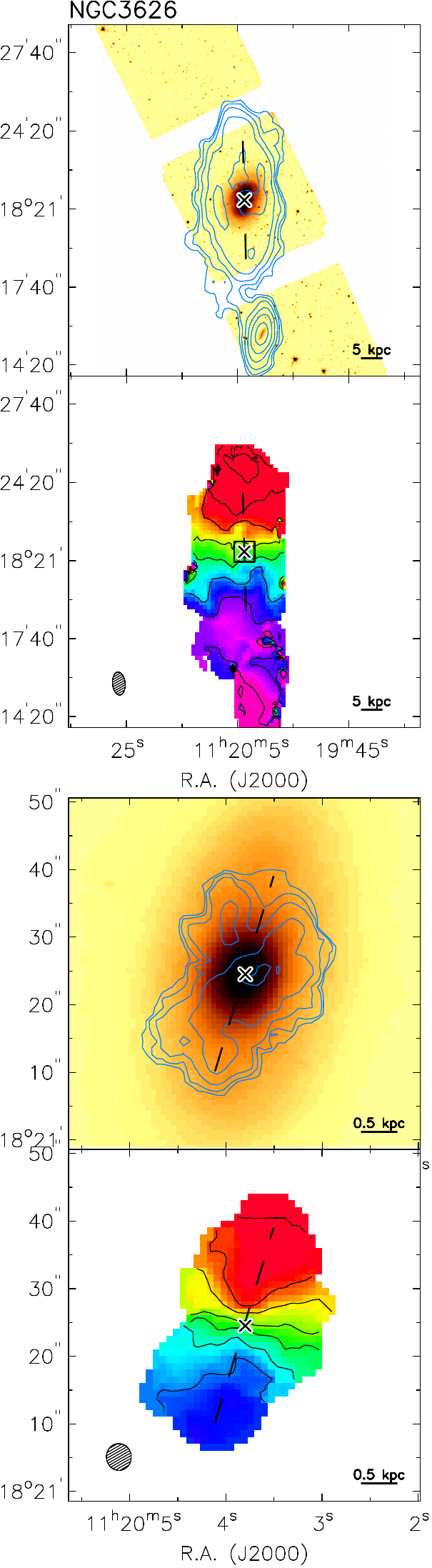}
	}
	\subfloat{\includegraphics[height=0.77\textheight]{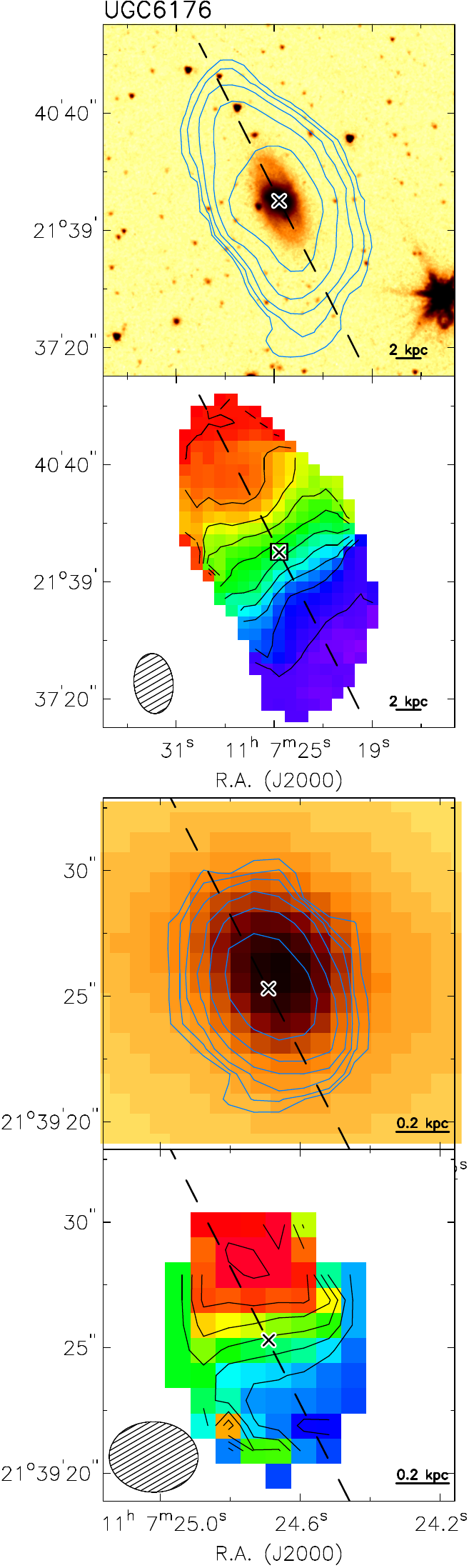}
	}
	\caption{Moment maps for NGC\,2824 (left), NGC\,3626 (middle), and UGC\,6176 (right). The top section shows the H{\footnotesize I} data, while the bottom section shows the CO data. In each section, the top row shows the Spitzer image at 3.6 $\mu$m overlaid with the gas intensity maps, while the bottom row shows the velocity field. The box in the H{\footnotesize I} velocity field depicts the size of the CO frame. The cross corresponds to the galaxy center; the dashed line indicates the kinematic major axis; the physical scale is illustrated in the bottom-right corner; the beam shape is shown in the bottom-left corner. The H{\footnotesize I} and CO density contours are multiples of the pseudo-3$\sigma$ value (0.5, 1, 2, 4,...) where 3$\sigma_{\textrm{map}}$ is given in Tab. \ref{Tab: cube params}.}
	\label{Fig: united map plot}
\end{figure*}

\begin{figure*}[!]
	\centering
	\includegraphics[width=0.9\textwidth]{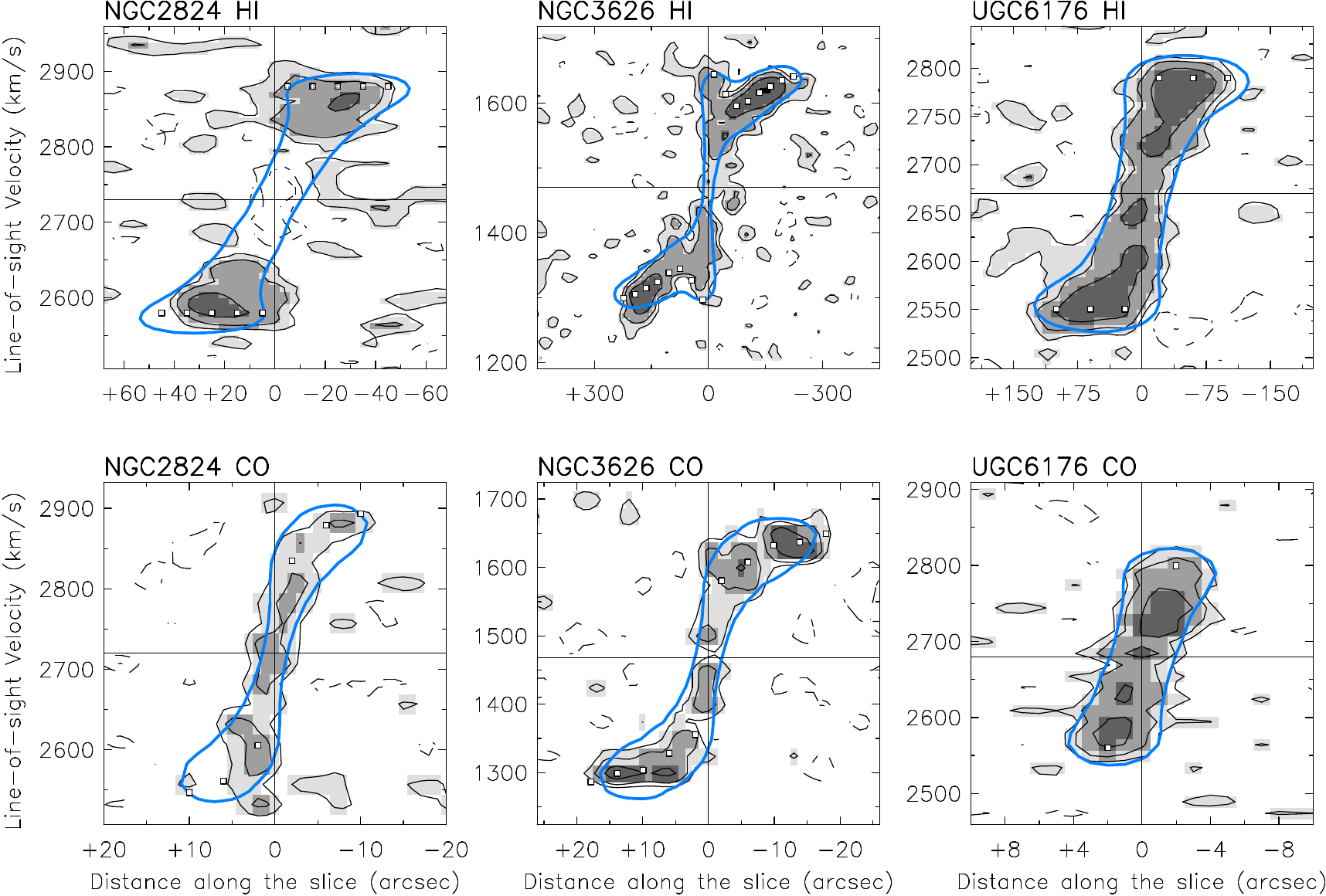}
	\caption{Major$-$axis PV diagrams from the observed H{\footnotesize I} cube (top row) and the CO one (bottom row), overlaid with the projected rotation curve (white squares) and model PV diagrams (blue contour corresponding to 1.5 $\sigma$ from Tab. \ref{Tab: cube params}).}
	\label{Fig: summary PVs}
\end{figure*}

Figure \ref{Fig: united map plot} shows the intensity maps superimposed on the Spitzer [3.6] images (top rows) and the velocity maps (bottom rows). For all three galaxies, the H{\footnotesize I} distribution is more extended than the stellar body, while the CO distribution is concentrated near the galactic center. Both H{\footnotesize I} and CO velocity fields display regular rotation with no clear signs of strong non-circular motions. This is also evinced by position-velocity (PV) diagrams extracted along the kinematic major axis of the disk, which are shown in Figure \ref{Fig: summary PVs}. The morphological position angle (PA) from the intensity maps agrees with the kinematic PA from the velocity maps. It also agrees with the orientation of the stellar bodies for NGC\,2824 and UGC\,6176. For NGC\,3626, instead, the PA suggested by the elongation of the stellar disk differs from the average H{\footnotesize I} major axis, pointing to a warped gas disk.

\begin{table*}[!]
	\small
	\caption{Properties of the analyzed datacubes.}
	\centering
	\begin{tabular}{ccccccc}
		\hline
		\hline
		Galaxy & Line & Vel. Res & Beam & PA$_{\text{beam}}$ & $\sigma$ & 3$\sigma_\text{map}$\\
		& & (km/s) & (arcsec$^2$) & ($^\circ$) & (Jy/beam) & (Jy/beam km/s)\\
		\hline
		NGC\,2824 & CO & 25.0 & 4.28 x 4.00 & -27.24 & 0.01019 & 0.3950\\ 
		& H{\footnotesize I} & 8.25 & 28.60 x 12.89 & 0.50 & 0.00015 & 0.0569\\
		NGC\,3626& CO & 25.0 & 3.95 x 3.68 & 4.67 & 0.00505 & 0.0375\\
		& H{\footnotesize I} & 8.25 & 59.36 x 32.38 & 5.15 & 0.00054 & 0.3242\\ 
		UGC\,6176& CO & 15.0 & 3.54 x 2.81 & 89.01 & 0.01071 & 0.0735\\ 
		& H{\footnotesize I} & 8.25 & 51.48 x 33.33 & 5.39 & 0.00053 & 0.0569\\
		\hline
	\end{tabular}
\tablefoot{For each CO and HI cube, we list the velocity resolution after smoothing, the synthesized beam, the PA of the beam, the rms noise $\sigma$, and the pseudo-3$\sigma$ contour in the total intensity maps 3$\sigma_\text{map}$ (estimated using the formulae from \protect\citealt{PseudoSigma1} and \protect\citealt{Lelli14}).}
\label{Tab: cube params}
\end{table*}

\section{Derivation of Rotation Curves} \label{sec: rot curves}

We derive rotation curves by fitting the velocity fields with a tilted-ring model to obtain initial estimates, which are subsequently refined by building kinematic 3D disk models (e.g., \citealp{Lelli12a, Lelli12b}). We also verified our results using the software $^{\text{3D}}$Barolo \citep{Barolo}, which directly fits the 3D cube bypassing the 2D velocity maps. 

In the tilted-ring fit, each ring is described by six free parameters: center ($x_0$, $y_0$), systemic velocity ($v_\text{sys}$), position angle (PA), inclination ($i$), and rotation velocity ($v_\text{rot}$). The width of the rings is chosen to be either equal to the width of the beam major axis, or $\sqrt{(b_{\text{maj}}\cdot b_{\text{min}})}$ if the beam is strongly elongated and not aligned with the PA of the galaxy. Here, $b_{\text{maj}}$ and $b_{\text{min}}$ are the major and minor axes of the beam, respectively. The number of rings is determined such that the major axis of the galaxy is fully covered. The pixels in the velocity maps are weighted using a cosine function to give more weight to points close to the major axis, which provide most information regarding the rotation velocity. We simultaneously fit both halves of the galaxy, neglecting possible asymmetries between the different sides. For NGC\,3626, however, the CO data do not cover the full gas emission at approaching velocities due to the limited bandwidth of the observations, so we fit only the receding side.

We perform a series of iterations using the GIPSY task \textit{rotcur} \citep{BegemanThesis}. In the first iteration all the parameters are left free, while in the second iteration we fix $x_0$, $y_0$, and $v_{\rm sys}$ using the mean values across the rings. The centers and systemic velocities from the CO and H{\footnotesize I} data are consistent with each other, as well as with optical determinations from the Nasa Extragalactic Database. In the next two iterations, the PA and inclination are fixed in arbitrary order. We checked that the order does not change the resulting rotation curve within the errors. For NGC\,2824 and UGC\,6176, the CO and H{\footnotesize I} disks show no strong indications of warps, so $i$ and PA are kept constant with radius. For both galaxies, however, the inner CO inclination significantly differ from the outer H{\footnotesize I} inclination. For NGC\,3626, we model an H{\footnotesize I} warp by varying both PA and $i$ with radius. The last run of \textit{rotcur} is performed leaving only $v_{\text{rot}}$ as a free parameter.

To check the resulting rotation curves and correct for beam smearing effects, we construct model cubes using the \textit{rotcur} results and the observed gas density profiles as inputs. We derive azimuthally-averaged H{\footnotesize I} and CO surface brightness profiles using the GIPSY task \textit{ellint}, adopting the best-fit geometric parameters. The CO surface brightnesses are converted into H$_2$ surface densities adopting the Milky-Way conversion factor $X_{\text{CO}} = 2 \times 10^{20} \text{cm}^{-2} (\text{K km s}^{-1})^{-1}$ \citep{Bolatto}. We create model cubes with the GIPSY task \textit{galmod} for the H{\footnotesize I} data and a custom-built task for the CO data. Since the latter task does no automatic conversion from H$_2$ surface density to CO surface brightness, we renormalize the total flux of the CO model to be equal to the observed flux. We assume that the gas disk has a velocity dispersion of 10 km s$^{-1}$ and a thickness of 100 pc.

\begin{figure*}[!]
	\begin{center}
	\includegraphics[width=0.95\textwidth]{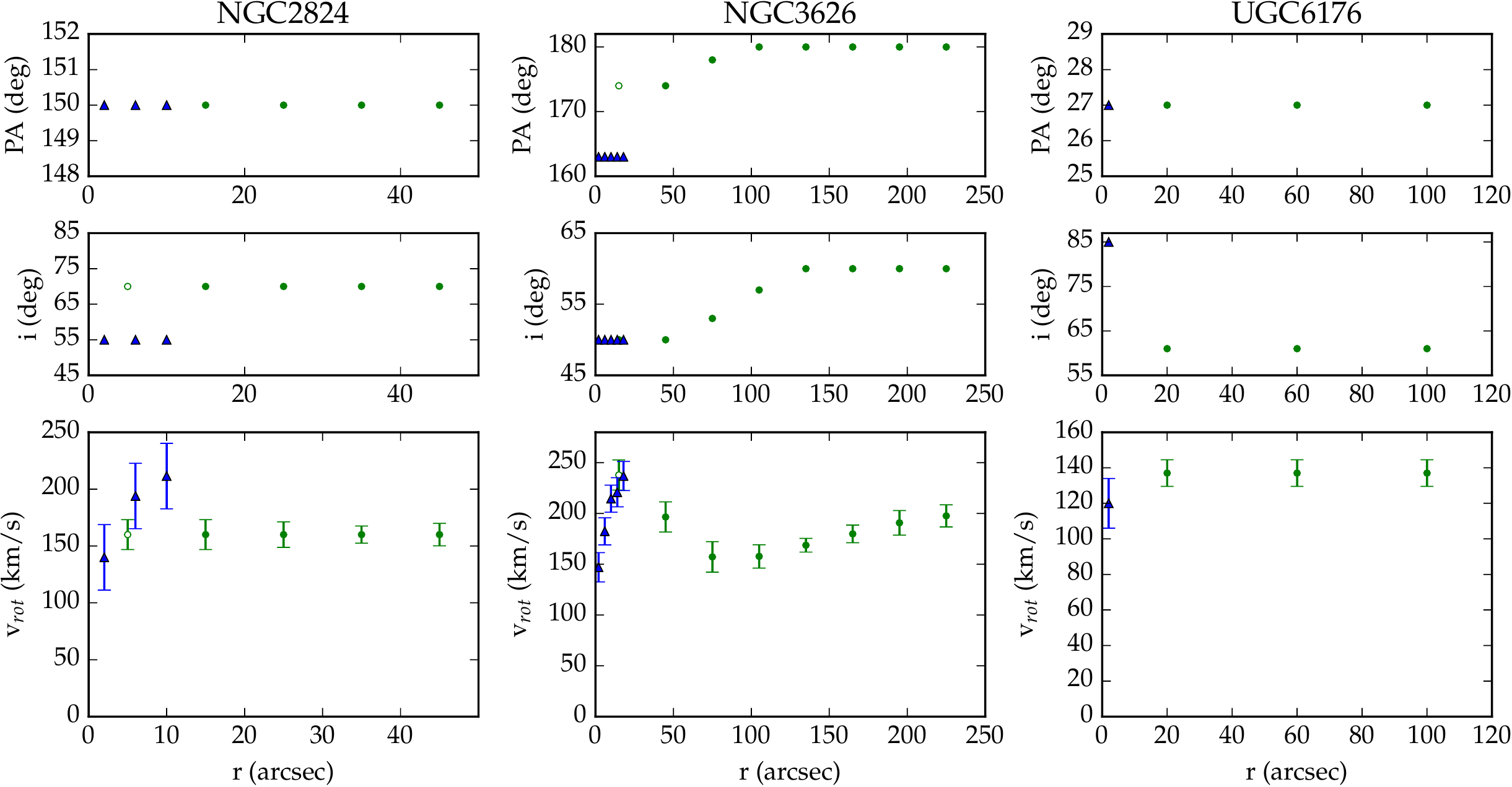}
	\end{center}
	\centering
	\caption{Position angle (top), inclination (middle), and final rotation curves (bottom) from H{\footnotesize I} (green points) and CO (blue triangles) observations. Open circles show omitted points in the combined rotation curves due to the overlap of CO and H{\footnotesize I} data.}
	\label{Fig: Barolo comparison}
\end{figure*}

We compare PV diagrams and channel maps from the model cubes with the observed ones. A disagreement between models and observations indicates that the input density profiles and/or rotation curves are affected by beam smearing. Beam smearing tends to flatten the surface density profile and cause an underestimation of the rotation velocity, especially in the innermost regions. We manually adjust individual points in the inner parts of the rotation curves and/or density profiles to minimize the residuals. In some cases, the PA and inclination angle are also manually adjusted by a few degrees to find the best compromise between the kinematic estimates from \textit{rotcur} and the observed intensity maps. The best-fit 3D models are illustrated in Fig.\,\ref{Fig: summary PVs} and presented in a more extensive way in Appendix \ref{App: Atlas}. The adopted geometric parameters are summarized in Tab.\,\ref{Tab: rot curve static params} and in Tab.\,\ref{Tab: PaInc N3} for the warp model of NGC\,3626. The uncertainties are estimated by running $^{\text{3D}}$Barolo with all the parameters fixed to the final value and only one parameter of interest left free. The final geometric parameters and rotation curves are shown in Figure \ref{Fig: Barolo comparison}.

In general, $^{\text{3D}}$Barolo returns similar rotation curves as our manual procedure with some exceptions. For NGC\,2824, the CO data have too low spatial resolutions and signal-to-noise ratio to ensure a good $^{\text{3D}}$Barolo fit, while the H{\footnotesize I} data shows a strong absorption feature at small radii that $^{\text{3D}}$Barolo cannot model. For NGC\,3626, the missing CO data at approaching velocities cannot be automatically modeled by $^{\text{3D}}$Barolo, whereas our manual procedure can take this shortcoming into account by focusing on the receding portion of the PV diagram. For these reasons, we prefer our manual procedure with respect to fully automatic fits with $^{\text{3D}}$Barolo.


We combine the CO and H{\footnotesize I} rotation curves into a single, optimal rotation curve. At radii where rotation velocities are available from both tracers, we favor the CO data since beam-smearing effects are less severe due to their higher spatial resolution. For NGC\,2824 and NGC\,3626, the combined rotation curves rise steeply in the central parts, decline at intermediate radii, and flattens in the outer regions. Similar rotation curves are often observed in massive spiral galaxies \citep{Casertano1991, Noordermeer07}. For UGC\,6176, the rotation curve is almost perfectly flat, similarly to spiral galaxies of lower masses \citep[e.g.,][]{BegemanThesis}. The shapes of the rotation curves are further discussed in Sect.\,\ref{sec: Discussion}.

\section{Mass Models}

We derive detailed mass models using Spitzer 3.6 $\mu$m images to estimate the stellar gravitational field, together with H{\footnotesize I} and CO maps to estimate the gas gravitational field (Fig. \ref{Fig: united map plot}). In Section \ref{sec: DM}, we will use these mass models and the observed rotation curves to study the DM halos of our three lenticular galaxies.

\subsection{Stellar and gaseous contributions}

The SPARC database \citep{SPARC} provides [3.6] luminosity profiles and non-parametric bulge-disk decompositions for several ETGs in Atlas$^{\rm 3D}$ including our three S0s (see \citealp{RAR}). Given the intrinsic ambiguity of bulge-disk decompositions for lenticular galaxies, we explored different decompositions' scheme to estimate their impact on the final mass models. After several trials, we decided to use a single, disk-like component for both NGC\,2824 and NGC\,3626 because the bulge contribution is small and uncertain. This is a conservative choice that avoids adding unnecessary free parameters to the mass model. All the observed features in the surface brightness profiles are preserved in calculating the stellar gravitational contribution, including the inner light concentrations. For NGC\,2824 and NGC\,3626, we simply assume that these light concentrations lie in a flattened structure with a similar mass-to-light ratio as the stellar disk, as it may be the case for a pseudo-bulge (e.g., \citealp{KormendyFreeman}). For UGC\,6176, we use the same bulge-disk decomposition as in \cite{RAR}.

\begin{table}
	\small
	\centering
	\caption{Geometric parameters of CO and HI disks}.
	\begin{tabular}{l c c c}
		\hline
		\hline
		& NGC 2824    & NGC 3626    & UGC 6176    \\
		\hline
		RA (hms)   & 09 19 2.23  & 11 20 4.15 & 11 07 24.70 \\
		DEC (dms)  & +26 16 12.0 & +18 21 24.5 & +21 39 25.9 \\
		$V_{\rm sys}$ (km s$^{-1}$) & 2725$\pm$ 4 & 1468$\pm$ 5 & 2680$\pm$ 3 \\
		PA ($^{\circ}$) & 160$\pm$9 & $-$ & 33$\pm$7 \\
		$i_{\rm CO}$ ($^{\circ}$) & 50$\pm$2 & $-$ & 85$\pm$1 \\
		$i_{\rm HI}$ ($^{\circ}$) & 70$\pm$2 & $-$ & 58$\pm$1 \\
		\hline
	\end{tabular}
	\tablefoot{The coordinates of the kinematic center (RA and DEC) are given in J2000. We adopt the same systemic velocity ($V_{\rm sys}$) and position angle (PA) for CO and HI disks, but we use different inclinations ($i$). The PA and $i$ of NGC\,3626 are radius-dependent and can be found in Table \ref{Tab: PaInc N3}.}
	\label{Tab: rot curve static params}
\end{table}

We calculate the gravitational contribution of each component using the GIPSY task \textit{rotmod}. We assume a spherical bulge (when present) and thick disks with a vertical exponential profile. The scale$-$heights of the H{\footnotesize I} and CO disks are set to 100 pc, while that of the stellar disk ($h_{\rm z}$) is assumed to correlate with the exponential scale$-$length ($h_{\rm R}$) as $h_{\rm z} = 0.196 h_{\rm R}^{0.633}$ \citep{DiskScaleHeight}. This relation has been derived for edge-on spiral galaxies and may be potentially inappropriate for lenticular galaxies (see, e.g., \citealp{deGrijs}). Luckily, the 3D geometry of baryons has a minor effect on the final mass models: the maximum variation in the peak circular velocities is about 20\% for axial ratios between zero (razor-thin disk) and one (sphere); see, e.g., Figure 3 in \citet{Noordermeer}. This effect is smaller than the current uncertainties in the stellar mass-to-light ratios from stellar population models (e.g., \citealp{SchombertMcGaughLelli}). Having computed the gravitational field of the individual components in the disk mid-plane, the baryonic rotation curve reads
\begin{equation} \label{eq: v compos}
v^2_{\text{bar}}(R) = \Upsilon_{\text{d}} v^2_{\text{d}}(R) + \Upsilon_{\text{b}} v^2_{\text{b}}(R) + 1.33\left[v^2_{\text{HI}}(R) + v^2_{\text{H2}}(R)\right],
\end{equation} 
where $v_d$ and $v_b$ account for the stellar contribution (disk and bulge, respectively), while $v_{\text{HI}}$ and $v_{\text{H2}}$ account for the gas contribution (atomic and molecular, respectively). The molecular contribution is computed assuming the standard conversion factor $X_\textrm{CO}= 2 \times 10^{20} \text{cm}^{-2} (\text{K km s}^{-1})^{-1}$ of the Milky Way. $\Upsilon$ denotes the mass-to-light ratio for the stellar disk and the bulge, indicated by the subscripts $d$ and $b$, respectively. The 1.33 factor in front of the gaseous components takes the mass contribution of primordial helium into account (\citealp{SPARC}, but see also \citealp{McGaugh2020}). Table \ref{Tab: all rot curves U} provides the circular velocities and the surface densities of the individual components at the same radii at which rotation velocities are measured.

\begin{table}
	\small
	\centering
	\caption{Warp model for NGC\,3626.}
	\begin{tabular}{cccc}
		\hline
		\hline
		R ($''$) & R (kpc) & $i$ ($^\circ$) & $PA$ ($^\circ$)\\
		\hline
		15 & 1.67 & 50 & 174\\
		45 & 4.50 & 50 & 174 \\
		75 & 8.33 & 53 & 178 \\ 
		105 & 11.66 & 57 & 180 \\
		135 & 14.99 & 60 & 180 \\
		165 & 18.32 & 60 & 180 \\
		195 & 21.65 & 60 & 180 \\
		225 & 24.99 & 60 & 180 \\
		\hline
	\end{tabular}
	\tablefoot{Sampled radii ($R$) in both arcsec and kpc, adopted inclination ($i$), and position angle (PA) for the warped disk of NGC\,3626.}
	\label{Tab: PaInc N3}
\end{table}

\begin{figure*}[!]
	\begin{center}
		\includegraphics[width =0.95\textwidth]{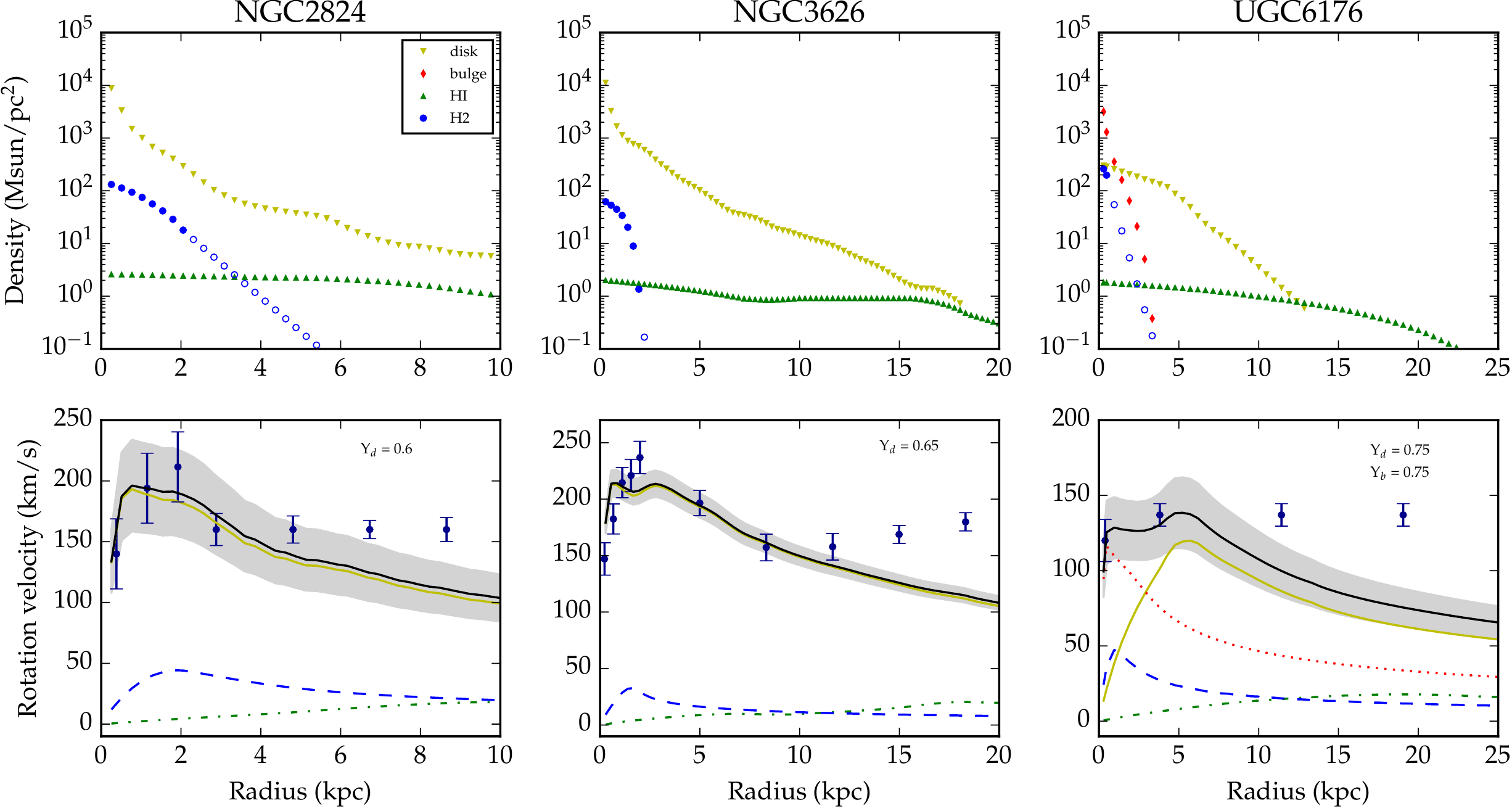}
	\end{center}
	\caption{Top panels: surface density profiles for the H$_2$ disk (blue circles), HI disk (green upward triangles), stellar disk (yellow downward triangles), and stellar bulge (red diamonds) when necessary. Open circles indicate that the surface density is extrapolated from an exponential fit. Bottom panels: the observed rotation curve (dots with $\pm$1$\sigma$ error bars) is fitted with a maximum-disk mass models (black solid line with $\pm$1$\sigma$ error band). The baryonic rotation curve is decomposed into the contributions of molecular gas (blue dashed line), atomic gas (green dash-dotted line), stellar disk (yellow solid line), and stellar bulge (red dotted line) when necessary. The maximum mass-to-light ratios are given in the top right corner.}
	\label{Fig: Max disk mine}
\end{figure*}

\subsection{Maximum-disk mass models} \label{subsec: Max disk}

The baryonic rotation curve depends on the adopted stellar mass-to-light ratios (Eq. \ref{eq: v compos}). Here we adjust their values to achieve a maximum-disk fit (\citealp{vanAlbadaSancisi}, \citealp{Starkman}), ensuring that the inner part of the observed rotation curve is fully explained by luminous matter. Our fits can be seen in Fig. \ref{Fig: Max disk mine} along with the maximal values of $\Upsilon_{\rm b}$ and $\Upsilon_{\rm d}$. Surprisingly, the maximum mass-to-light ratios are relatively small for red and dead ETGs, which are expected to be in the range 0.8-1.0 $M_\odot/L_\odot$ \citep{Schombert14}, but are near the high-end boundary of the expectations for star-forming LTGs \citep{SchombertMcGaughLelli}. This is further discussed in Sect.\,\ref{sec: Discussion}. The baryonic rotation curves do not agree with the observed ones at large radii. This indicates the presence of the DM effect in the galaxies, as we investigate in the next Section. 

\begin{sidewaystable*}  
	\begin{center}
		\caption{Rotation curves and mass models for three lenticular galaxies}
		\begin{tabular}{ cccccccccccc }
			Galaxy & $R$ [kpc] & $v_\text{obs}$ [km/s] & $e_{v,\text{obs}}$ [km/s] & $v_{\text{H}_2}$ [km/s] & $v_{\text{HI}}$ [km/s] & $v_\text{disk}$ [km/s] & $v_\text{bulge}$ [km/s] & $\Sigma_{\text{H}_2}$ [L$_\odot$/pc$^2$] &  $\Sigma_\text{HI}$ [L$_\odot$/pc$^2$] &  $\Sigma_\text{disk}$ [L$_\odot$/pc$^2$] &  $\Sigma_\text{bulge}$ [L$_\odot$/pc$^2$]\\
			\hline
			\hline
			& 0.38 & 140 & 29 & 17.82 & 0.95 & 222.95 & $-$ & 120.89 & 2.61 & 8422.37 & $-$\\
			& 1.15 & 194 & 29 & 39.08 & 2.83 & 243.31 & $-$ & 60.77 & 2.54 & 1238.20 & $-$\\
			& 1.92 & 211 & 29 & 44.22 & 4.48 & 234.68 & $-$ & 18.87 & 2.46 & 506.32 & $-$\\
			NGC\,2824 & 2.88 & 160 & 10 & 38.55 & 6.39 & 209.06 & $-$ & 3.96 & 2.37 & 139.56 & $-$\\
			& 4.81 & 160 & 11 & 28.95 & 10.31 & 169.19 & $-$ & 0.19 & 2.25 & 59.68 & $-$\\
			& 6.73 & 160 & 8 & 23.76 & 14.80 & 151.28 & $-$ & 0.01 & 1.90 & 17.84 & $-$\\
			& 8.65 & 160 & 10 & 20.70 & 17.82 & 132.94 & $-$ & 0.00 & 1.28 & 10.12 & $-$\\
			\hline
			& 0.22 & 147 & 15 & 7.48 & -0.22 & 199.14 & $-$ & 63.93 & 2.04 & 23094.69 & $-$\\
			& 0.67 & 182 & 15 & 19.16 & 1.75 & 265.53 & $-$ & 49.70 & 1.97 & 3566.60 & $-$\\
			& 1.11 & 214 & 15 & 28.68 & 2.78 & 258.78 & $-$ & 34.04 & 1.90 & 1747.42 & $-$\\
			& 1.55 & 220 & 30 & 32.82 & 3.70 & 251.56 & $-$ & 13.32 & 1.82 & 1233.41 & $-$\\
			& 2.00 & 237 & 30 & 29.53 & 4.55 & 254.81 & $-$ & 0.74 & 1.75 & 1048.37 & $-$\\
			& 4.99 & 197 & 15 & 16.26 & 8.79 & 239.53 & $-$ & 0 & 1.26 & 157.55 & $-$\\
			NGC\,3626 & 8.32 & 157 & 12 & 9.44 & 10.36 & 199.06 & $-$ & 0 & 0.87 & 36.76 & $-$\\
			& 11.65 & 158 & 11 & 10.50 & 11.29 & 173.70 & $-$ & 0 & 0.92 & 13.83 & $-$\\
			& 14.98 & 169 & 7 & 9.25 & 15.36 & 154.07 & $-$ & 0 & 0.92 & 3.18 & $-$\\
			& 18.31 & 180 & 9 & 8.37 & 20.41 & 138.64 & $-$ & 0 & 0.49 & 0 & $-$\\
			& 21.64 & 191 & 12 & 7.70 & 18.97 & 124.74 & $-$ & 0 & 0.21 & 0 & $-$\\
			& 24.97 & 198 & 11 & 7.15 & 17.64 & 115.31 & $-$ & 0 & 0.09 & 0 & $-$\\ 
			
			
			\hline
			& 0.38 & 120 & 14 & 30.50 & 0.74 & 20.39 & 125.74 & 229.04 & 1.83 & 390.16 & 2782.91\\
			UGC\,06176 & 3.81 & 137 & 8 & 27.18 & 6.53 & 87.10 & 154.91 & 0.06 & 1.55 & 175.01 & 0.04\\
			& 11.43 & 137 & 8 & 15.15 & 14.76 & 56.18 & 50.28 & 0 & 0.87 & 1.92 & 0\\
			& 19.05 & 137 & 8 & 11.77 & 17.81 & 41.33 & 38.95 & 0 & 0.29 & 0 & 0\\
		\end{tabular}
		\tablefoot{Sampled radii ($R$), observed rotation velocities ($v_{\rm obs}$), and relative errors ($e_{v, \rm{obs}}$) from CO and HI datacubes. Contributions to the circular velocity from the molecular gas disk ($v_{\text{H}_2}$), atomic gas disk ($v_{\rm HI}$), stellar disk ($v_{\rm d}$), and -if present- stellar bulge ($v_{\rm b}$), together with the corresponding surface densities at $R$. For convenience, the velocity contributions and surface density profiles of the stellar disk and the bulge are normalized to a mass-to-light ratio of 1 at 3.6 $\mu$m.}
		\label{Tab: all rot curves U}
	\end{center}
\end{sidewaystable*}


\section{Dark matter halo fits} \label{sec: DM}

To model the discrepancy between the observed and baryonic rotation curves at large radii (Fig. \ref{Fig: Max disk mine}), we add a DM halo to the mass models. We fit three different halo profiles: NFW \citep{NFW}, Einasto \citep{Einasto}, and DC14 \citep{DiCintio}. While the NFW and Einasto profiles are suggested by DM-only cosmological simulations \citep[e.g.,][]{DuttonMaccio}, the DC14 profile is derived from hydro-dynamical simulations of galaxy formation, taking into account the effects of the stellar feedback on the inner DM halo.

Rotation-curve fits with the DC14 profile are described in \cite{Katz} and \cite{LiDM, Li20}. In short, the shape of the DC14 profile depends on the quantity $X = \log(M_\star/M_{200})$ where $M_\star$ is the present-day stellar mass and $M_{200}$ is the halo mass, defined as the enclosed mass at which the halo density equals 200 times the critical density of the Universe. For $-4<X<-2$, the DC14 profile is shallower than the NFW profile due to the formation of central DM cores by stellar feedback. For higher and lower values of $X$, the DC14 reduces to the NFW profile because the supernovae energy is too small to overcome the galaxy potential well and drive a significant mass redistribution. Fox $X>-1.3$, however, the feedback from active galactic nuclei (AGN), which is not included in the simulations of \citet{DiCintio}, is expected to become important, making the DC14 profile potentially inaccurate. After performing the fits we check that $X<-1.3$ for all galaxies, so the use of the DC14 profile is justified. Specifically, we find $-1.37$ for NGC\,2824, $-1.99$ for NGC\,3626, and $-1.44$ for UGC\,6176. These stellar-to-halo mass ratios are rather high, so the DC14 results are expected to be similar to the NFW and Einasto ones.

\begin{figure*}[!] 
\centering
\includegraphics[width=0.95\textwidth]{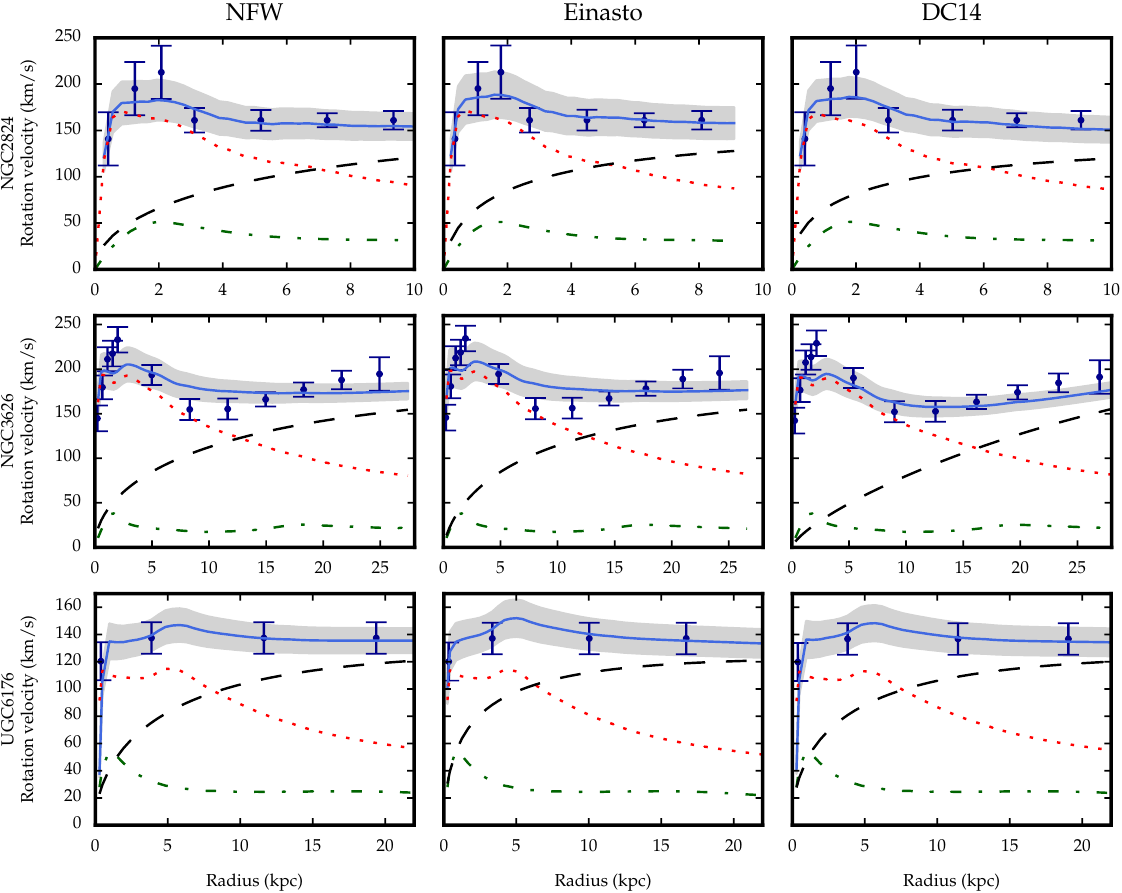}
	\caption{The observed rotation curve (blue dots with $\pm$1$\sigma$ error bars) is fitted with the NFW (left), Einasto (middle), and DC14 (right) halo models. The model rotation curve (blue solid line with $\pm$1$\sigma$ error band) is decomposed into the contributions of stars (bulge plus disk; red dotted line), gas (atomic plus molecular; green dash-dotted line), and DM halo (black dashed line).}
	\label{Fig: summary DM}
\end{figure*}

\begin{table*}[h]
	\centering
	\small
	\caption{Maximum-probability values with $\Lambda$CDM constraints for the fit parameters.}
	\begin{tabular}{ cccccccccc }
		\hline
		\hline
		Galaxy & Halo profile & D [Mpc] & $i$ [$^\circ$] & $\Upsilon_\text{d}$ & $\Upsilon_\text{b}$ & V$_{200}$ [km/s] & C$_{200}$& $\alpha_\epsilon$\\
		\hline
		& NFW & 43 $\pm$ 7 & 59 $\pm$ 9 & 0.47 $\pm$ 0.09 & $-$ & 130 $\pm$ 28 & 8 $\pm$ 2&$-$\\ 
		NGC\,2824 & Einasto & 37 $\pm$ 8 & 59 $\pm$ 9 & 0.47 $\pm$ 0.10 & $-$ & 114 $\pm$ 21 & 11 $\pm$ 3 & 0.18 $\pm$ 0.07\\ 
		& DC14 & 41 $\pm$ 7 & 59 $\pm$ 8 & 0.46 $\pm$ 0.09 & $-$ & 113 $\pm$ 28 & 11 $\pm$ 6& $-$\\ 
		\hline
		& NFW & 23 $\pm$ 2 & 62 $\pm$ 8 & 0.54 $\pm$ 0.10 & $-$ & 169 $\pm$ 33 & 5 $\pm$ 1& $-$\\ 
		NGC\,3626 & Einasto & 22 $\pm$ 2 & 61 $\pm$ 8 & 0.56 $\pm$ 0.11 & $-$ & 160 $\pm$ 56 & 5 $\pm$ 2 & 0.3 $\pm$ 0.1\\ 
		& DC14 & 25 $\pm$ 2 & 64 $\pm$ 7 & 0.52 $\pm$ 0.08 & $-$ & 228 $\pm$ 44 & 7 $\pm$ 1 & $-$\\ 
		\hline
		& NFW & 40 $\pm$ 5 & 56 $\pm$ 2 & 0.49 $\pm$ 0.10 & 0.69 $\pm$ 0.14 & 110 $\pm$ 10 & 8 $\pm$ 2&$-$\\ 
		UGC\,6176 & Einasto & 35 $\pm$ 5 & 56 $\pm$ 2 & 0.48 $\pm$ 0.10 & 0.70 $\pm$ 0.15 & 102 $\pm$ 9 & 10 $\pm$ 3 & 0.17 $\pm$ 0.05\\ 
		& DC14 & 40 $\pm$ 6 & 56 $\pm$ 2 & 0.47 $\pm$ 0.10 & 0.69 $\pm$ 0.14 & 109 $\pm$ 26 & 10 $\pm$ 4&$-$\\
		\hline
	\end{tabular}
	\label{Tab: DM fitting params U}
\end{table*}

\begin{figure*}[!]
\centering
\includegraphics[width=0.95\textwidth]{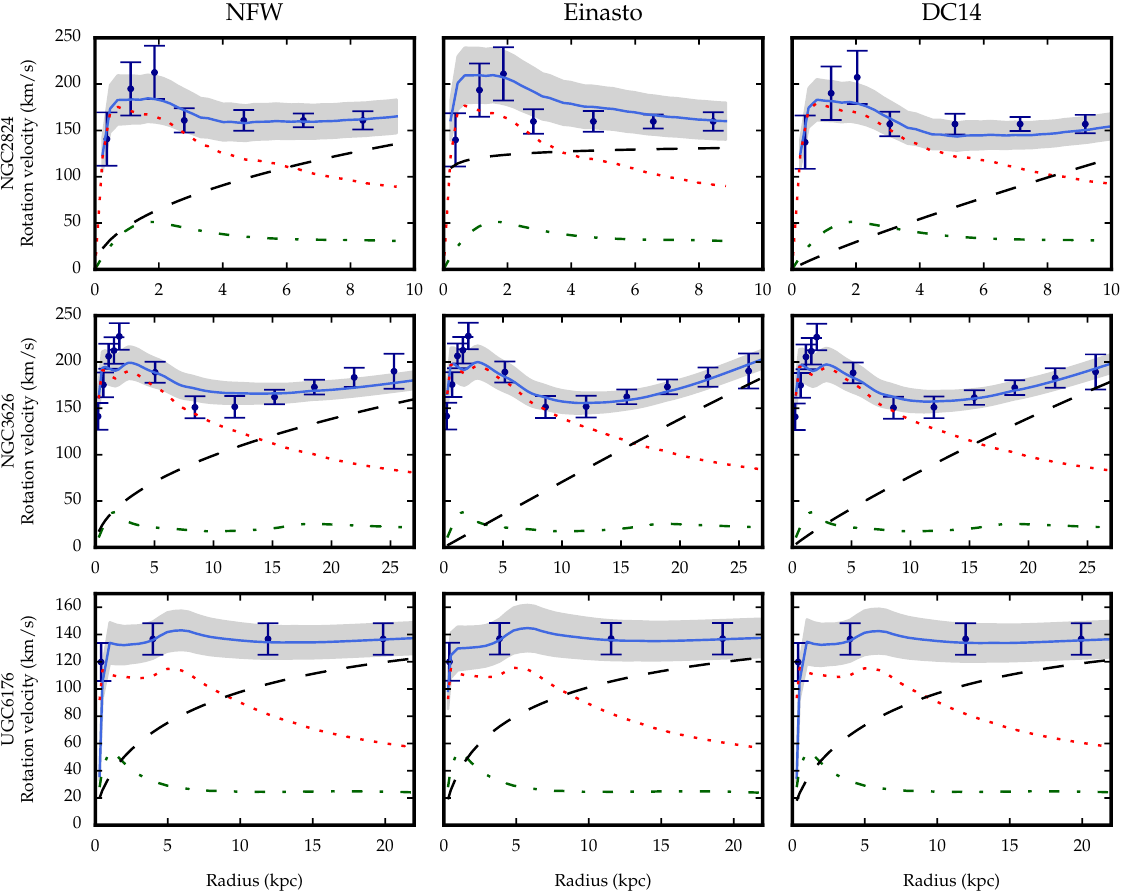}
	\caption{Same as Fig. \ref{Fig: summary DM}, without cosmological priors imposed on the DM halo parameters in the fitting procedure.}
	\label{Fig: summary DM flat}
\end{figure*}

\subsection{$\Lambda$CDM priors} \label{sec: LCDM priors}

Rotation-curve fits with DM halos are subject to a number of degeneracies due to the significant number of fitting parameters. The availability of Spitzer [3.6] photometry greatly alleviates the so-called disk-halo degeneracy \citep{Albada1985}, but additional ``priors'' in a Bayesian context are needed to fully break the problem degeneracies. Thus, it is sensible to constrain the free parameters to domains that are in agreement with cosmological expectations. Specifically, we impose two relations expected from $\Lambda$CDM cosmological simulations, following the same procedures as in \cite{LiDM, Li20}.

\cite{DuttonMaccio} used cosmological DM-only simulations to study the relation between the mass ($M_{200}$) and concentration ($C_{200}$) of the halo. This is given by
\begin{equation}
\log{c_{200}} = a - b \log{(M_{200}/[10^{12}h^{-1}M_\odot])},
\end{equation}
with an intrinsic scatter of 0.11 dex. Here $a$ and $b$ are cosmology-dependent parameters, while $h = H_0/100$ where $H_0$ is the Hubble constant. We assume $H_0 = 73$ km s$^{-1}$ Mpc$^{-1}$ to be consistent with the local SPARC distance scale \citep{SPARC}.

\cite{Moster} derived a stellar-to-halo mass relation by matching the number of observed galaxies in the local Universe to the predicted number of DM halos from N-body simulations. The functional form of the stellar-to-halo-mass (SHM) relation is assumed to be
\begin{equation} 
\frac{M_\star}{M_{200}} = 2N \Big[ \Big(\frac{M_{200}}{M_1}\Big)^\beta 
+ \Big(\frac{M_{200}}{M_1}\Big)^\gamma \Big]^{-1},
\end{equation}
where $\log(M_1) = 11.59$, $N = 0.0351$, $\beta = 1.376$ and $\gamma = 0.608$.

Without imposing these two relationships, the parameters of the DM halos are poorly constrained (see Appendix \ref{App:DM}), so in the following we will mostly focus on rotation-curve fits with log-normal $\Lambda$CDM priors.

\subsection{Fitting procedure}

To fit the DM halos, we follow \cite{LiDM, Li20} and use their Markov-Chain-Monte-Carlo (MCMC) code, which maps the posterior distributions of the fitting parameters using the python package emcee \citep{emcee}. The code is supplied the observed rotation curve and the circular velocities of the individual baryonic components. The free parameters are the halo mass and halo concentration (plus the shape parameter $\alpha_{\epsilon}$ for the Einasto profile), the stellar mass-to-light ratios, the disk inclination ($i$), and the galaxy distance ($D$). For convenience, we perform the fits using $V_{200} = [10\,G\,H_0 M_{200}]^{1/3}$ \citep{McGaugh2012}, where $G$ is the gravitational constant. We assume Gaussian priors on $i$ and $D$ centered at the estimated values and with standard deviations given by the estimated uncertaintes. Since our galaxies show radial variations of $i$, we assume that the inclination uncertainty is equal to half of the total variation. We adopt log-normal priors on the mass-to-light ratios, centered at 0.5 for the disk and 0.7 for the bulge with a standard deviation of 0.1 dex.

\begin{figure*}[!]
	\centering
	\includegraphics[width=0.9\textwidth]{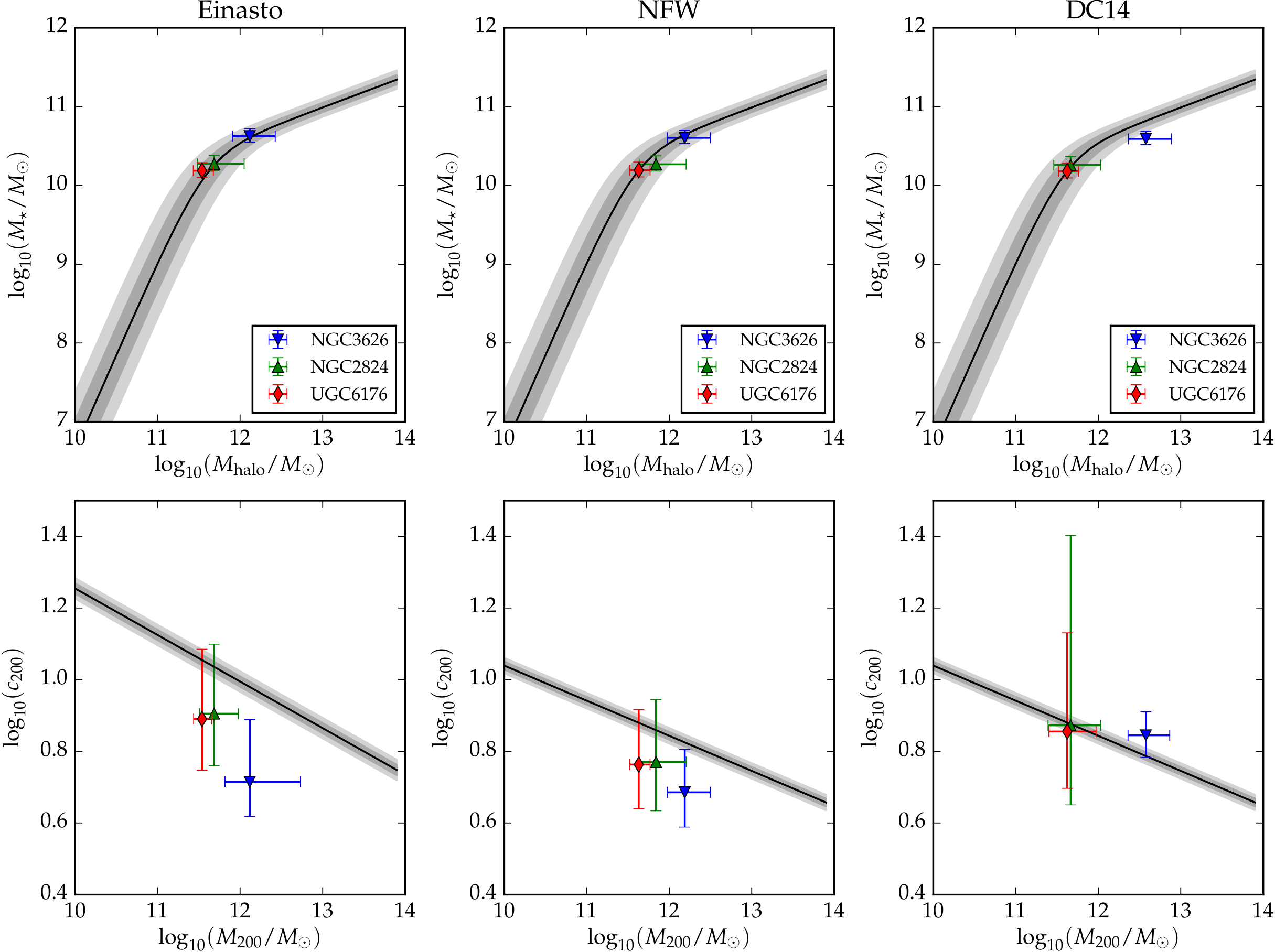}
	\caption{Stellar-to-halo-mass relation (top) and halo mass-concentration relation (bottom). The black solid line and the grey shaded areas show the relations expected in $\Lambda$CDM together with their intrinsic scatter \protect\citep{Moster, DuttonMaccio}. The solid points with $\pm$1 sigma errorbars show the maximum-probability values from rotation-curve fits with $\Lambda$CDM priors.}
	\label{Fig: LCDM priors}
\end{figure*}
For the halo parameters, we use loose boundaries to keep them in a physically sensible range: $10 < V_{200} < 500$ km/s, $0 < c_{200} < 100$, and for the Einasto profile $0 < \alpha_{\epsilon} < 2$. When we apply the $\Lambda$CDM priors, we need to rescale the concentration of the DC14 profile since it is defined in a slightly different way from that of the NFW and Einasto profiles \citep{DiCintio}. Following \citet{Katz}\footnote{There is a typo in Eq.(3) of \citet{Katz} since the factor 10$^{-5}$ appears inside the exponential. The equation written here is the correct one, which was used in the analysis of \citet{Katz}.}, we adopt
\begin{equation}
c_{200}^\text{rescaled} = c_{200}^\text{DC14}\times\big[1 + 10^{-5}~\exp(3.4 X+ 15.3)\big]^{-1}.
\end{equation}
To impose the mass-concentration relation, we adopt $a = 0.830$ and $b = -0.098$ for the NFW and DC14 profiles, assuming a WMAP5 cosmology. This is the best compromise between the WMAP3 cosmology assumed in the simulations of \cite{DiCintio} and the WMAP7 cosmology assumed in the SHM relation of \cite{Moster}. However, as stated in \cite{LiDM}, the difference induced by different cosmological parameters is nearly negligible and will mostly play a role in the final distance estimates. For the Einasto profile, we adopt $a = 0.977$ and $b = -0.130$ assuming a Planck cosmology since this is the only available calibration. We also impose an additional constraint on $\alpha_\epsilon$ describing the dependence of this parameter on halo mass. According to \cite{DuttonMaccio}, this is given by
\begin{equation}
\alpha_\epsilon = 0.0095\nu^2 + 0.155,
\end{equation}
with $\log\nu = -0.11+0.146m+0.0138m^2+0.00123m^3$, $m = \log(M_{\text{halo}}/10^{12}h^{-1}M_\odot)$, and a vertical scatter of 0.16 dex. 

For each fit, we perform 2000 burn-in steps and 1000 iterations after that. Then, we re-initialize the walkers around the maximum-probability values and perform 1000 more iterations. We visually checked that the chains have converged. In cases when the posterior distributions have complex shapes, we checked that changing the number of the burn-in steps and/or walkers does not improve the results. This generally happens for the posterior distributions of fits without $\Lambda$CDM priors. Imposing $\Lambda$CDM priors breaks intrinsic degeneracies between halo parameters, leading to better behaved posterior distributions (see Appendix \ref{App:DM}). 

\subsection{Fitting results}

Figures \ref{Fig: summary DM} and \ref{Fig: summary DM flat} illustrate that the observed rotation curves are reproduced at both small and large radii. All fits return close to maximal stellar disks with $\Upsilon_{\rm d}\simeq0.5$ $M_\odot/L_\odot$. The inner galaxy regions are largely dominated by baryons. Table \ref{Tab: DM fitting params U} gives the maximum-probability values for the fitting parameters, while Appendix \ref{App:DM} illustrates their full posterior distributions.

When the $\Lambda$CDM priors are imposed, different halo profiles give nearly indistinguishable results for all galaxies. When we adopt uniform priors on the halo parameters, however, we find significant differences for NGC\,3626. For this object, the rotation curve displays a strong decline at intermediate radii driven by the stars, followed by a nearly solid-body rise driven by the halo. The latter cannot be reproduced by a cuspy DM halo, like the NFW or the Einasto and DC14 with $\Lambda$CDM constraints, but it is well described by a cored DM halo, like Einasto with low $\alpha$ values and DC14 with intermediate $X$ values. Thus, the rotation curve of NGC\,3626 suggests that DM cores are not only a prerogative of low-mass galaxies, but they may exist in high-mass galaxies too, providing a challenge for current stellar-feedback models \citep{DiCintio} and possibly pointing to black-hole feedback shaping the inner DM halo \citep{Maccio2020}.

Figure \ref{Fig: LCDM priors} (top panels) illustrates that the maximum-probability halo parameters adhere well to the SHM relation \citep{Moster} once it is imposed as a lognormal prior. For the DC14 profile, however, NGC\,3626 is slightly offset to the right of the SHM relation because its rotation curve is pushing the fit towards low $M_\star/M_{\rm 200}$ values in order to have a less cuspy DM profile. Figure \ref{Fig: LCDM priors} (bottom panels) shows that the halo parameters are also consistent with the imposed mass$-$concentration relation \citep{DuttonMaccio} within the large uncertainties.

\section{Discussion}\label{sec: Discussion}

\subsection{Radial acceleration relation}

LTGs follow a tight radial acceleration relation (RAR; \citealp{McGaugh2016}, \citealp{RAR}): the observed centripetal acceleration at each radius correlates with that predicted by the baryonic components. This implies that, to a first approximation, the rotation-curve shape can be predicted by the baryon distribution, and vice versa. The scatter around the RAR is remarkably small for astronomical standards and is largely dominated by observational scatter, leaving little space for intrinsic scatter in the relation (\citealp{Li18}, but see also \citealp{Stone}). In addition to the LTGs from the SPARC sample, \cite{RAR} also considered 25 ETGs and 62 dwarf spheroidals, finding that they follow the same relation as LTGs (with the possible exception of ultra-faint dwarfs for which the observational situation is more uncertain). The RAR might represent a universal law of nature with severe implications on cosmological models (see \citealt{RAR} for a discussion).

\begin{figure}[!]
	\centering
	\includegraphics[width=0.45\textwidth]{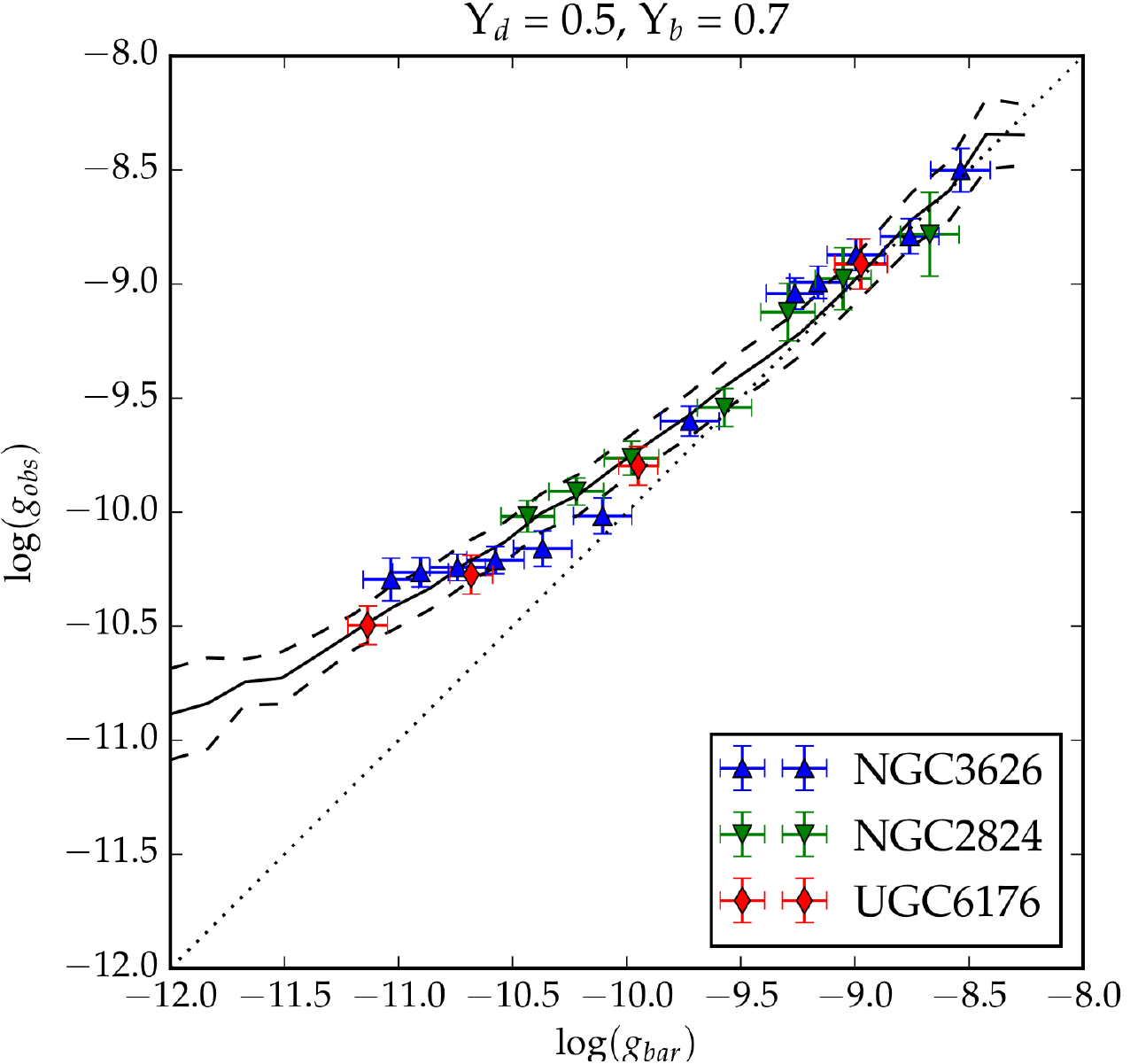}                                                                                                                                        
	\caption{The RAR for our ETGs using $\Upsilon_{\rm d} =0.5$ and $\Upsilon_{\rm b}=0.7$ as in \protect\citet{RAR}. Black and dashed lines correspond, respectively, to the mean and standard deviation of the RAR for LTGs in the SPARC database. The dotted line shows the 1:1 relation.}
	\label{Fig: RAR}
\end{figure}
ETGs are the least represented galaxy type in the RAR. The sample of \cite{RAR} included 9 so-called slow rotators and 16 fast rotators (\citealp{Cappellari11}). The slow rotators are generally giant boxy ellipticals, in which the observed centripetal acceleration can be measured using X-ray halos in hydrostatic equilibrium \citep{Buote}. The fast rotators comprise both disky ellipticals and lenticular galaxies: their inner circular velocities were estimated fitting Jeans axisymmetric models to the stellar kinematics \citep{Cappellari11}, while the outer circular velocities were measured using H{\footnotesize I} observations \citep{denHeijerETGs}. Therefore, the mass models of these fast rotators were limited to two measured points only. Having derived full, extended rotation curves for rotating ETGs, we can more rigorously investigate their location on the RAR.

\begin{figure}[!] 
	\centering
	\includegraphics[width= 0.5\textwidth]{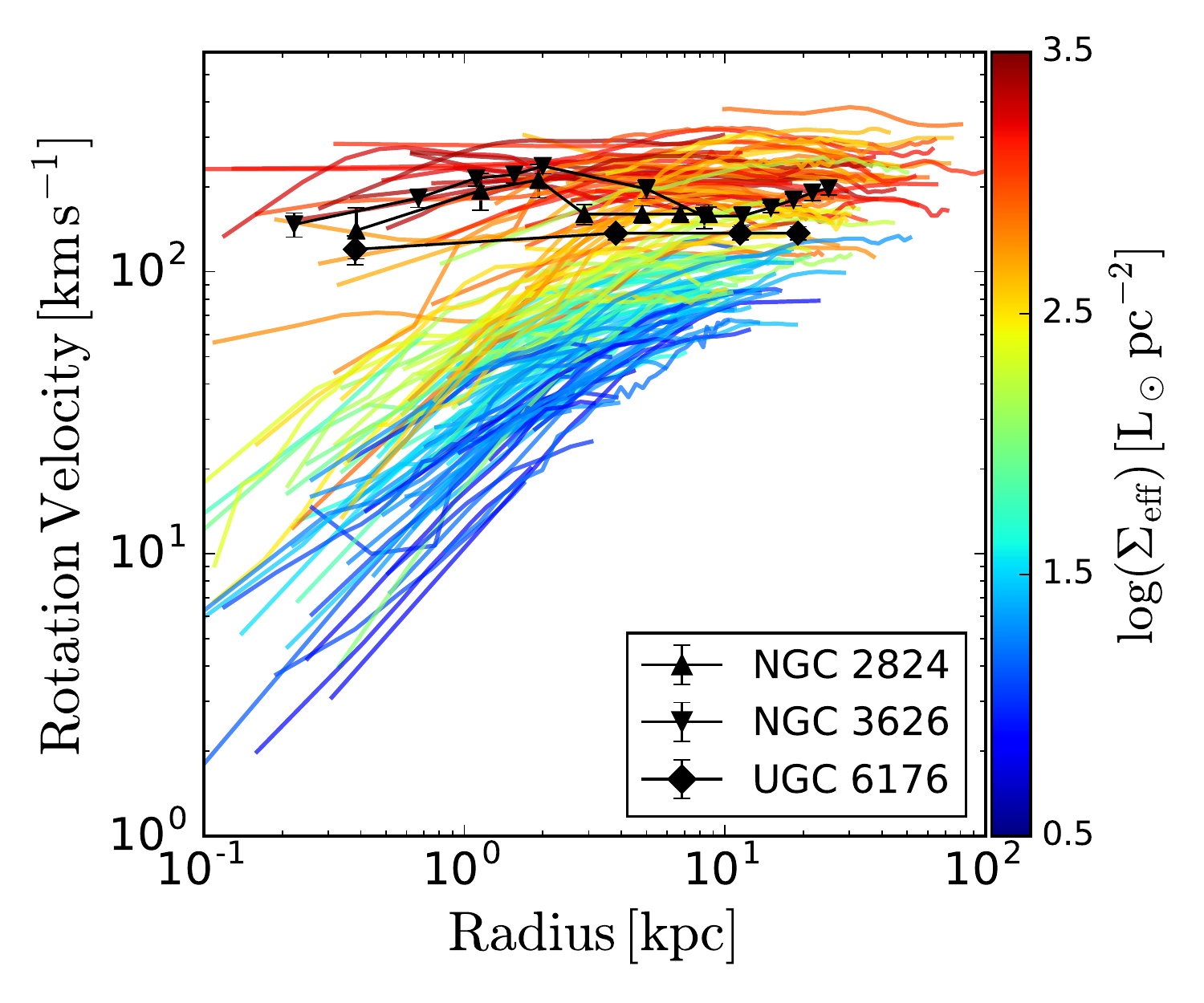}
	\caption{Rotation curves of LTGs from the SPARC sample \protect\citep{SPARC} overlaid with the rotation curves of our ETGs. The LTG rotation curves are color-coded by their effective surface brightness at 3.6 $\mu$m. The location of the ETGs is consistent with their high effective surface brightnesses of $\sim$10$^{3}$ $L_\odot$ pc$^{-2}$.}
	\label{Fig: ETGs on SPARC}
\end{figure}
Following \cite{RAR}, we calculate the observed centripetal acceleration as $g_{\text{obs}} = v^2_{\text{obs}}(R)/R$ and the one predicted from the baryon distribution as $g_{\text{bar}} = v^2_{\text{bar}}(R)/R$ with $\Upsilon_{\rm d} = 0.5$ and $\Upsilon_{\rm b} = 0.7$. Figure \ref{Fig: RAR} shows that our ETGs lie on top of the RAR derived for LTGs within 1$\sigma$. Although our sample is too small to make statistically robust conclusions, this result suggests that galaxies follow the same RAR independently of their morphology and evolutionary stage. Thus, the rotation curves of ETGs can be predicted from the baryonic mass distribution using the same relation as for LTGs. This points to the possible universality of the RAR, in line with the predictions of Milgromian dynamics \citep[MOND,][]{MOND}. A bigger ETG sample with both CO and H{\footnotesize I} data is needed to reinforce this result.

\subsection{Evolution of early-type galaxies}

The dichotomy of early-type galaxies, dividing them into slow and fast rotators \citep[e.g.][]{Emsellem}, is well established but the formation and evolution of the two galaxy types is subject of debate. The galaxies studied in this paper are rotating lenticulars, so we focus on their formation and evolution. 

Several authors (e.g., \citealp{KormendyBender12}; \citealp{Laurikainen}) proposed that lenticulars evolve from spirals through secular quenching of star formation, possibly driven by gas stripping and/or starvation (i.e., lack of gas accretion). These conclusions are motivated by the fact that the stellar disks and bulges of S0s follow the same photometric scaling relations as those of spirals. The fading scenario is backed from comparative population studies, showing that the fraction of LTGs decreases as a function of environmental density with nearly the same slope as the fraction of rotating ETGs increases \citep{Cappellari11}. Numerical simulations, however, show that lenticulars formed via major mergers may have coupled bulge-disk components that follow the same scaling relations as spirals \citep{Querejeta}, thus photometric data alone may not be sufficient to distinguish between secular fading and major mergers. In this evolutionary context, kinematical and dynamical studies acquire primary importance.
	
\begin{figure*}[!]
	\centering
	\includegraphics[width = 0.9\textwidth]{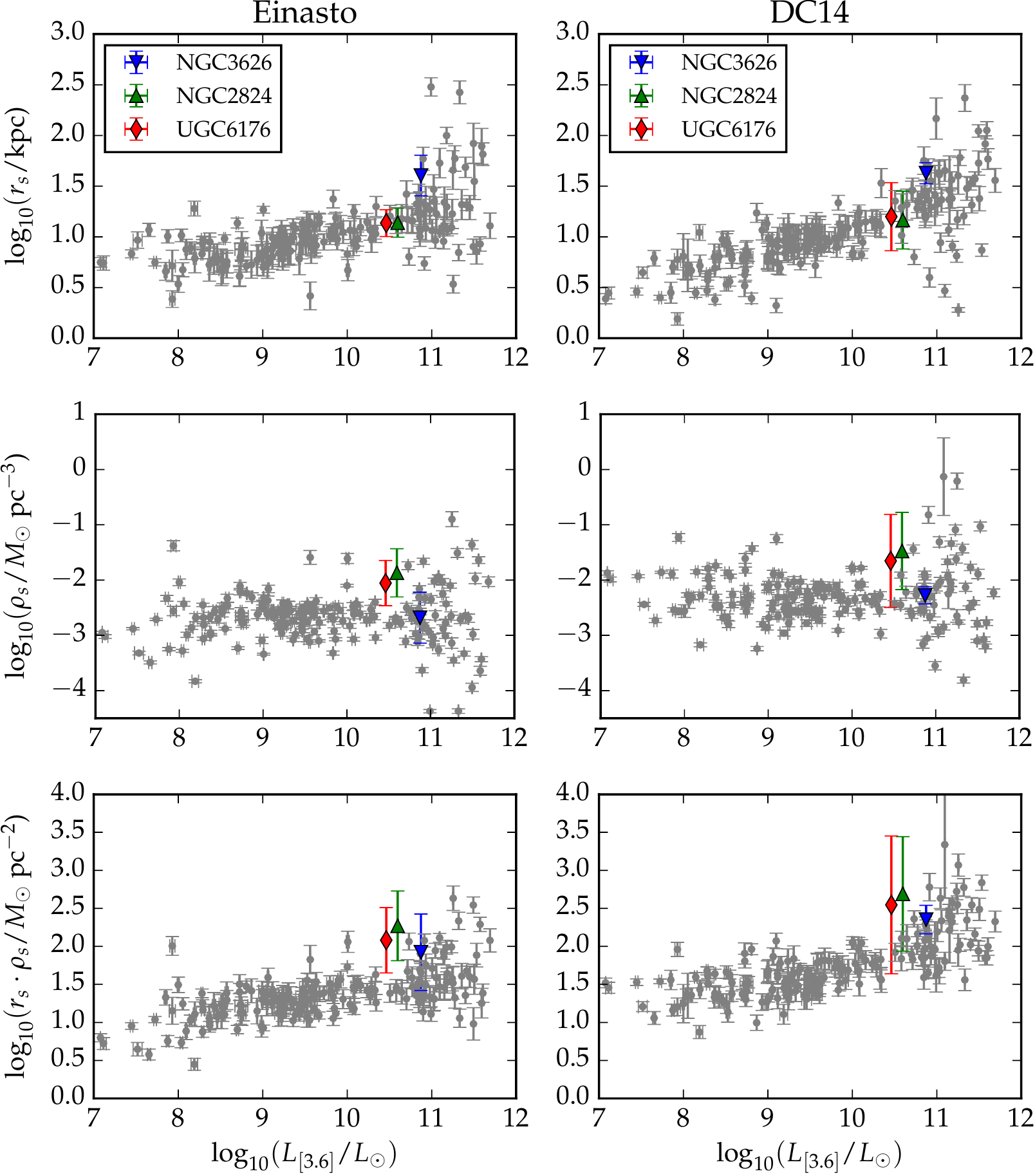}
	\caption{DM halo parameters versus [3.6] luminosity for our three ETGs and the LTGs in the SPARC sample \protect\citep[grey dots from][]{LiDM}: halo scale radius (top panels), characteristic volume density (middle panels), and their product (bottom panels).}
	\label{Fig: comparison to Li}
\end{figure*}

\cite{Rizzo18} performed kinematic bulge-disk decompositions of unbarred S0s, separating the light contributions from bulge and disk at each spatial position of the optical spectrum. They found that the stellar disks of S0s have the same specific angular momentum as those of spirals, concluding that major mergers are not the dominant formation mechanisms for S0s in the group and field environments. Diametrically different conclusions are drawn by \cite{Falcon}, which, however, did not perform a kinematic bulge-disk decomposition of optical spectra but included bulges in their estimate of the specific stellar angular momentum. The results of our work, based on gas kinematics that is exempt from highly non-trivial spectral decompositions, give further support to the scenario in which S0s are forming from aging spiral galaxies in a fluent transition.

Figure \ref{Fig: ETGs on SPARC} compares the rotation curves of ETGs with those of LTGs from SPARC \citep{SPARC}. The rotation-curve shapes of S0s are similar to spirals of comparable effective surface brightness and flat rotation velocity, hence of comparable baryonic mass (stars plus gas) since HI-rich ETGs follow the same baryonic Tully-Fisher relation as LTGs \citep{denHeijerETGs}. This implies that the shape of the total gravitational potential is similar in spirals and lenticulars. Moreover, the amount of specific angular momentum in the gas disk of S0s is clearly similar to that of spirals, in analogy to the specific angular momentum in the stellar disk \citep{Rizzo18}. Finally, our values for the mass-to-light ratios (0.5$-$0.7 $M_\odot/L_\odot$ at 3.6 $\mu$m) are lower than expected for old and passive ETGs with exponentially declining star-formation histories (0.8$-$1.0 $M_\odot/L_\odot$, see Fig. 11 in \citealt{Schombert14}), but are in the typical range of star-forming LTGs \citep{SchombertMcGaughLelli}. This suggests that the star-formation activity in our three ETGs must have dropped (if not stopped) only recently, which allows us to classify them as faded spirals.

Our findings are consistent with those of \cite{Yildiz}, who used a sample of H{\footnotesize I}-rich and H{\footnotesize I}-poor ETGs from the ATLAS$^{3\text{D}}$ survey to study their star-formation properties. They compared the UV colors of the two samples and found that H{\footnotesize I}-rich ETGs are bluer than H{\footnotesize I}-poor ETGs. The most gas-rich ETGs are as blue as the outer parts of typical LTGs. \cite{Yildiz} concluded that gas-rich ETGs host young stellar populations in their outskirts. They also found evidence of low star formation efficiency in these regions. Although our galaxies are not part of Yildiz's sample, their H{\footnotesize I} mass places them among the gas-rich ETGs (NGC\,2824 was not considered by \citealt{Yildiz} due to the weak H{\footnotesize I} detection in the Atlas$^{\rm 3D}$ data, while NGC\,3626 and UGC\,6176 had no available UV observations). NGC\,3626 likely has some ongoing star formation, possibly triggered by a recent interaction with its companion galaxy, as suggested by the stellar and gas rings visible in its outer disk.

Further support for a faded-spiral scenario comes from comparing our derived DM halo parameters to the ones for the LTGs in the SPARC sample \citep{LiDM}. This is shown in Fig. \ref{Fig: comparison to Li}, where the halo scale length and characteristic densities are plotted versus the galaxy luminosity at 3.6 $\mu$m. The DM halos of our lenticular galaxies follow the same relations of SPARC galaxies within the errors. Thus, these ETGs live in similar DM halos as LTGs, indicating that they may be fading spirals. This also suggests that the evolution from spirals to lenticulars do not necessarily involve major structural changes in the DM halo, as one may expect after a major merger. Indeed, we can exclude major mergers in the recent past of our galaxies ($1-2$ Gyr). The orbital times in their outermost regions range from 0.3 to 0.8 Gyr: since the velocity fields are reasonably symmetric, the gas disk must have remained relatively unperturbed for several orbital times.

\section{Conclusion}

We analyzed the dynamics of three S0s that have inner CO emission and outer, extended H{\footnotesize I} disks. This special feature allowed us to trace the rotation curve from the innermost parts of the galaxies out to 10-20 effective radii. We derived mass models using Spitzer photometry at 3.6 $\mu$m and performed DM halo fits. Then, we compared the properties of our ETGs with those of LTGs from the SPARC sample to constrain the evolutionary history of lenticular galaxies. We found the following results:
   \begin{enumerate}
      \item The rotation-curve shapes of our ETGs are similar to those of LTGs of similar mass and surface brightness. This is consistent with a smooth transition from spirals to lenticulars.
      \item The dynamically inferred values of the stellar mass-to-light ratio are relatively small for passive ETGs. The retrieved values are similar to those of LTGs (0.5$-$0.7 $M_\odot/L_\odot$). This suggests recent star formation events in these galaxies, consistent with the picture of spirals slowly fading into lenticulars.
      \item The DM halo parameters follow the same scaling relations with galaxy luminosity as those of LTGs. This suggests that the transition from LTGs to ETGs happened without altering the halo structure of the galaxies, e.g., via a violent event like a major merger.
      \item The rotation curve of NGC\,3626 has a significant inner decline driven by the stars and a subsequent rise driven by the halo, which is poorly fitted by cuspy DM profiles. This suggests that DM cores may exist in high-mass galaxies.
      \item We confirm that ETGs follow the same radial acceleration relation as LTGs, pointing to the universality of this relation for galaxies of different morphologies and evolutionary stages. 
   \end{enumerate}

Our results paint a consistent picture of the evolution of lenticular galaxies, which is in agreement with previous works. A bigger sample of gas-bearing ETGs is needed to make more general statements. Such a sample might be accessible in the near future thanks to new CO surveys with ALMA and NOEMA, as well as H{\footnotesize I} surveys with the SKA and its pathfinders.

\begin{acknowledgements}
We thank the anonymous referee for a constructive report that improved the clarity of this paper.
A.S. is grateful for the hospitality and financial support of Harvard University and the European Southern Observatory (Germany Headquarters), where most of this work was carried out. We are in debt to Paolo Serra for providing us with a deep H{\footnotesize I} cube of NGC\,2824, Pengfei Li for sharing his rotation-curve fitting codes, and Enrico di Teodoro for crucial help with $^{\text{3D}}$Barolo. We also thank Ralph Bender, Michele Cappellari, Timothy Davis, Stacy McGaugh, and James Schombert for useful discussions. We acknowledge the Atlas$^{\rm 3D}$ survey team for making their datacubes publicly available. This research has made use of the NASA/IPAC Extragalactic Database (NED) which is operated by the Jet Propulsion Laboratory, California Institute of Technology, under contract with the National Aeronautics and Space Administration.
\end{acknowledgements}

%

%

\bibliography{biblist.bib}


\begin{appendix} 
\section{Notes on individual galaxies} \label{App: Atlas}
In this Appendix, we present in detail the results of our 3D kinematic modeling with GALMOD and $^\text{3D}$BAROLO (see Sect.\,\ref{sec: rot curves}). For each galaxy, we describe intensity maps, position-velocity (PV) diagrams, and channel maps for both H{\footnotesize I} and CO emissions. The figures are structured in a uniform way.
\subsection{NGC\,2824} \label{sec: NGC2824}

\graphicspath{{./Pictures/}}

\begin{figure*}[!]
	\begin{center}
		\includegraphics[width = \textwidth]{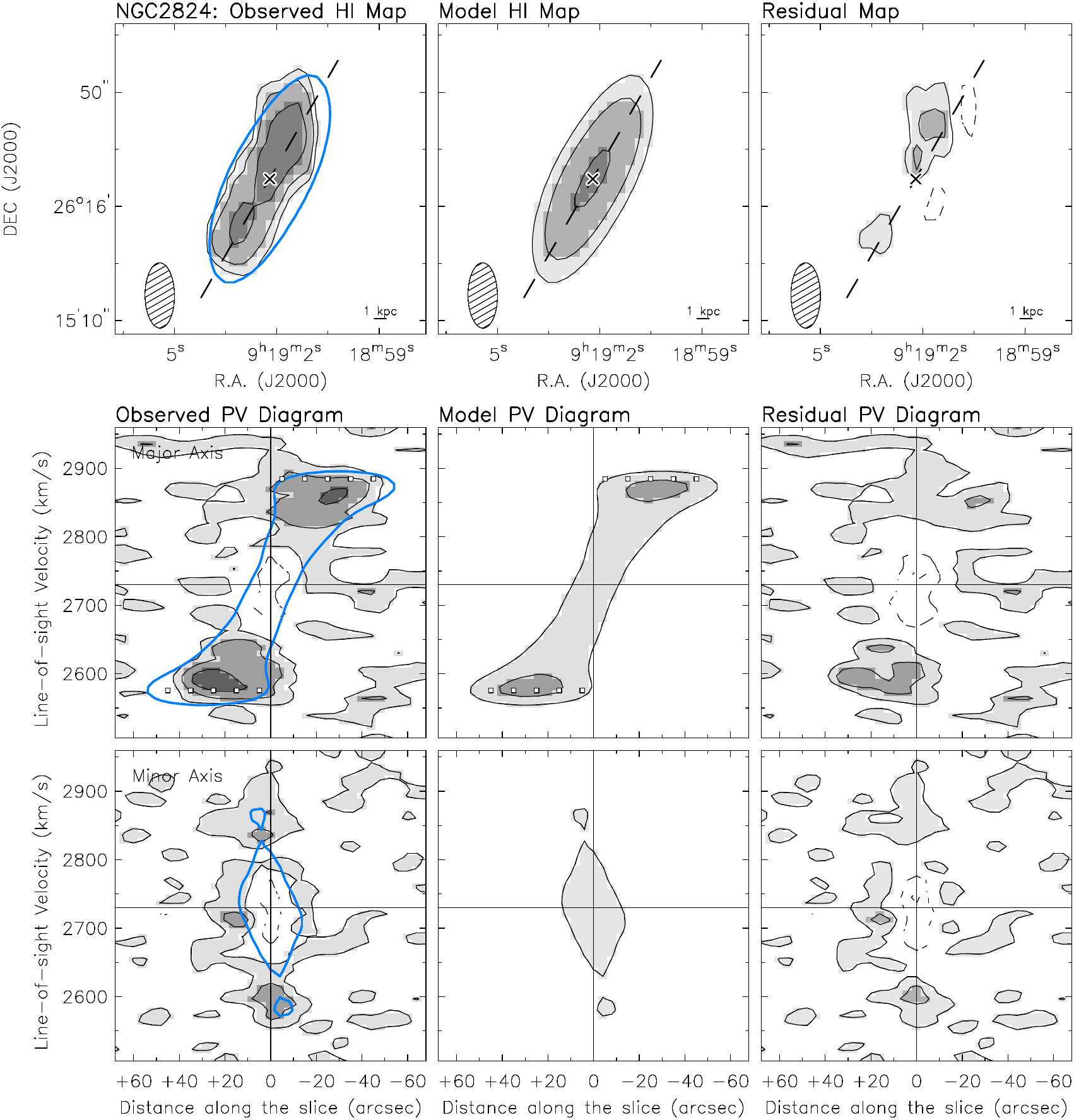}
	\end{center}
	\caption{Top panels: H{\footnotesize I} intensity maps from the observed cube (left), model cube (middle) and residual cube (right) of NGC\,2824. The cross and the dashed line indicate the galaxy center and position angle, respectively. The beam is shown in the bottom left corner. The physical scale is illustrated in the bottom right corner. The contours correspond to 3$\sigma_{\text{pseudo}}$, 6$\sigma_{\text{pseudo}}$, and so on. The outermost contour of the model is plotted in blue over the observed intensity map for comparison. Bottom panels: position-velocity diagrams taken along the major and minor axes from the observed cube (left), model cube (middle) and residual cube (right). The white squares show our derived rotation curve. The outermost contour of the model is plotted in blue over the observed PV diagrams for comparison. The density contours in the PV diagrams correspond to -1.5$\sigma$, -3$\sigma$ (dashed) and 1.5$\sigma$, 3$\sigma$, 6$\sigma$, and so on (solid), where $\sigma$ is the measured noise in the cube (see Tab. \ref{Tab: cube params}).}
	\label{N2 HI}
\end{figure*}

\begin{figure*}[!]
	\begin{center}
		\includegraphics[width = \textwidth]{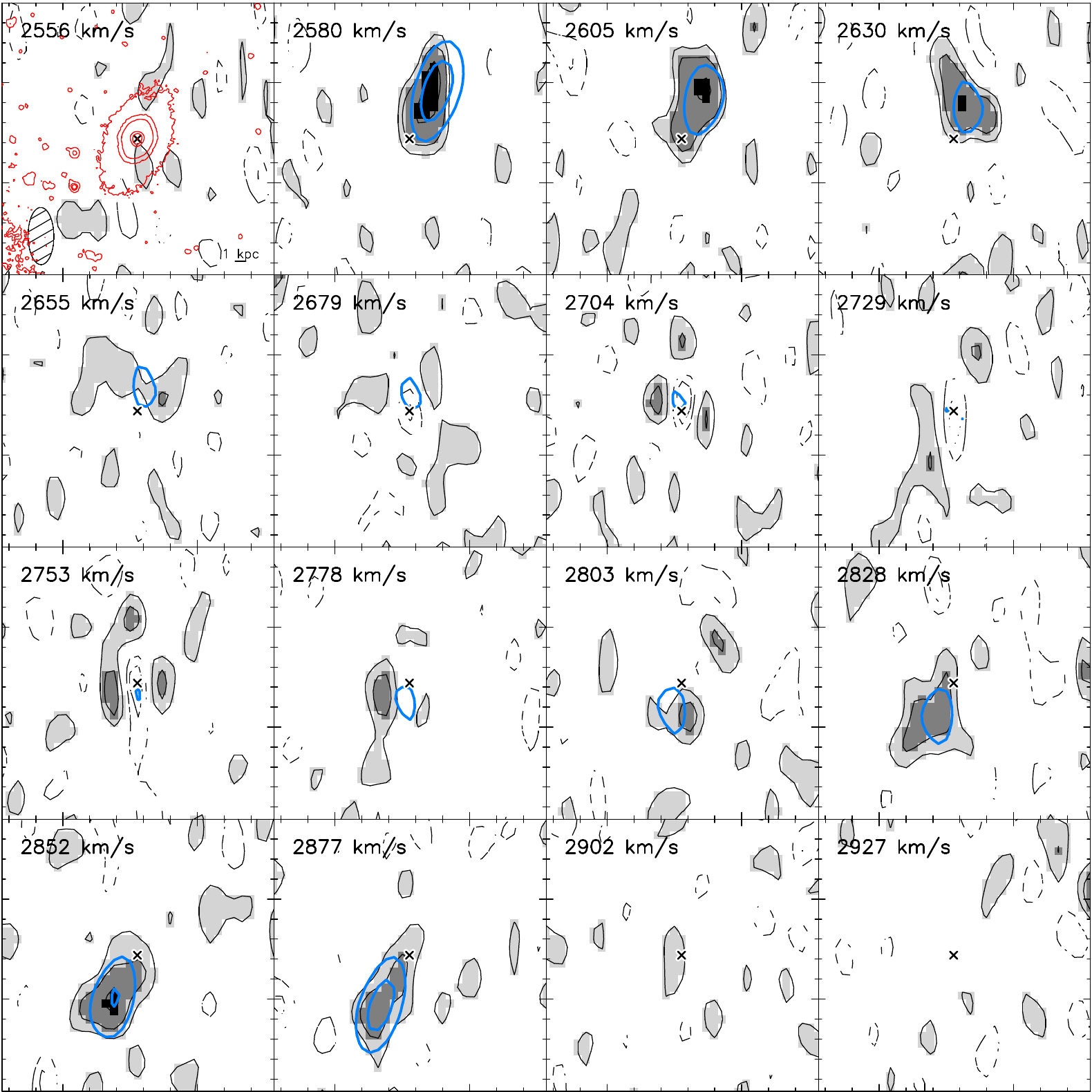}
	\end{center}
	\caption{H{\footnotesize I} channel maps for NGC\,2824 from the observed cube (grey scale) and the model cube (blue contours). The line-of-sight velocity is shown in the upper left corner. The cross indicates the galaxy center. The contours correspond to -3$\sigma$, -1.5$\sigma$ (dashed) and 1.5$\sigma$, 3$\sigma$, 6$\sigma$, and so on (solid). In the first panel, the red contours correspond to arbitrary isophotes of the Spitzer image at 3.6 $\mu$m, the beam is shown in the bottom-left corner, and the physical scale is given in the bottom-rigth corner.}
	\label{N2 HI chan}
\end{figure*}

\begin{figure*}[!] 
	\centering
	\includegraphics[width = \textwidth]{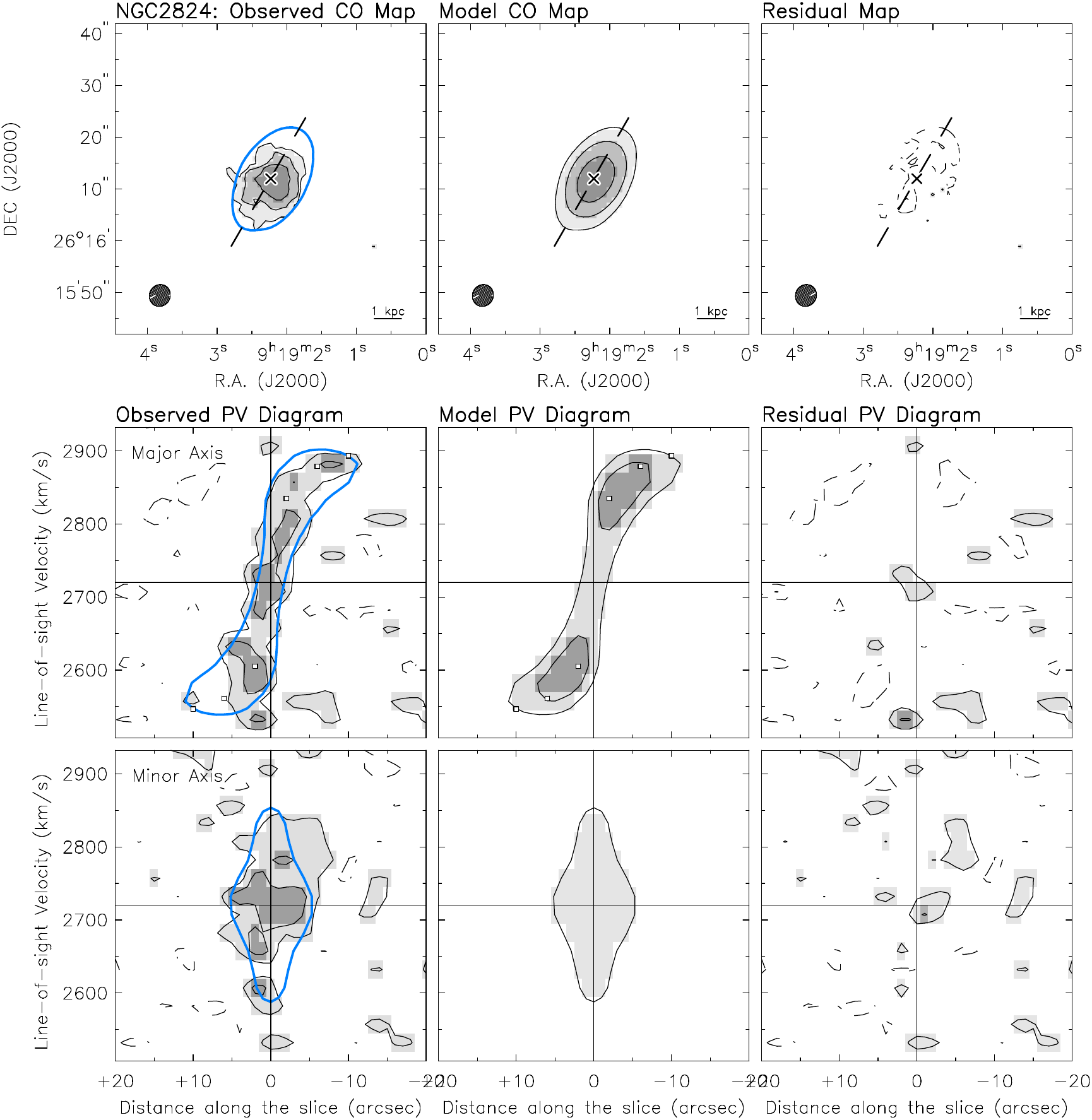}
	\caption{Same as Fig. \ref{N2 HI}, but for the CO data of NGC\,2824.}
	\label{N2 CO}
\end{figure*}

\begin{figure*}[!] 
	\begin{center}
		\includegraphics[width = \textwidth]{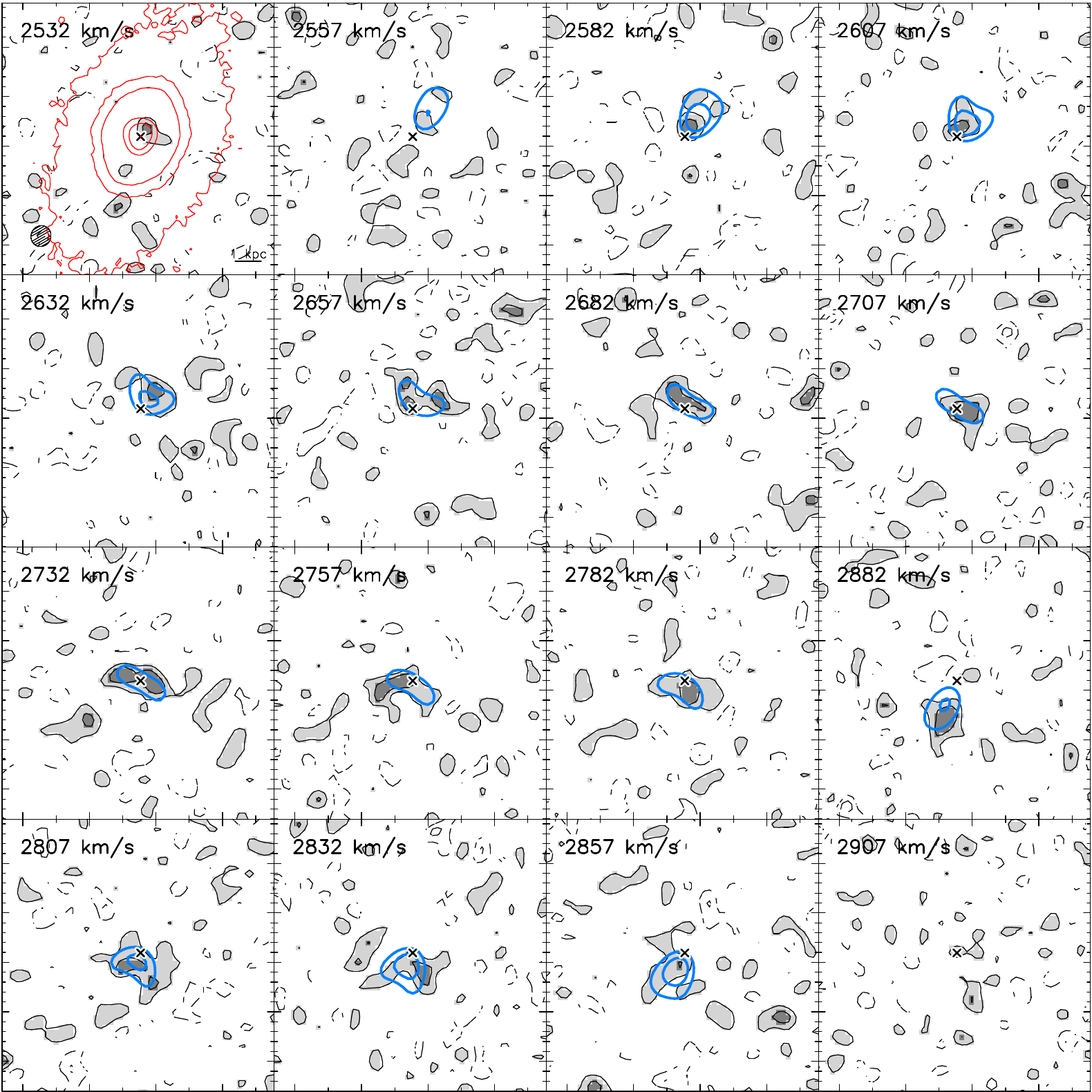}
	\end{center}
	\caption{Same as Fig. \ref{N2 HI chan}, but for the CO data of NGC\,2824.}
	\label{N2 CO chan}
\end{figure*}

The observed H{\footnotesize I} intensity map is well reproduced by our model (Fig.\,\ref{N2 HI}, top panels) apart from an over-denstiy in the northern half of the disk, visible in the residual map, that cannot be reproduced by an axisymmetric disk. The observed PV diagrams display negative signal towards the galaxy center, associated to H{\footnotesize I} absorption, that our model cannot reproduce (Fig.\,\ref{N2 HI}, bottom panels). Apart from these shortcomings, the H{\footnotesize I} kinematics are well reproduced. This is confirmed by Fig.\,\ref{N2 HI chan} comparing H{\footnotesize I} channel maps from the observed and model cubes.

The observed CO density maps indicates that the molecular disk is lopsided, extending farther towards the approaching part of the galaxy. This feature cannot be fully reproduced by our axisymmetric model (Figure \ref{N2 CO}, top panels). This also causes asymmetries in the PV diagrams (Figure \ref{N2 CO}, bottom panels). In our modeling, we give more weight to the extended, approaching half of the CO disk. Figure\,\ref{N2 CO chan} shows that the observed CO channel maps are well reproduced by our model.

\subsection{NGC\,3626} \label{sec: NGC3626}


\begin{figure*}[!]
	\begin{center}
		\includegraphics[width = \textwidth]{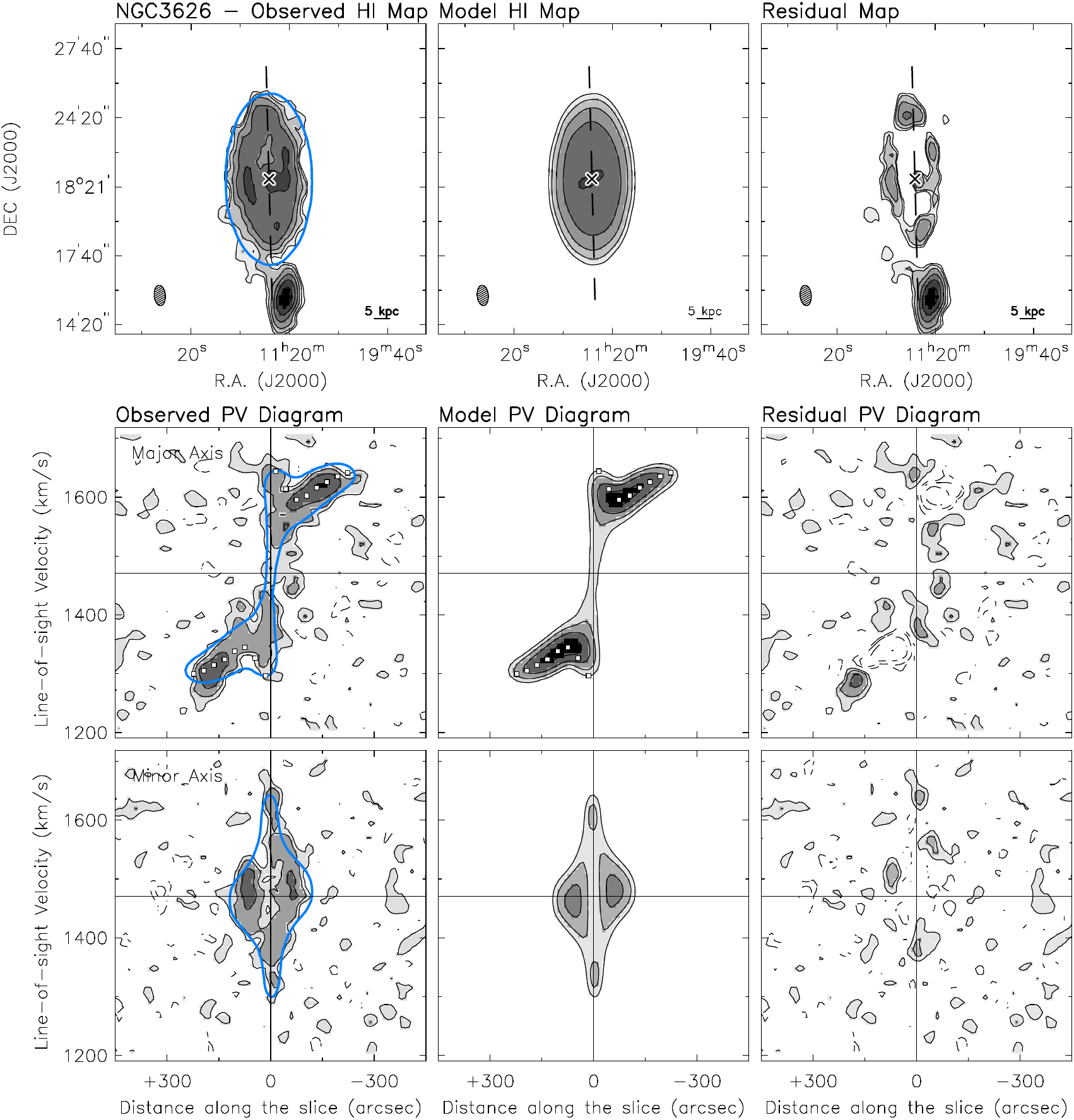}
	\end{center}
	\caption{Same as Fig. \ref{N2 HI}, but for the H{\footnotesize I} data of NGC\,3626.}
	\label{N3 HI}
\end{figure*}

\begin{figure*}[!]
	\begin{center}
		\includegraphics[width = \textwidth]{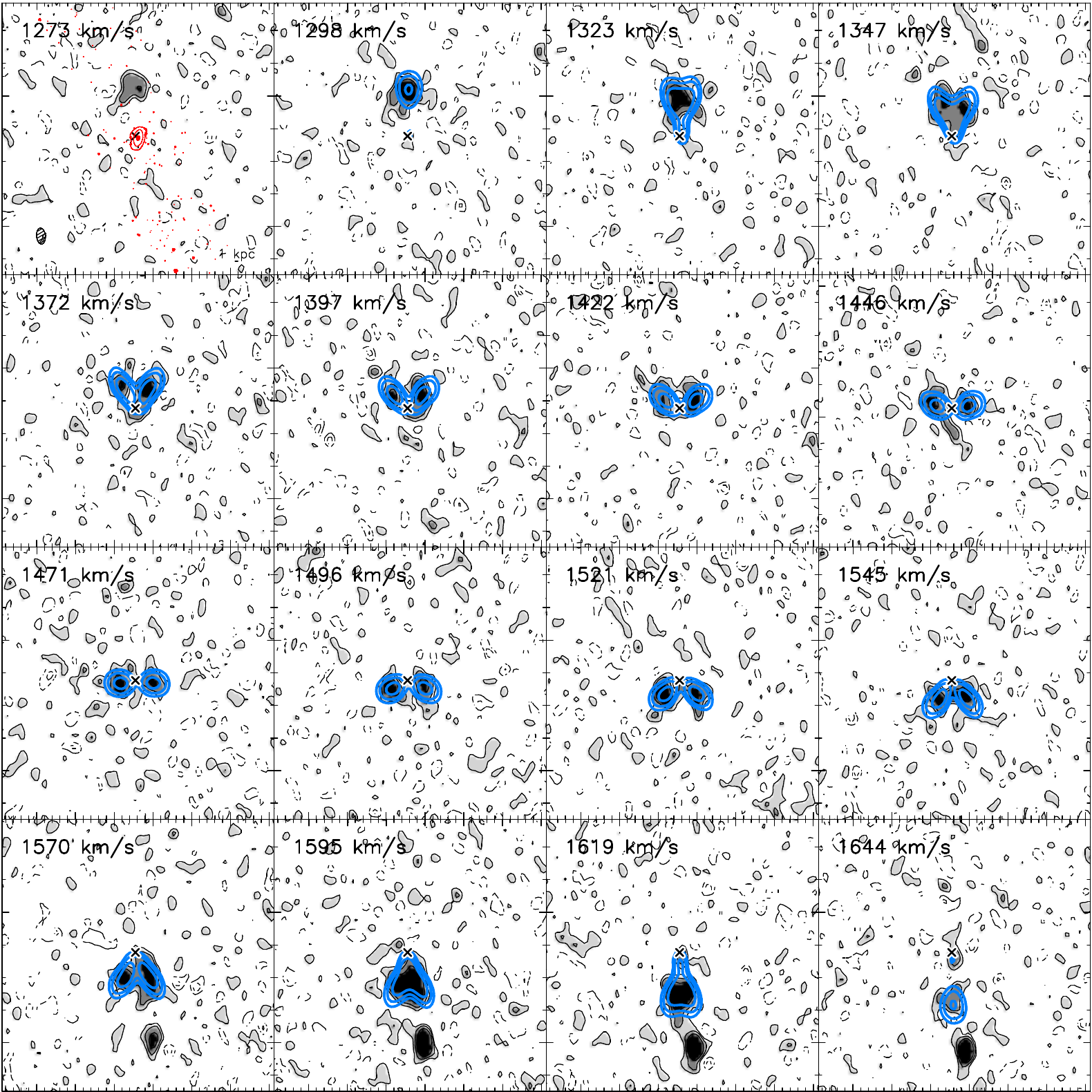}
	\end{center}
	\caption{Same as Fig. \ref{N2 HI chan}, but for the H{\footnotesize I} data of NGC\,3626.}
	\label{N3 HI chan}
\end{figure*}

\begin{figure*}[!] 
	\begin{center}
		\includegraphics[width = \textwidth]{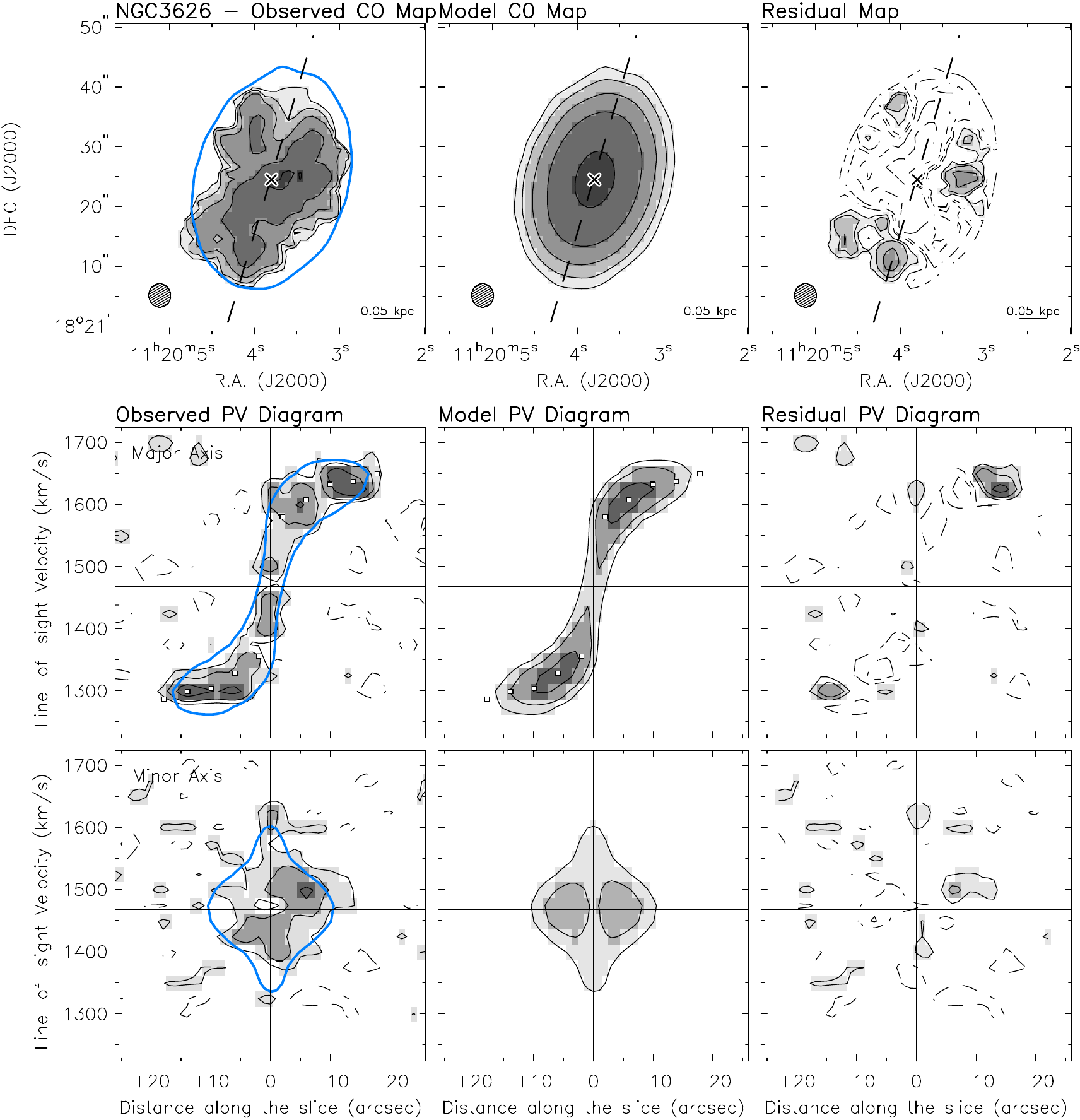}
	\end{center}
	\caption{Same as Fig. \ref{N2 HI}, but for the CO data of NGC\,3626.}
	\label{N3 CO}
\end{figure*}

\begin{figure*}[!] 
	\begin{center}
		\includegraphics[width = \textwidth]{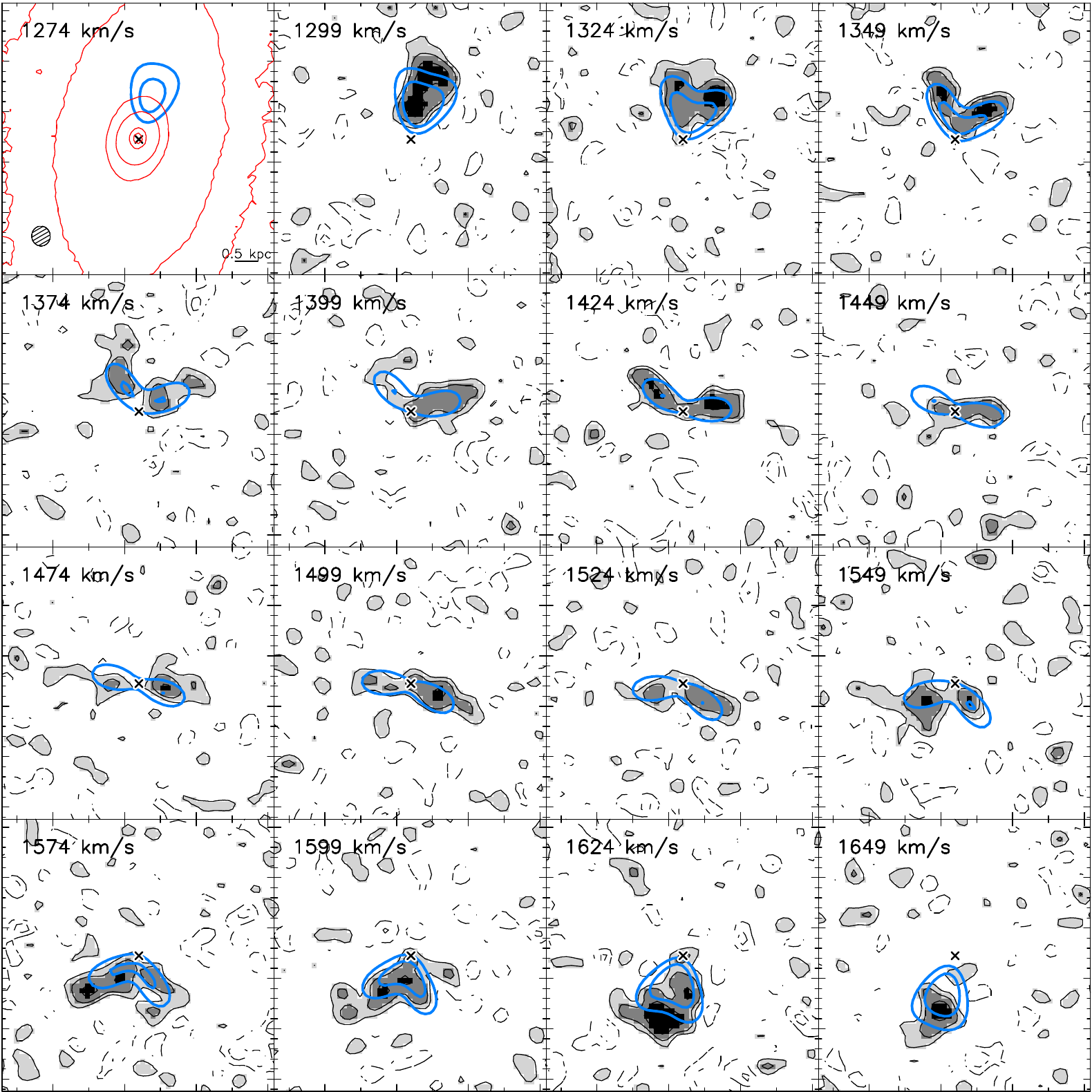}
	\end{center}
	\caption{Same as Fig. \ref{N2 HI chan}, but for the data CO of NGC\,3626.}
	\label{N3 CO chan}
\end{figure*}

 NGC\,3626 is connected to the companion galaxy UGC 6341 by a diffuse H{\footnotesize I} bridge (Fig.\,\ref{N3 HI}, top panels). We model only the H{\footnotesize I} emission associated to the disk of NGC\,3626. The residual map shows that our axisymmetric model reproduces the H{\footnotesize I} disk very well, except for a ring feature at $R\simeq3'$. The major-axis PV diagram shows a prominent dip (Fig.\,\ref{N3 HI}, bottom panels): since this feature is observed on both sides of the galaxy, we consider it an intrinsic feature of the rotation curve rather than an effect of non-circular motions. Overall, the H{\footnotesize I} kinematics is well reproduced, as shown by the channel maps in Fig.\,\ref{N3 HI chan}.

The CO distribution is quite asymmetric (Fig.\,\ref{N3 CO}, top panels), but this is likely due to the fact that the CO observations did not cover the full frequency range of galaxy emission, as can be appreciated by the sharp cutoff of the PV diagrams at low velocities (Fig.\,\ref{N3 CO}, bottom panels). In our modeling, we manually adjust the rotation curve to reproduce the receding part of the galaxy. The resulting model also reproduces the approaching half, including the non-measured part of it. The channels maps in Fig.\,\ref{N3 CO chan} further demonstrate that the CO kinematics is well reproduced by our model.

The CO data confirms the strong decline in the inner parts of the rotation curve. Even in the extreme case of an edge-on CO disk (which is clearly not the case as seen in Fig. \ref{N3 CO}), the CO rotation velocities at small radii are higher than the H{\footnotesize I} rotation velocities at large radii. Our rotation curve is about 50 km/s higher in the inner parts than the one presented in \cite{Mazzei}. We stress the H{\footnotesize I} disk is warped and the variations of $i$ and PA with radius are not well constrained by the data. At small radii, we adopt the values of $i$ and PA suggested by the CO disk, then we assume a simple linear rise to the higher values of $i$ and PA suggested by the H{\footnotesize I} disk at large radii. Clearly, a different modeling of the warp may slightly change the shape of the rotation curve and the resulting stellar mass-to-light ratios.

\subsection{UGC\,6176} \label{sec: UGC6176}

\begin{figure*}[!]
	\begin{center}
		\includegraphics[width = \textwidth]{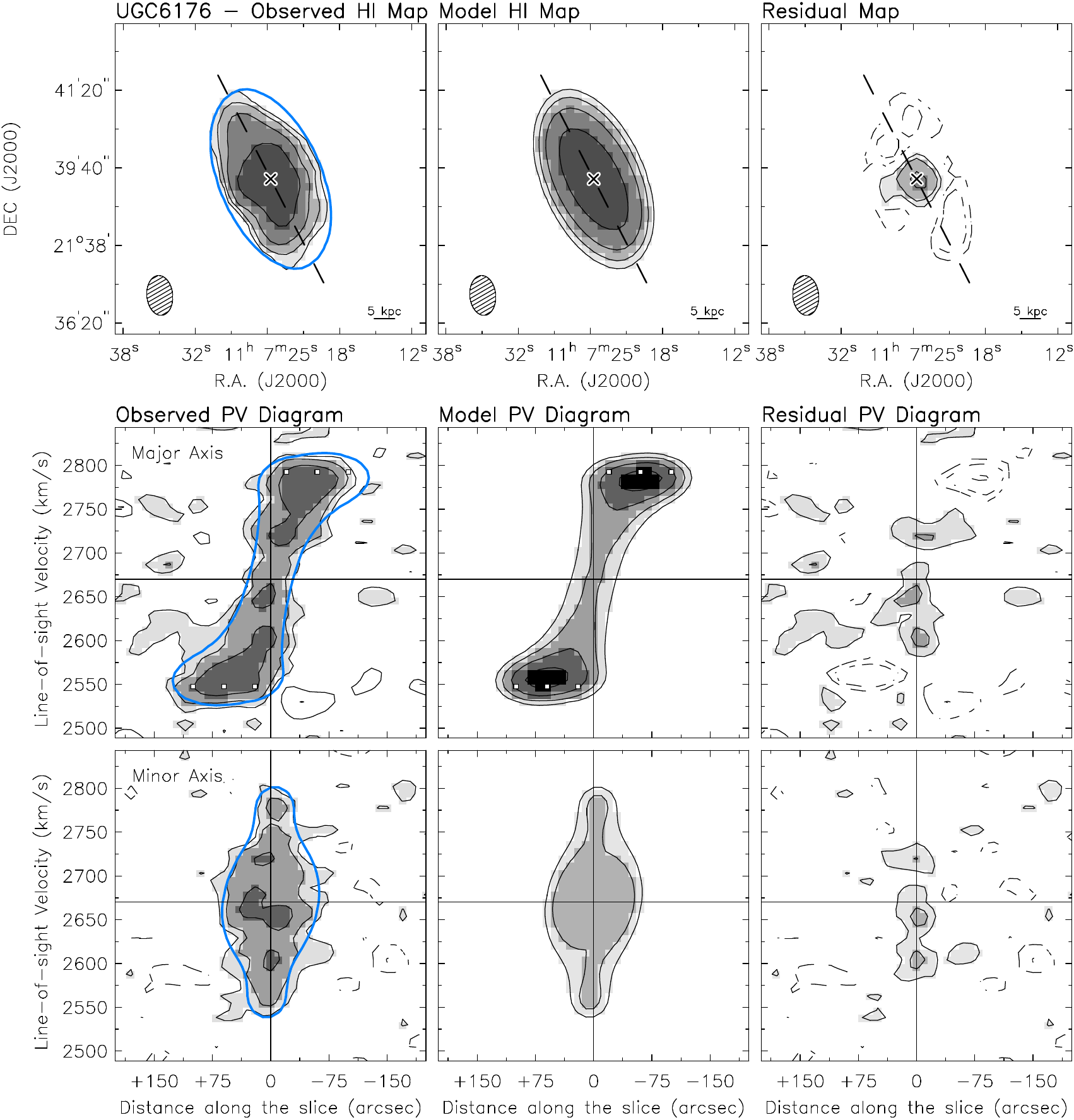}
	\end{center}
	\caption{Same as Fig. \ref{N2 HI}, but for the H{\footnotesize I} data of UGC\,6176.}
	\label{U HI}
\end{figure*}

\begin{figure*}[!]
	\begin{center}
		\includegraphics[width = \textwidth]{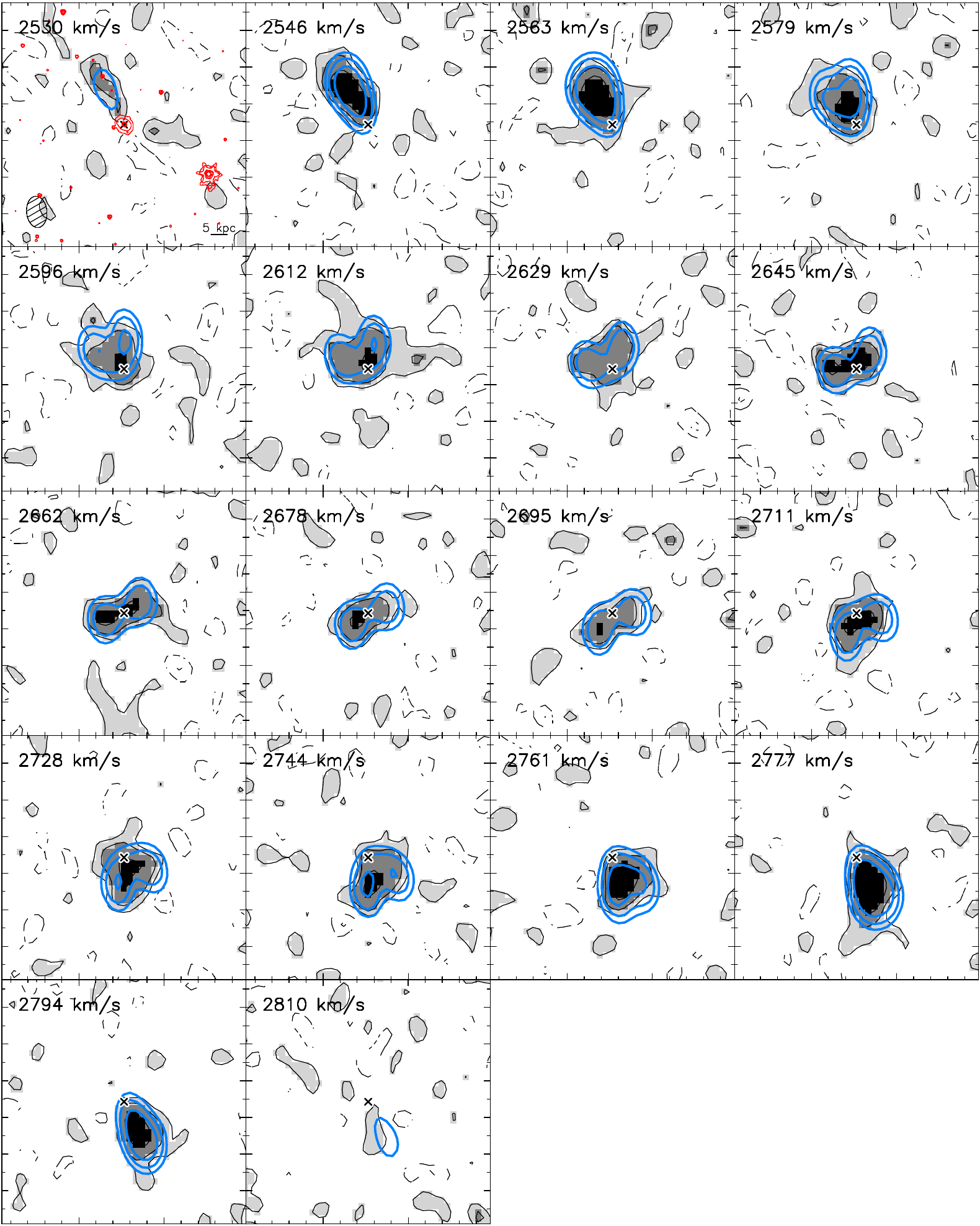}
	\end{center}
	\caption{Same as Fig. \ref{N2 HI chan}, but for the H{\footnotesize I} data of UGC\,6176.}
	\label{U HI chan}
\end{figure*}

\begin{figure*}[!]
	\begin{center}
		\includegraphics[width = \textwidth]{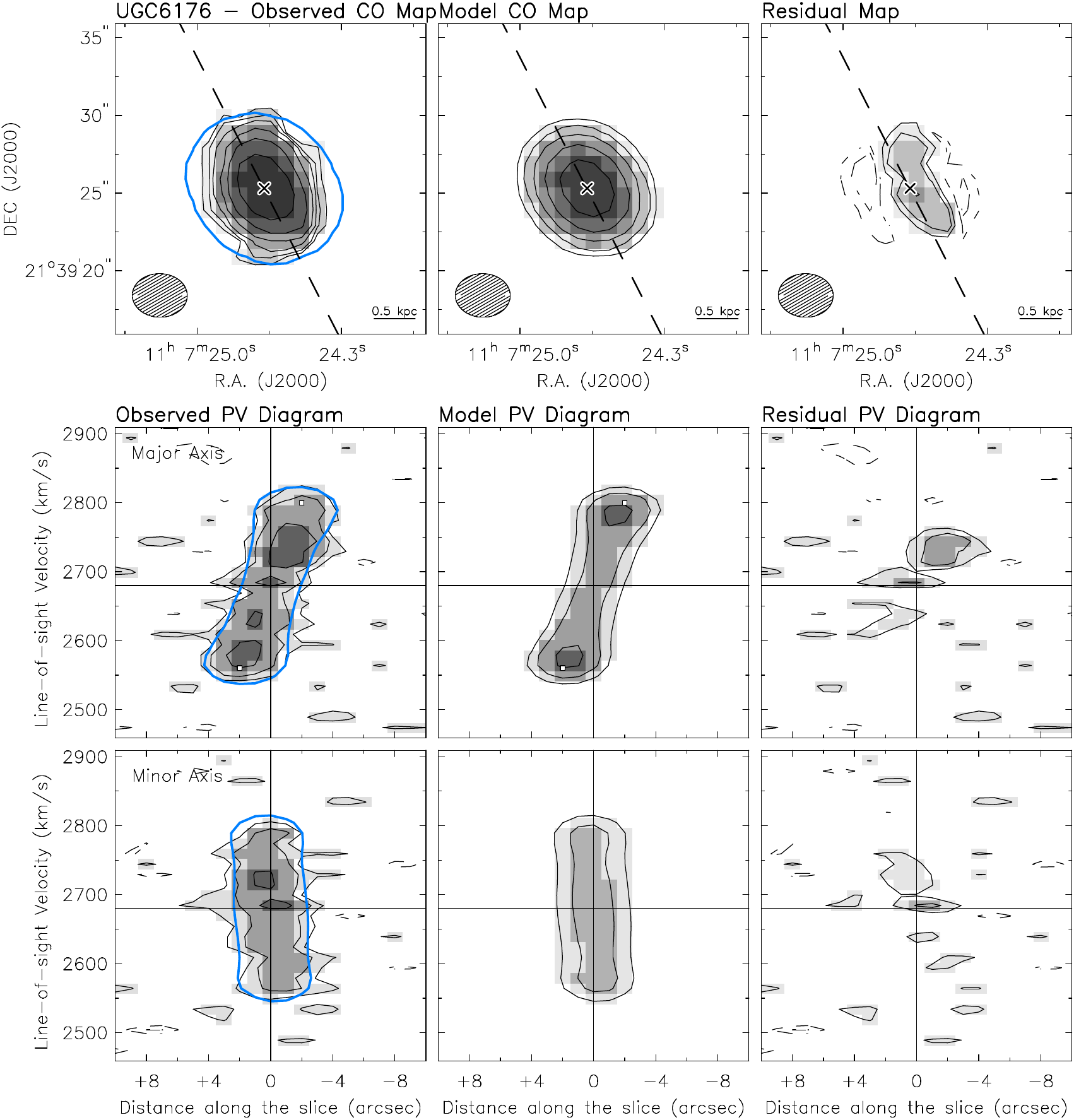}
	\end{center}
	\caption{Same as Fig. \ref{N2 HI}, but for the CO data of UGC\,6176.}
	\label{U CO}
\end{figure*}

\begin{figure*}[!]
	\begin{center}
		\includegraphics[width = \textwidth]{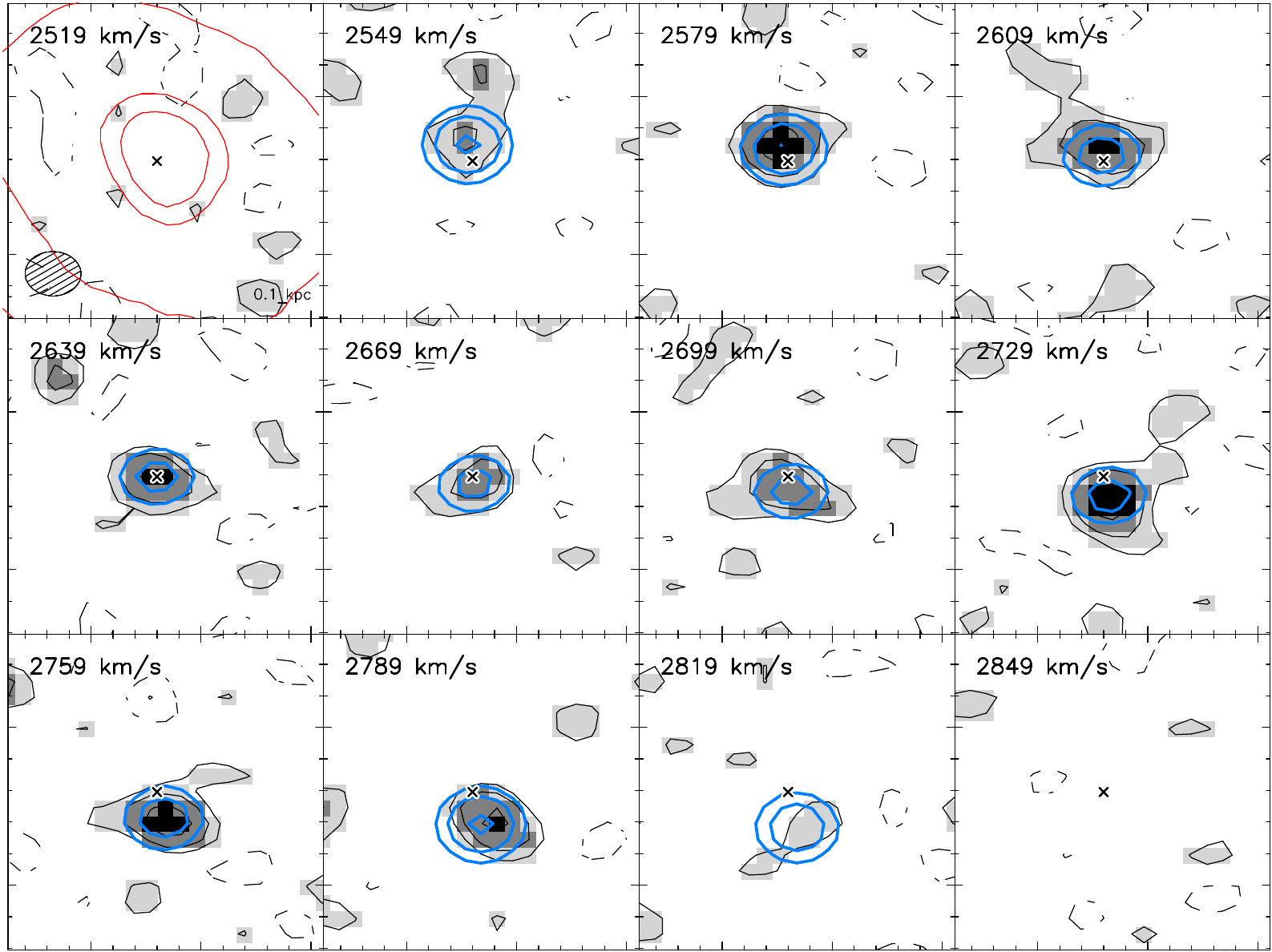}
	\end{center}
	\caption{Same as Fig. \ref{N2 HI chan}, but for the CO data of UGC\,6176.}
	\label{U CO chan}
\end{figure*}

The H{\footnotesize I} distribution and kinematics are regular and symmetric, so they are well described by our axisymmetric model (Figures \ref{U HI} and \ref{U HI chan}). We note, however, that the H{\footnotesize I} disk is poorly resolved and beam-smearing effects are severe, leading to a very broad PV diagram along the minor axis.

The CO disk is even less resolved than the H{\footnotesize I} emission, so we have only one ring in our model (Figure \ref{U CO}). The beam is elongated along the East-West direction leading to non-trivial beam-smearing effects. We could not properly reproduce the axial ratio of the observed CO map for any value of the disk inclination. We suspect that the CO beam is not Gaussian, so our model cannot reproduce such extreme beam-smearing effects. In the end, we adopted a relatively high inclination of 85$^{\circ}$, which gives the closest match to the observed CO map. At any rate, the observed PV diagrams are well described by our model. The CO channel maps are also reasonably reproduced (Fig.\,\ref{U CO chan}).

Although the ATLAS$^{\text{3D}}$ collaboration denotes UGC\,6176 as a warped galaxy, we find no strong evidence for it. The PA of the H{\footnotesize I} and the CO disks are consistent with each other and show no clear variation with radius, while the inclination of the CO disk cannot be properly assessed from the available data.

\section{Posterior distributions of DM halo fits} \label{App:DM}

In this Appendix, we show the posterior distributions of the free parameters from DM halo fits with and without $\Lambda$CDM priors (see section \ref{sec: LCDM priors}). When imposing $\Lambda$CDM priors (left panels), the posterior distributions are well-behaved and reasonably symmetric. Without imposing $\Lambda$CDM priors (right panels), the posterior distributions of the halo parameters become very complex, so their maximum-probability values are not fully trustworthy.

\begin{figure*}[h]
	\centering
	\subfloat[NFW]{%
		\includegraphics[width=0.38\textwidth]{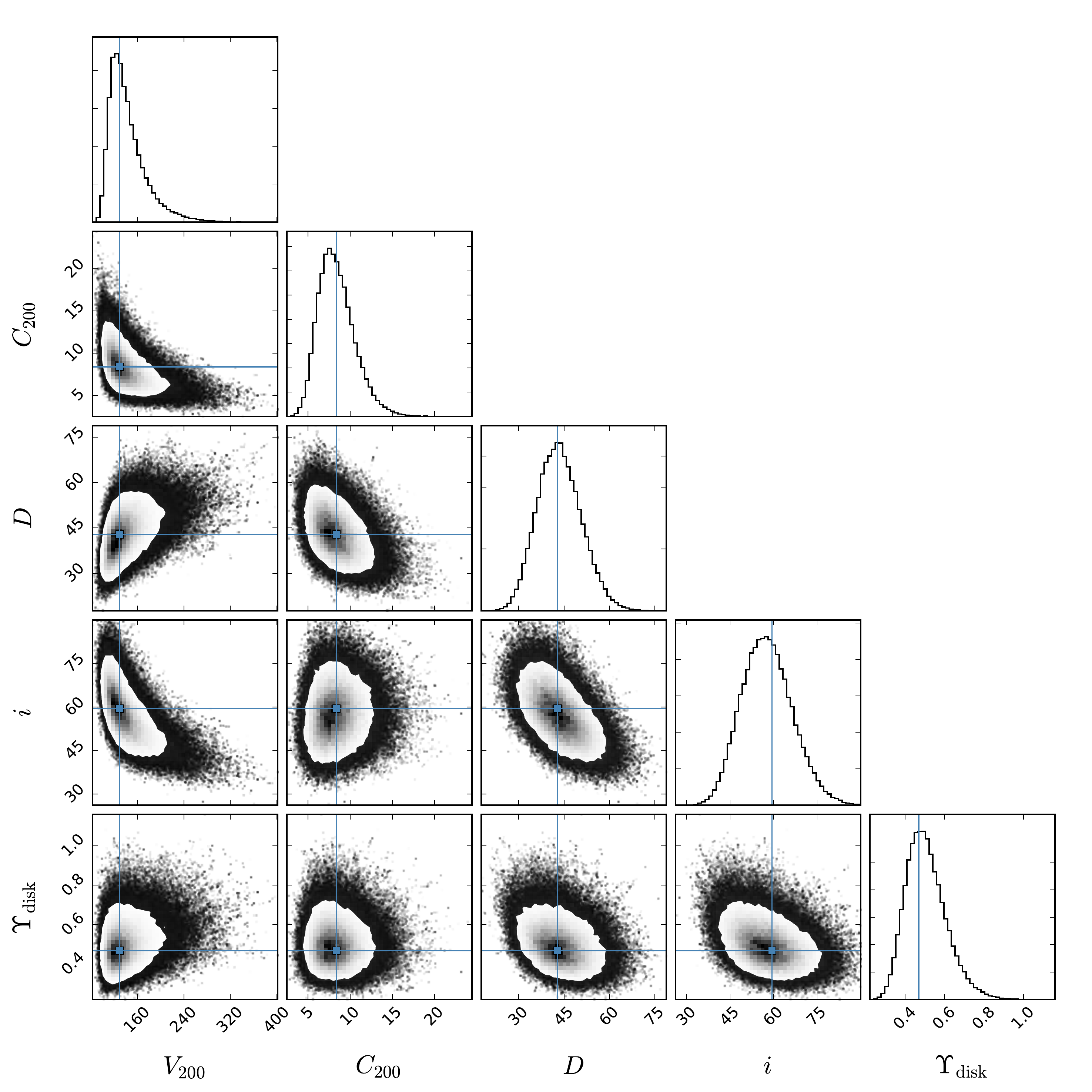}
	}
	\hspace{1cm}
	\subfloat[NFW (no priors)]{%
		\includegraphics[width=0.38\textwidth]{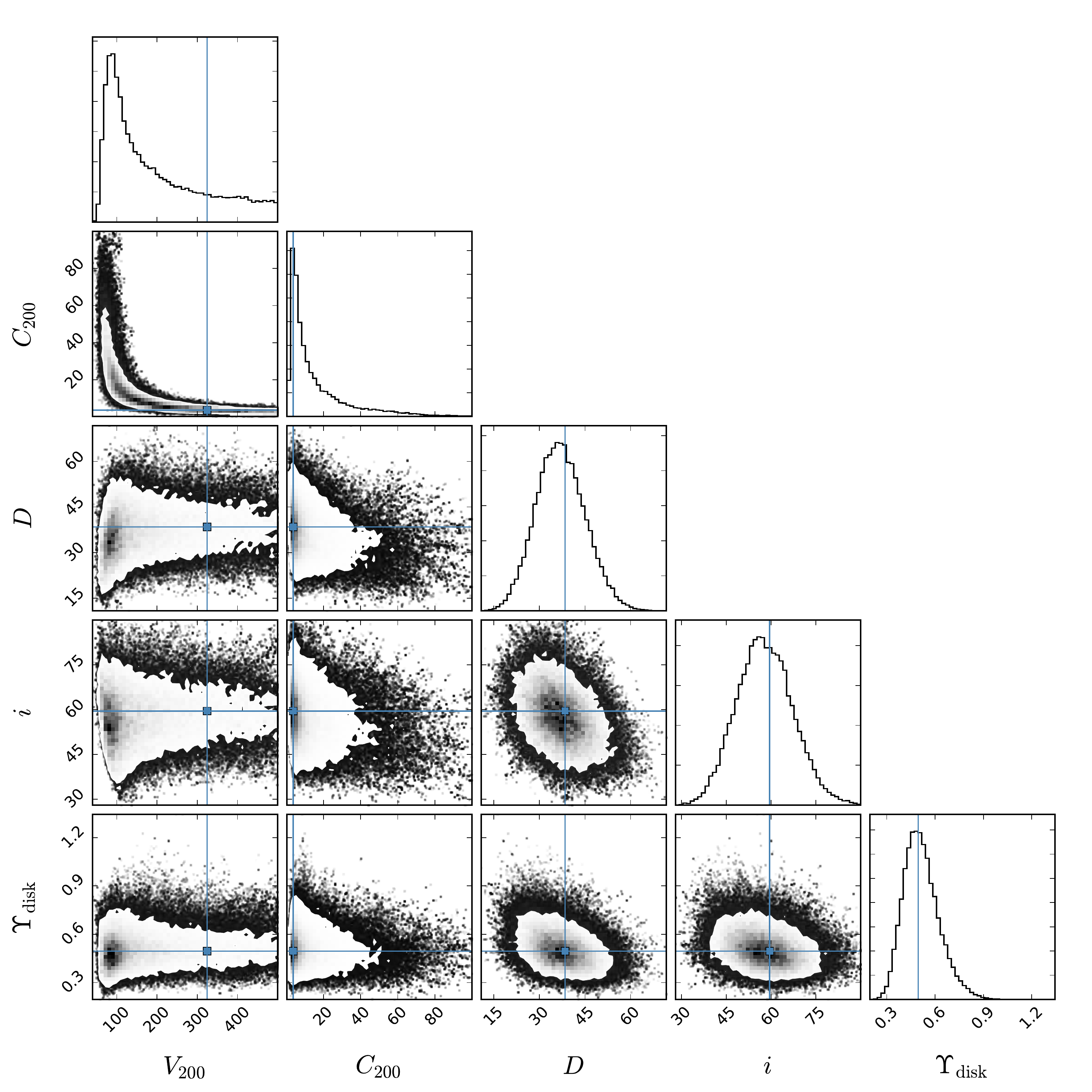}
	}
	\\
	\subfloat[Einasto]{%
		\includegraphics[width=0.38\textwidth]{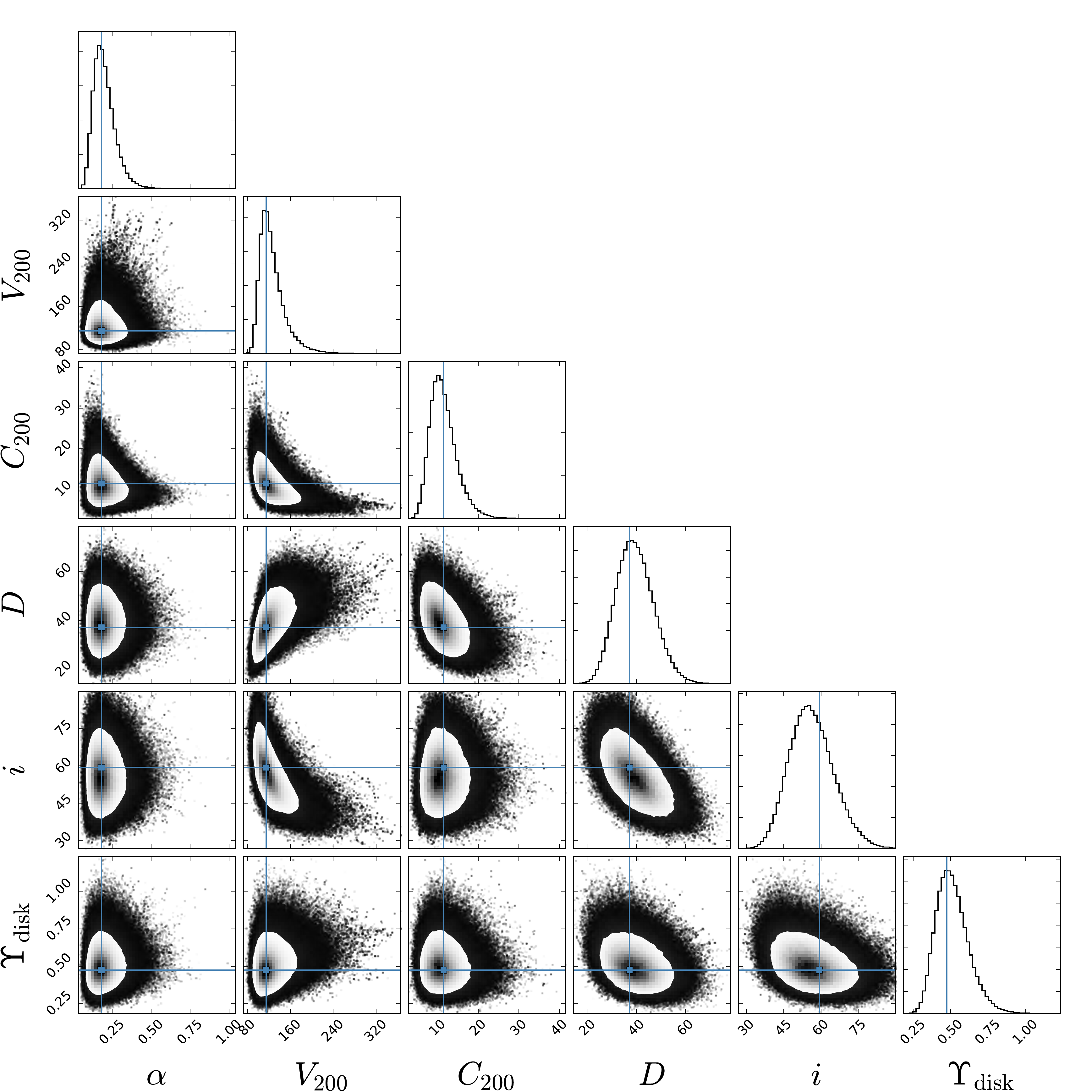}
	}
	\hspace{1cm}
	\subfloat[Einasto (no priors)]{%
		\includegraphics[width=0.38\textwidth]{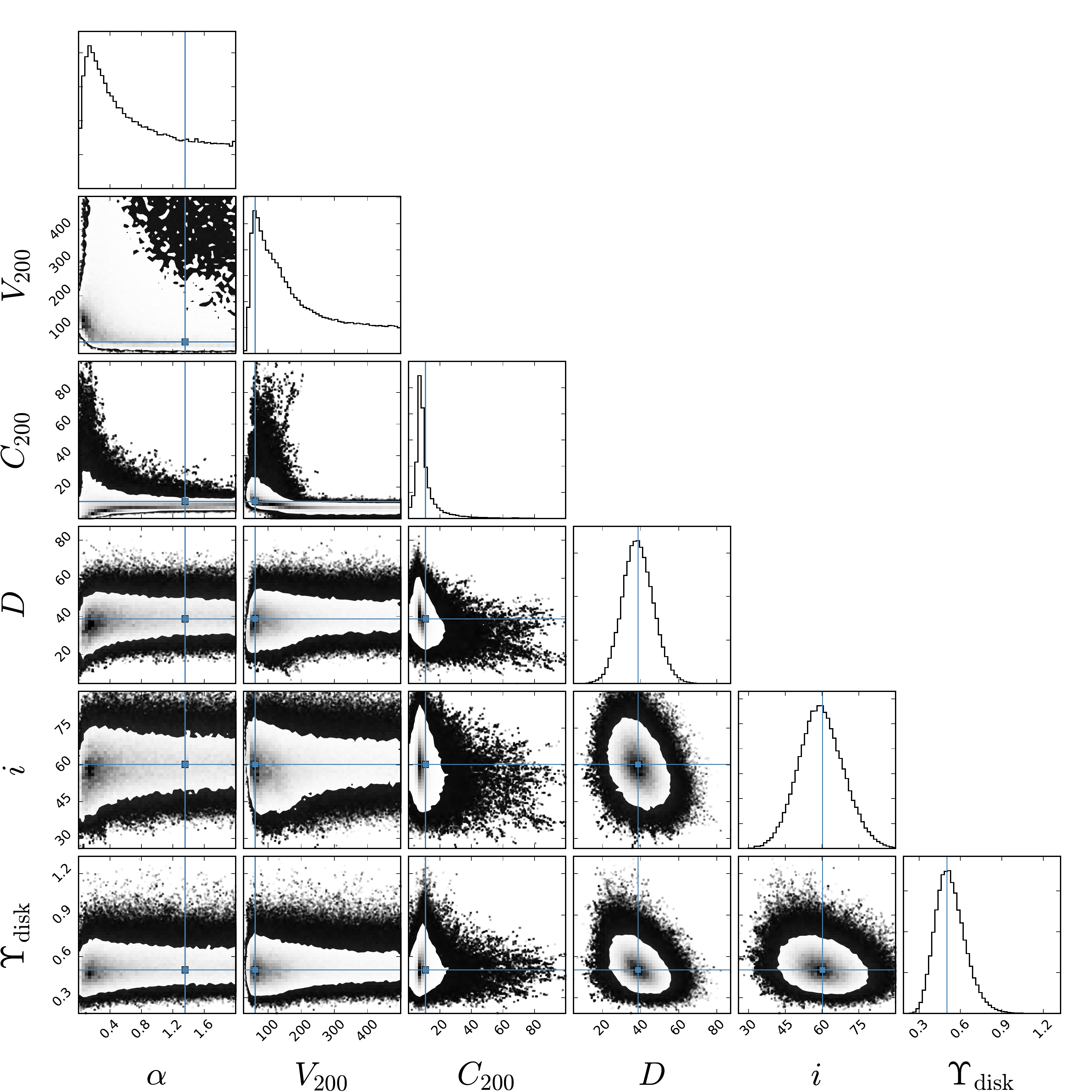}
	}
	\\
	\subfloat[DC14]{%
		\includegraphics[width=0.38\textwidth]{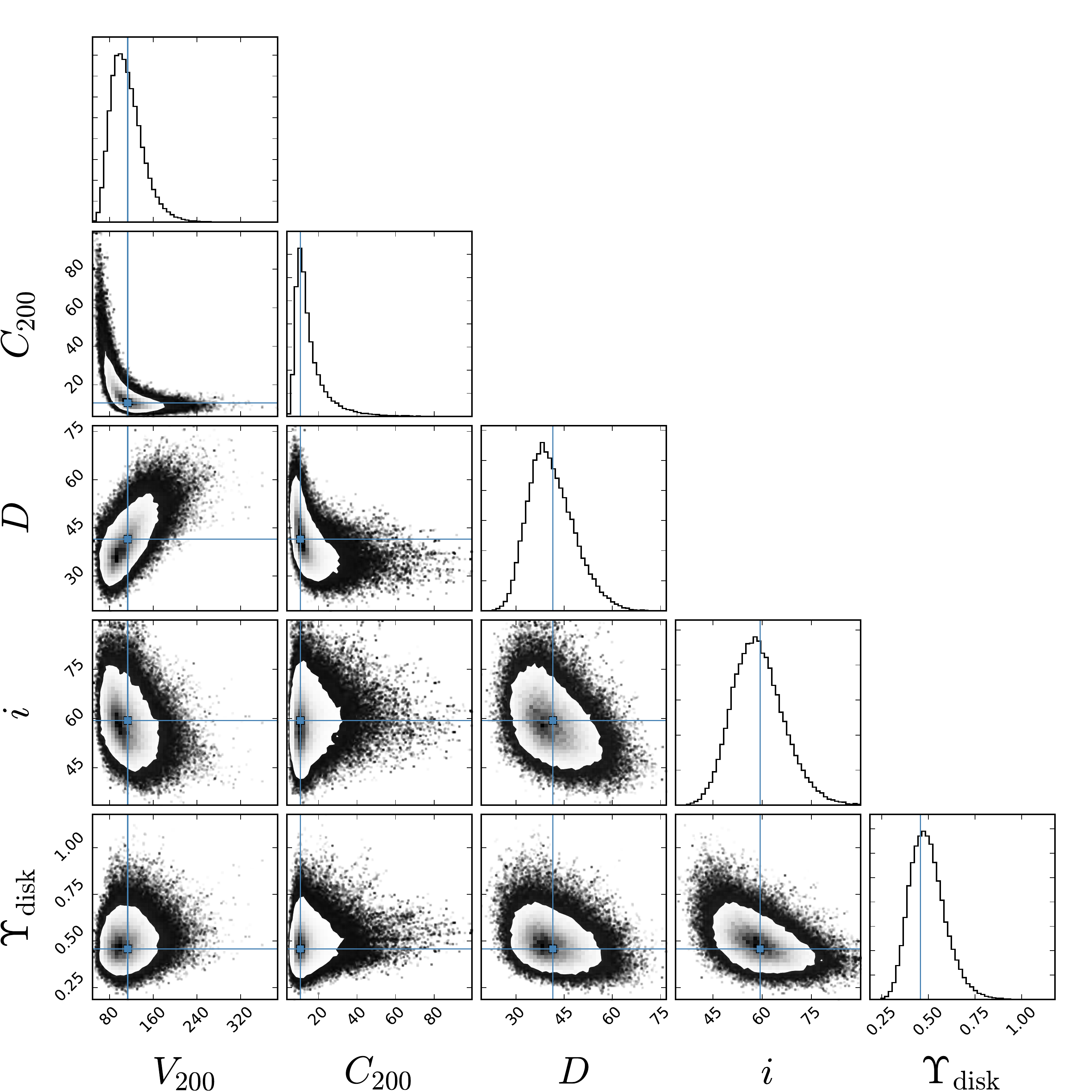}
	}
	\hspace{1cm}
	\subfloat[DC14 (no priors)]{%
		\includegraphics[width=0.38\textwidth]{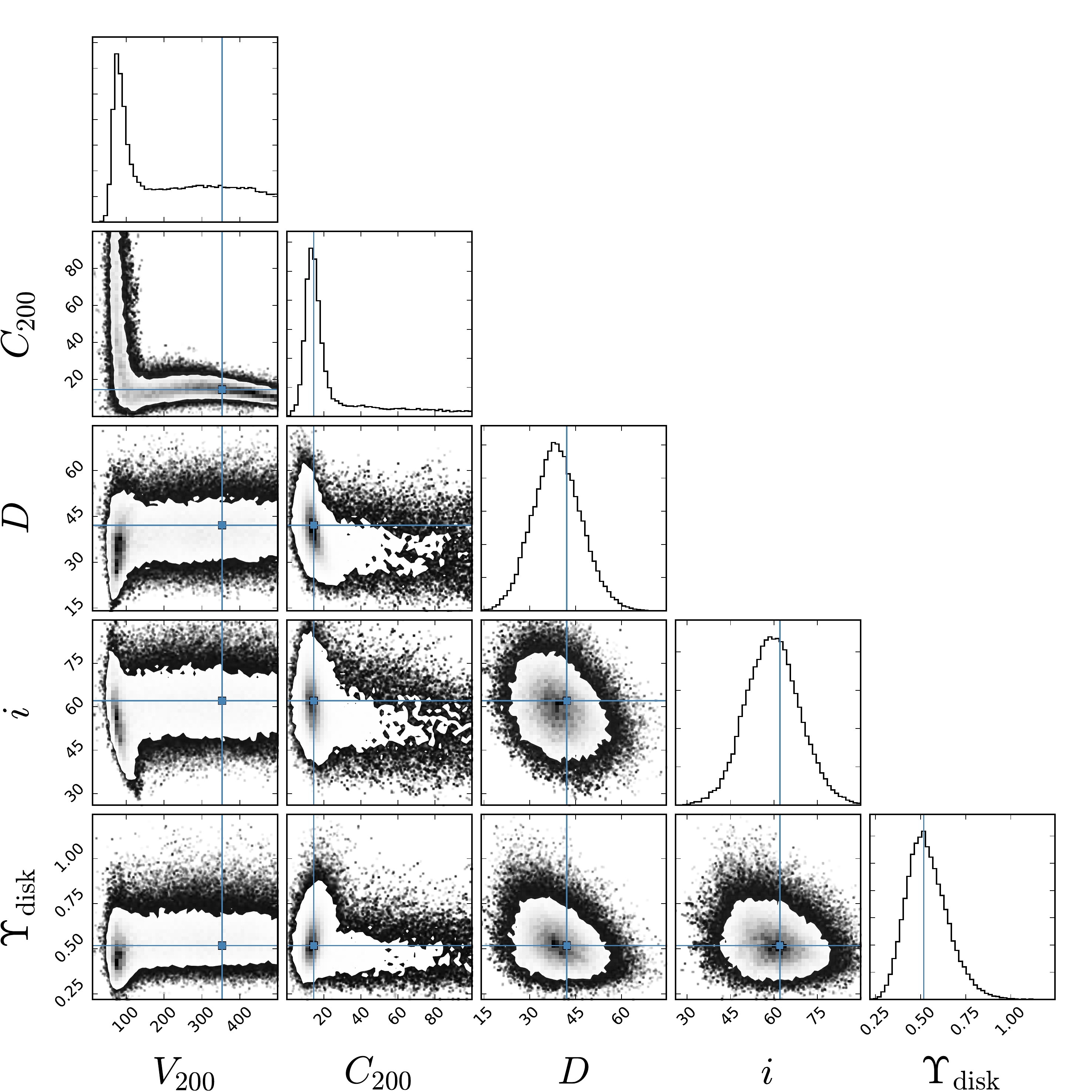}
	}
	\caption{Posterior distributions for the DM halo fits of NGC\,2824, with (left column) and without (right column) $\Lambda$CDM priors (see Sect. \ref{sec: LCDM priors} for details).}
	\label{Fig: N2 corner plots}
\end{figure*}

\begin{figure*}[!]
	\centering
	\subfloat[NFW]{%
		\includegraphics[width=0.38\textwidth]{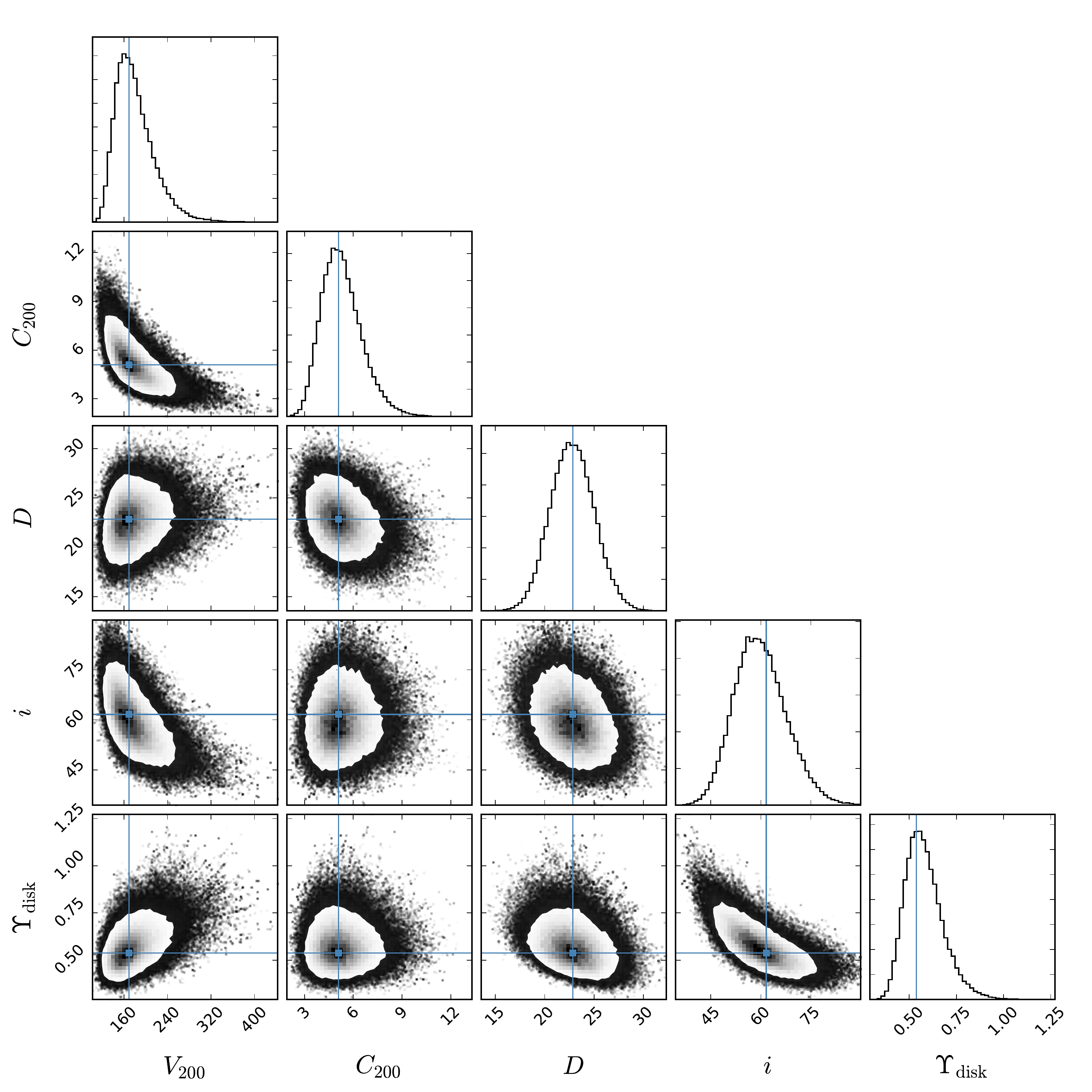}
	}
	\hspace{1cm}
	\subfloat[NFW (no priors)]{%
		\includegraphics[width=0.38\textwidth]{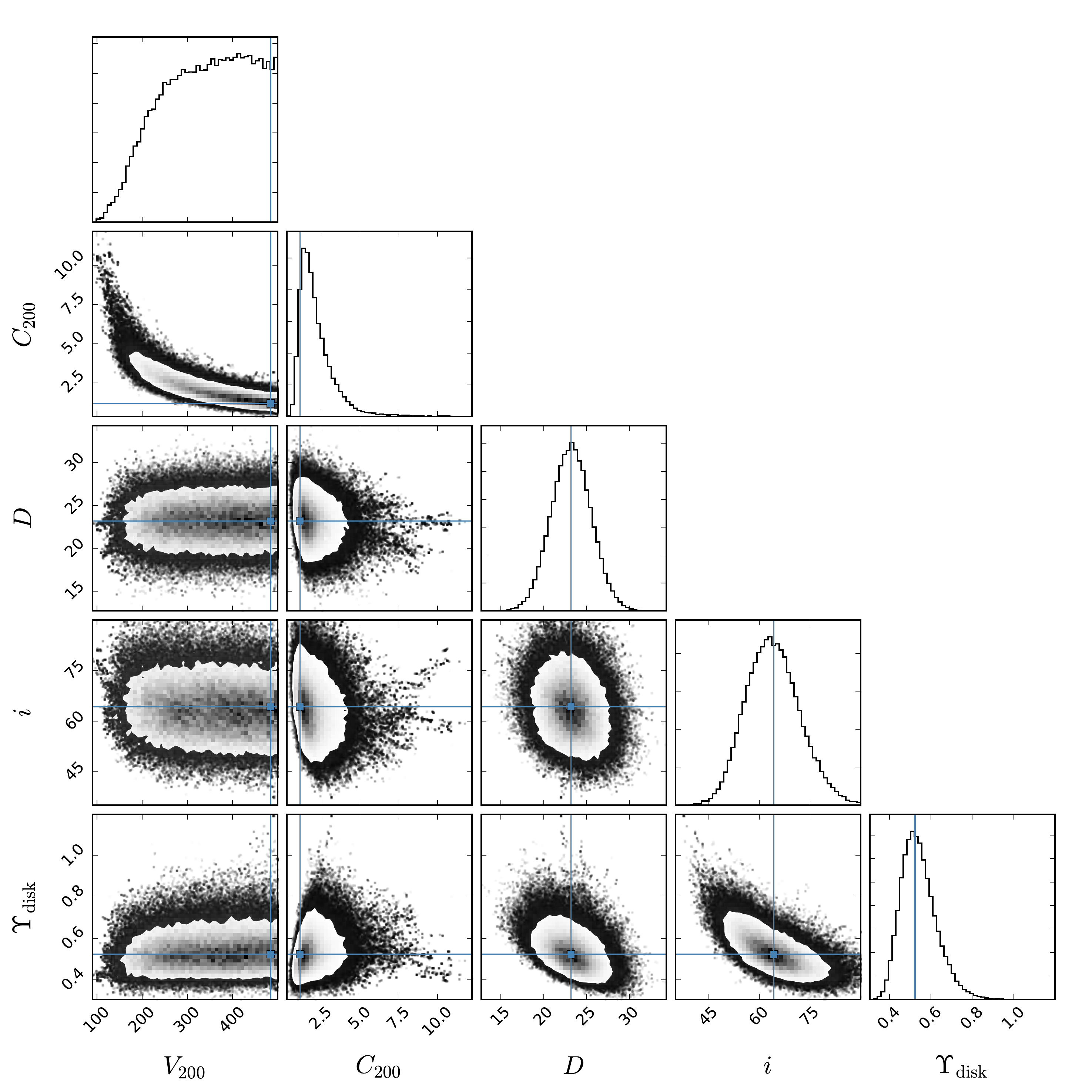}
	}
	\\
	\subfloat[Einasto]{%
		\includegraphics[width=0.38\textwidth]{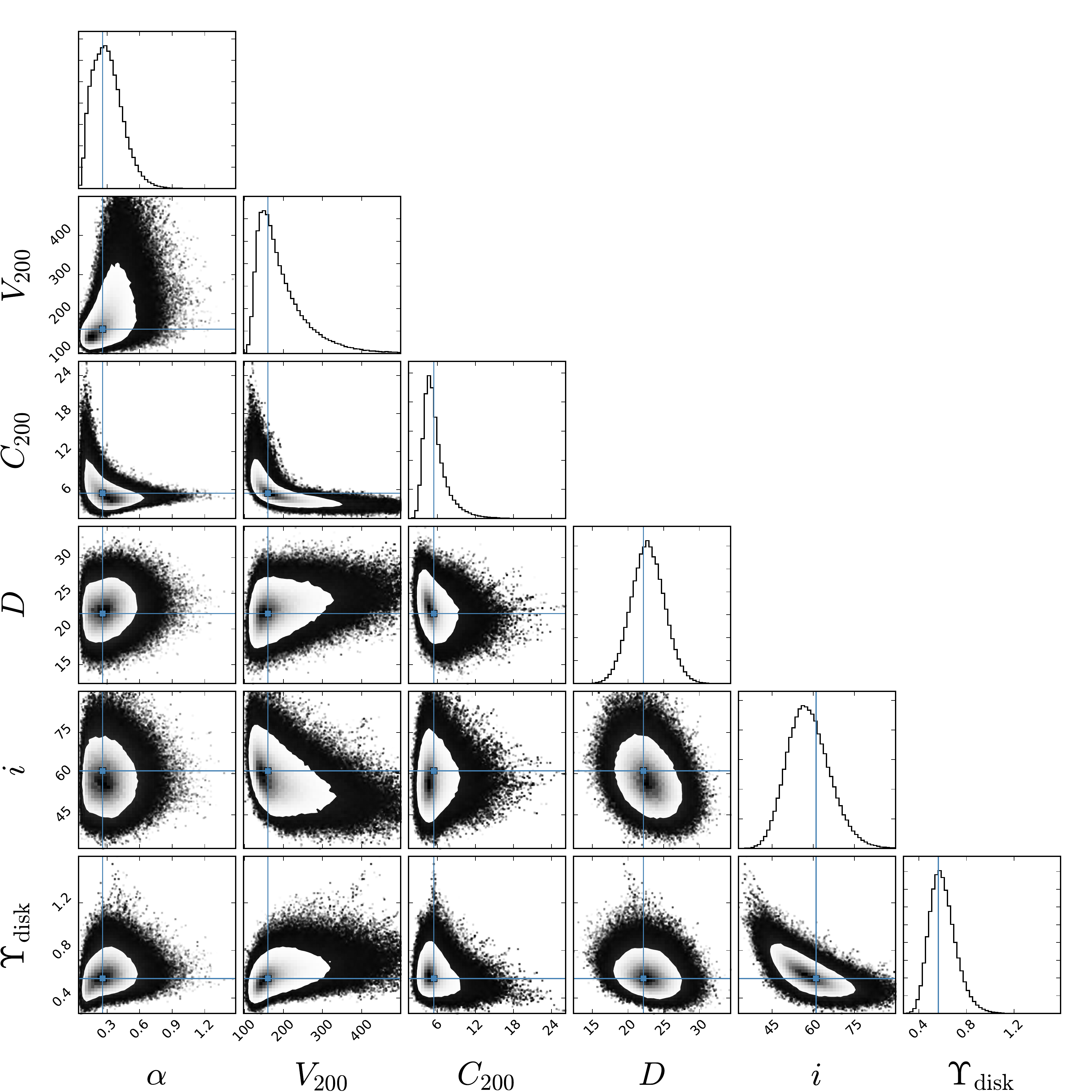}
	}
	\hspace{1cm}
	\subfloat[Einasto (no priors)]{%
		\includegraphics[width=0.38\textwidth]{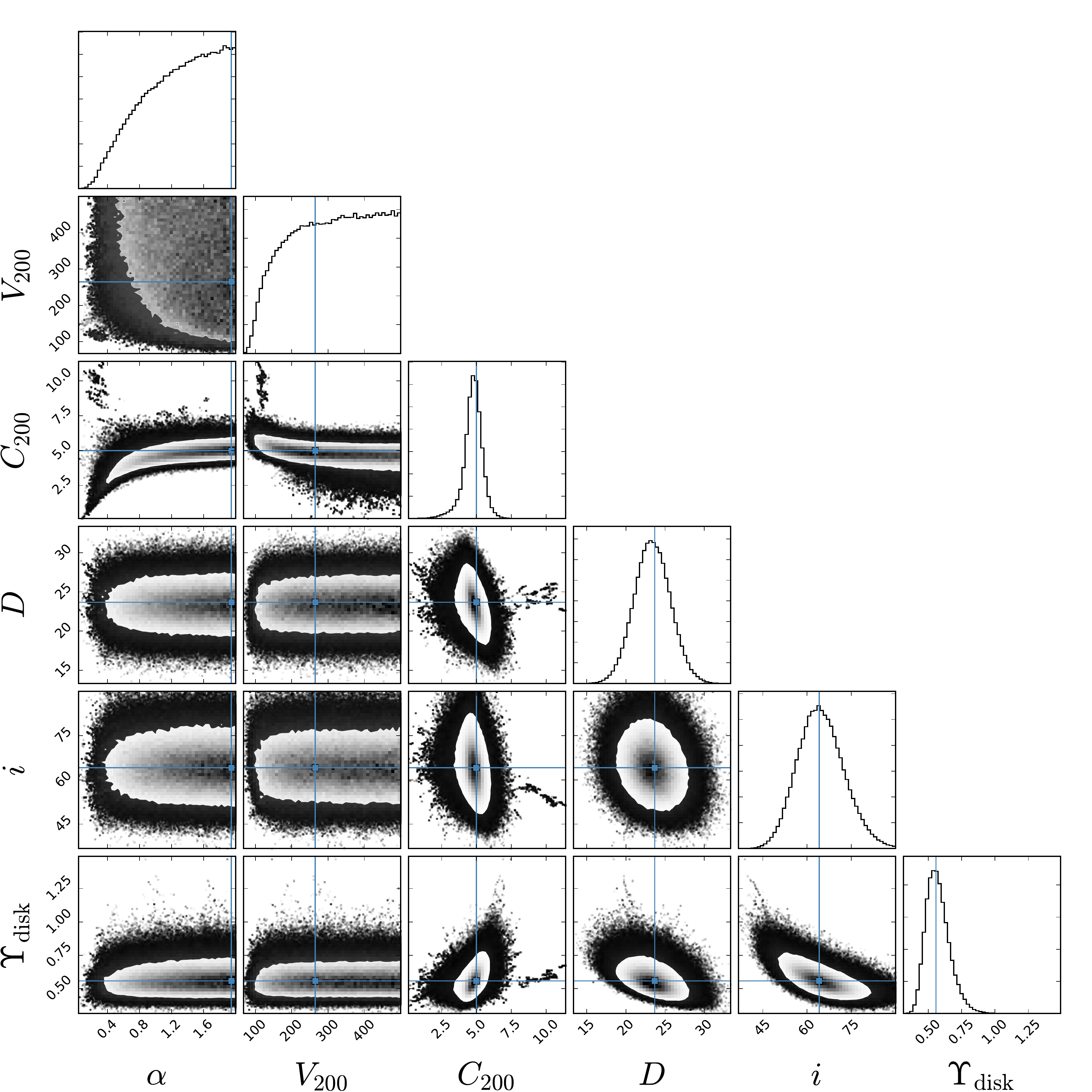}
	}
	\\
	\subfloat[DC14]{%
		\includegraphics[width=0.38\textwidth]{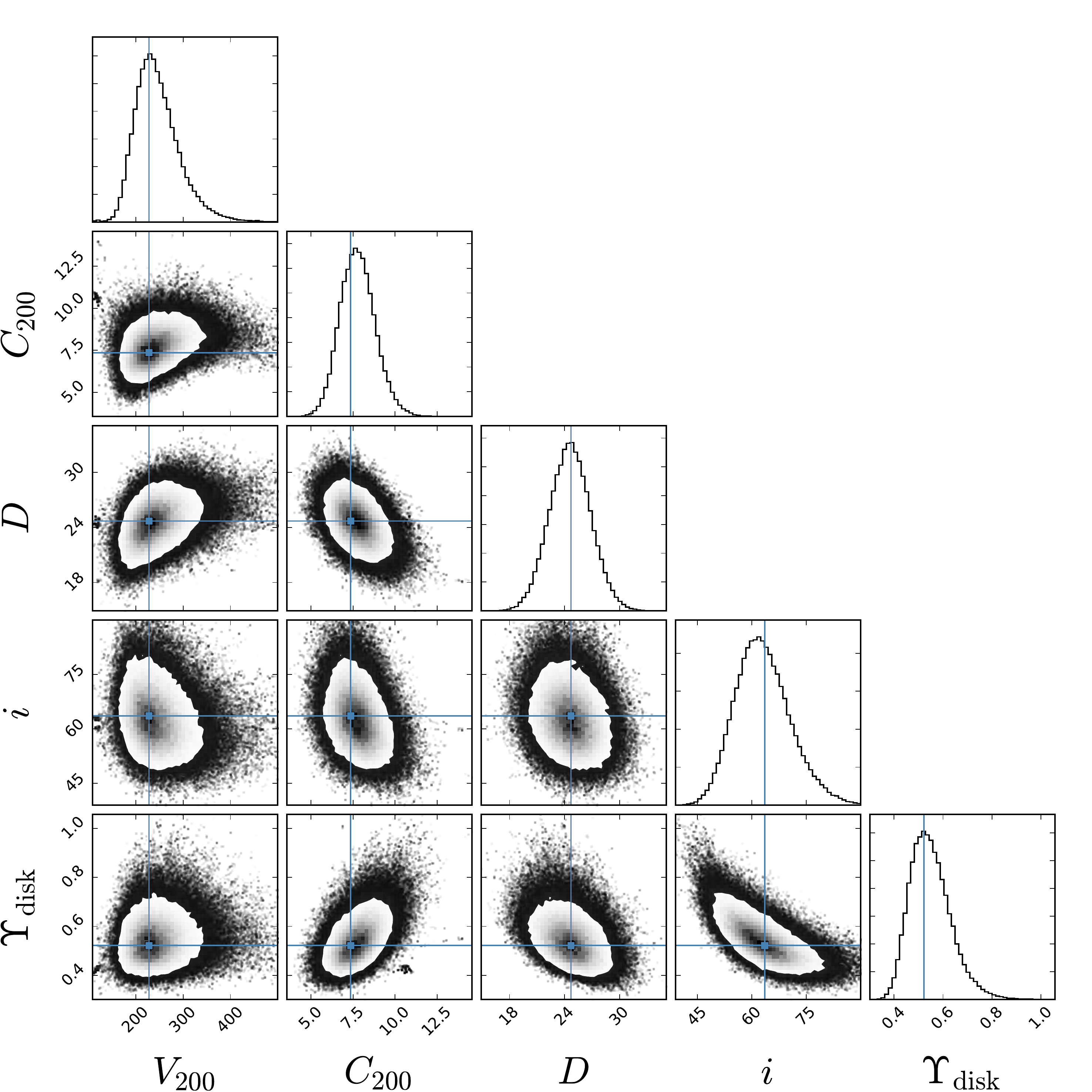}
	}
	\hspace{1cm}
	\subfloat[DC14 (no priors)]{%
		\includegraphics[width=0.38\textwidth]{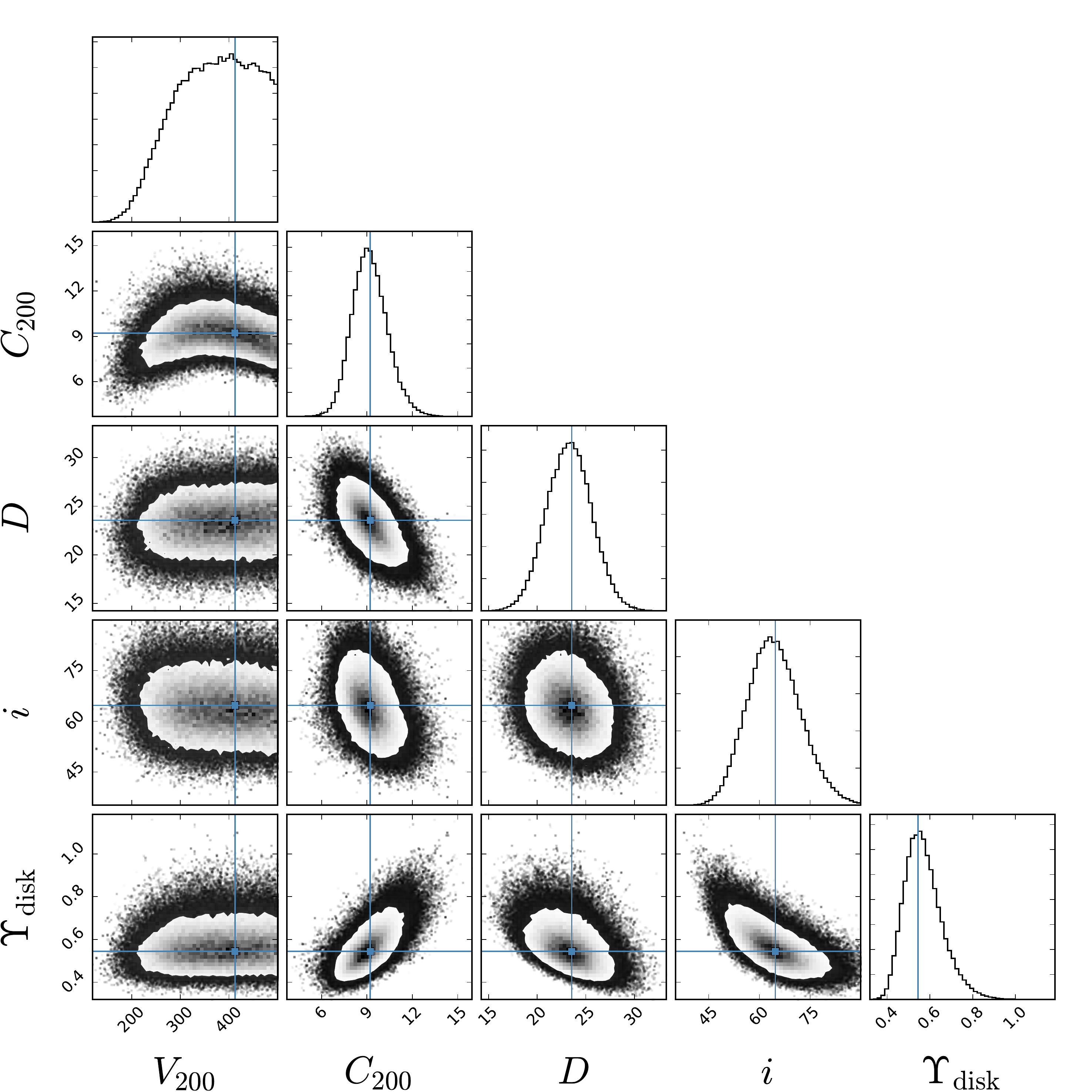}
	}
	\caption{Same as Fig. \ref{Fig: N2 corner plots}, but for NGC\,3626.}
	\label{Fig: N3 corner plots}
\end{figure*}

\begin{figure*}[!]
	\centering
	\subfloat[NFW]{%
		\includegraphics[width=0.38\textwidth]{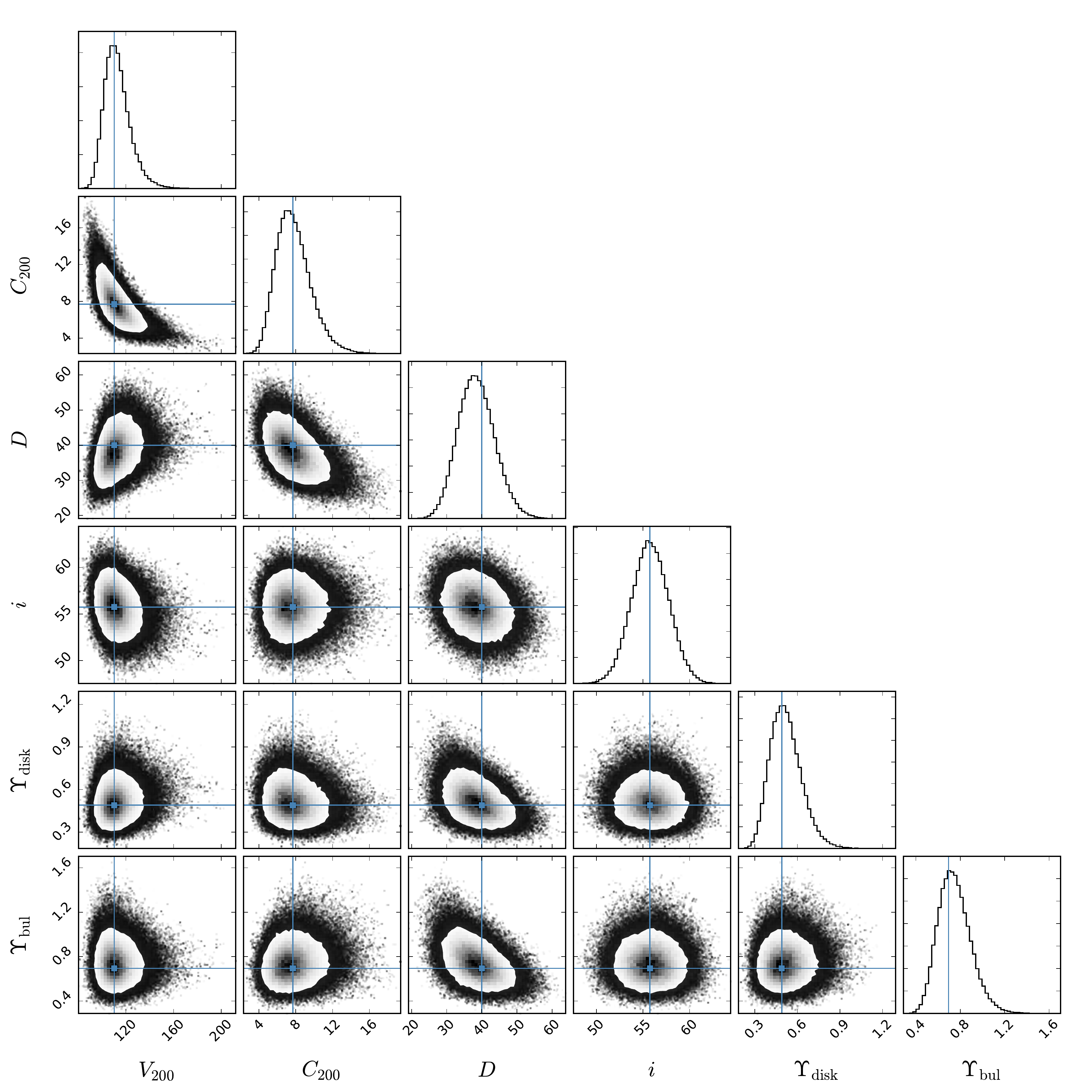}
	}
	\hspace{1cm}
	\subfloat[NFW (no priors)]{%
		\includegraphics[width=0.38\textwidth]{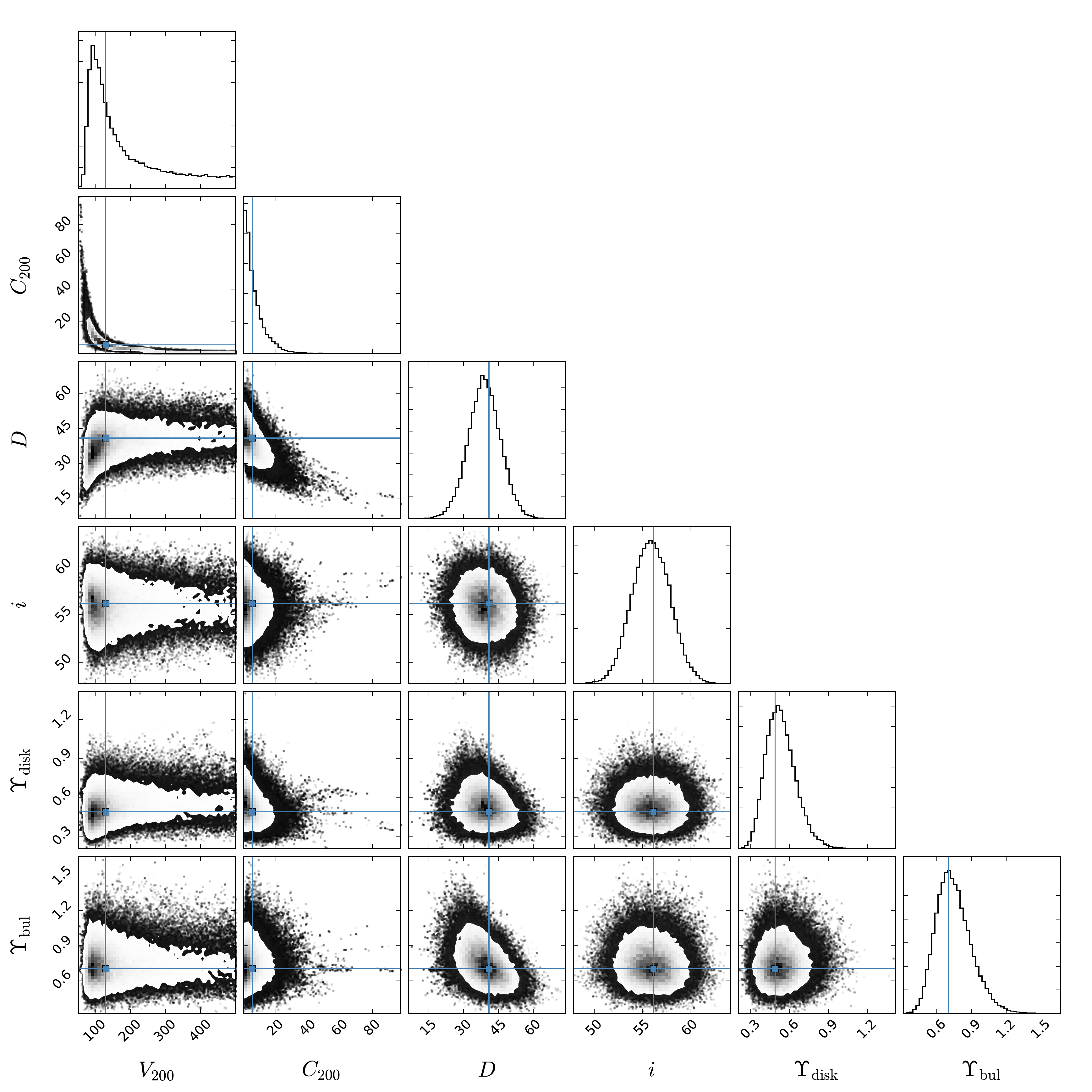}
	}
	\\
	\subfloat[Einasto]{%
		\includegraphics[width=0.38\textwidth]{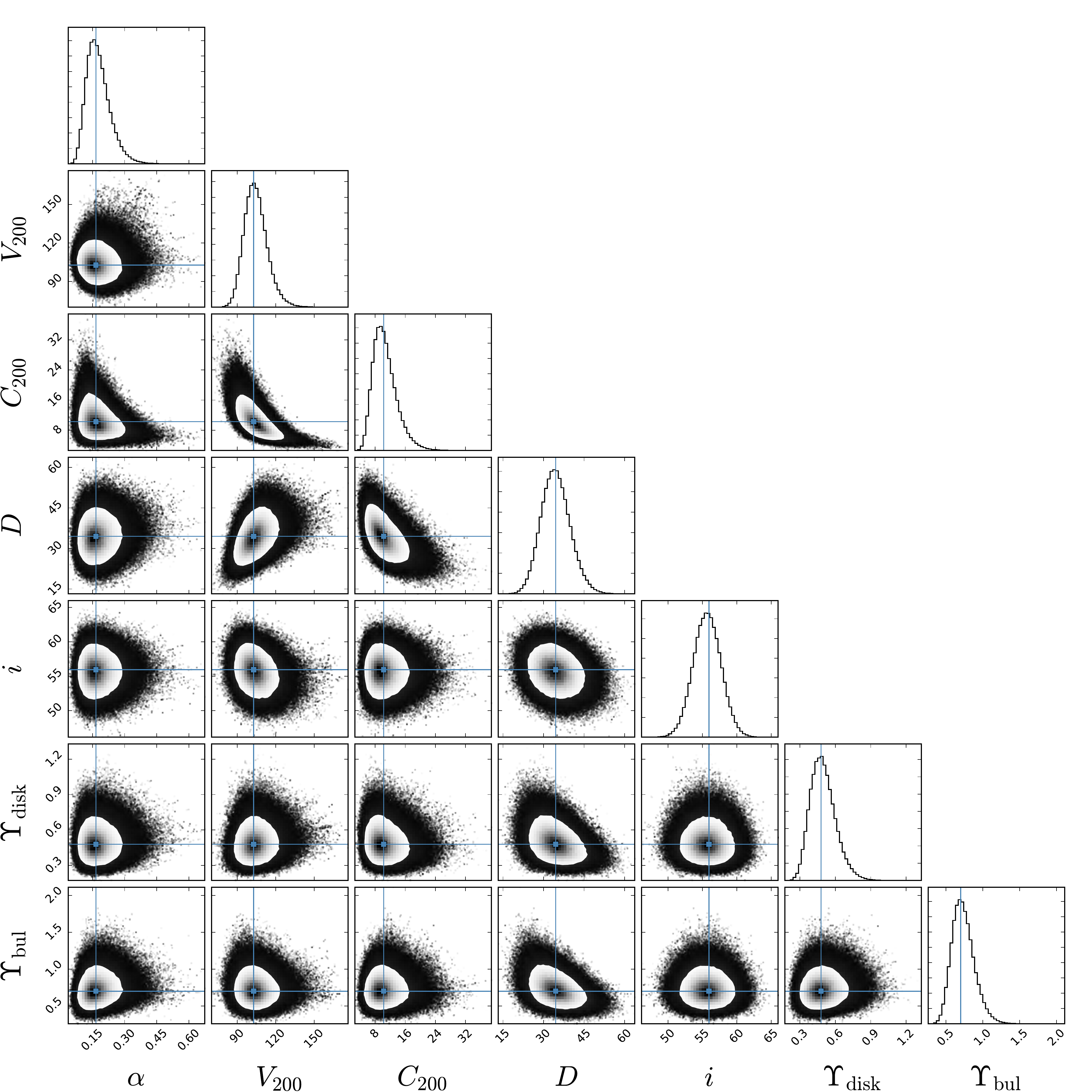}
	}
	\hspace{1cm}
	\subfloat[Einasto (no priors)]{%
		\includegraphics[width=0.38\textwidth]{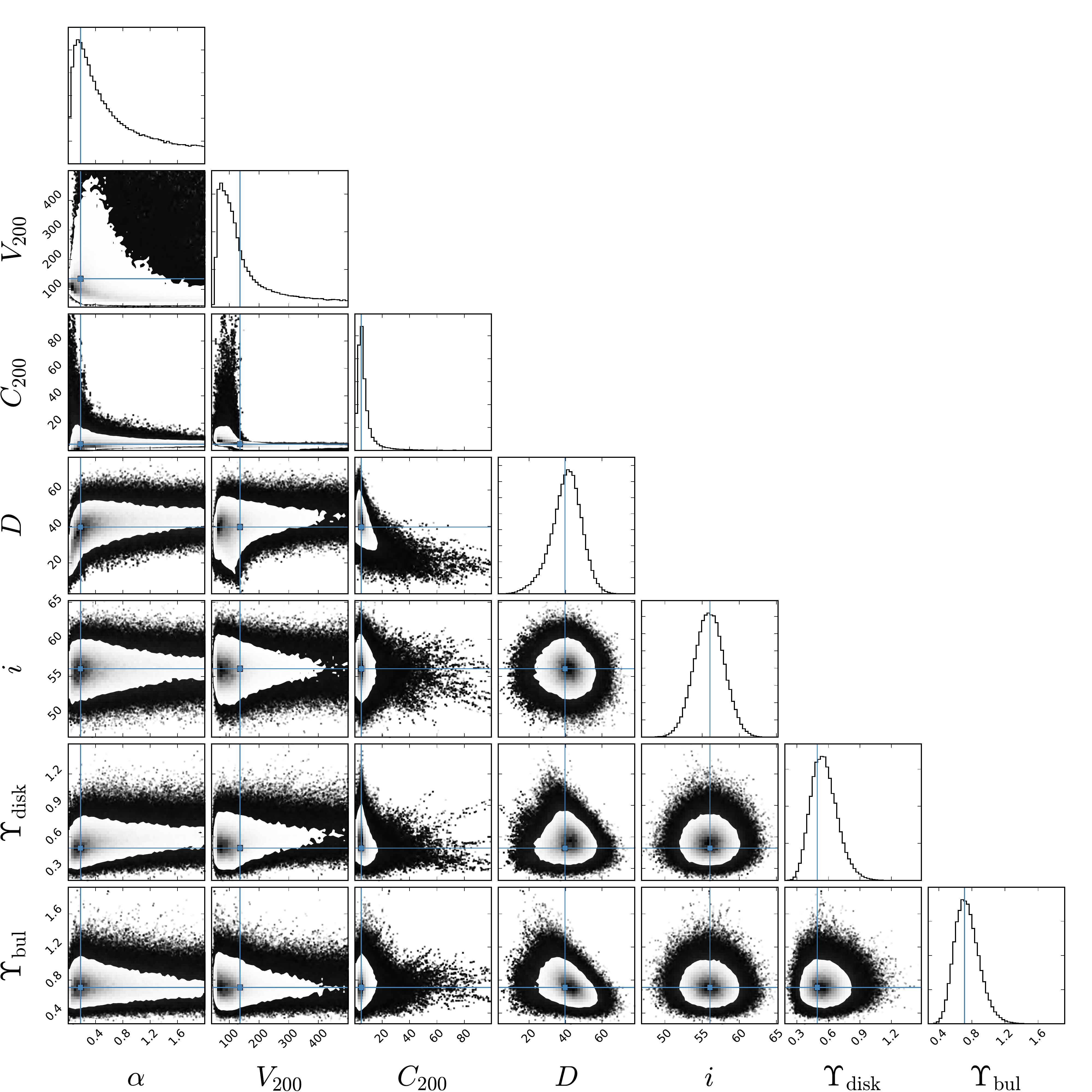}
	}
	\\
	\subfloat[DC14]{%
		\includegraphics[width=0.38\textwidth]{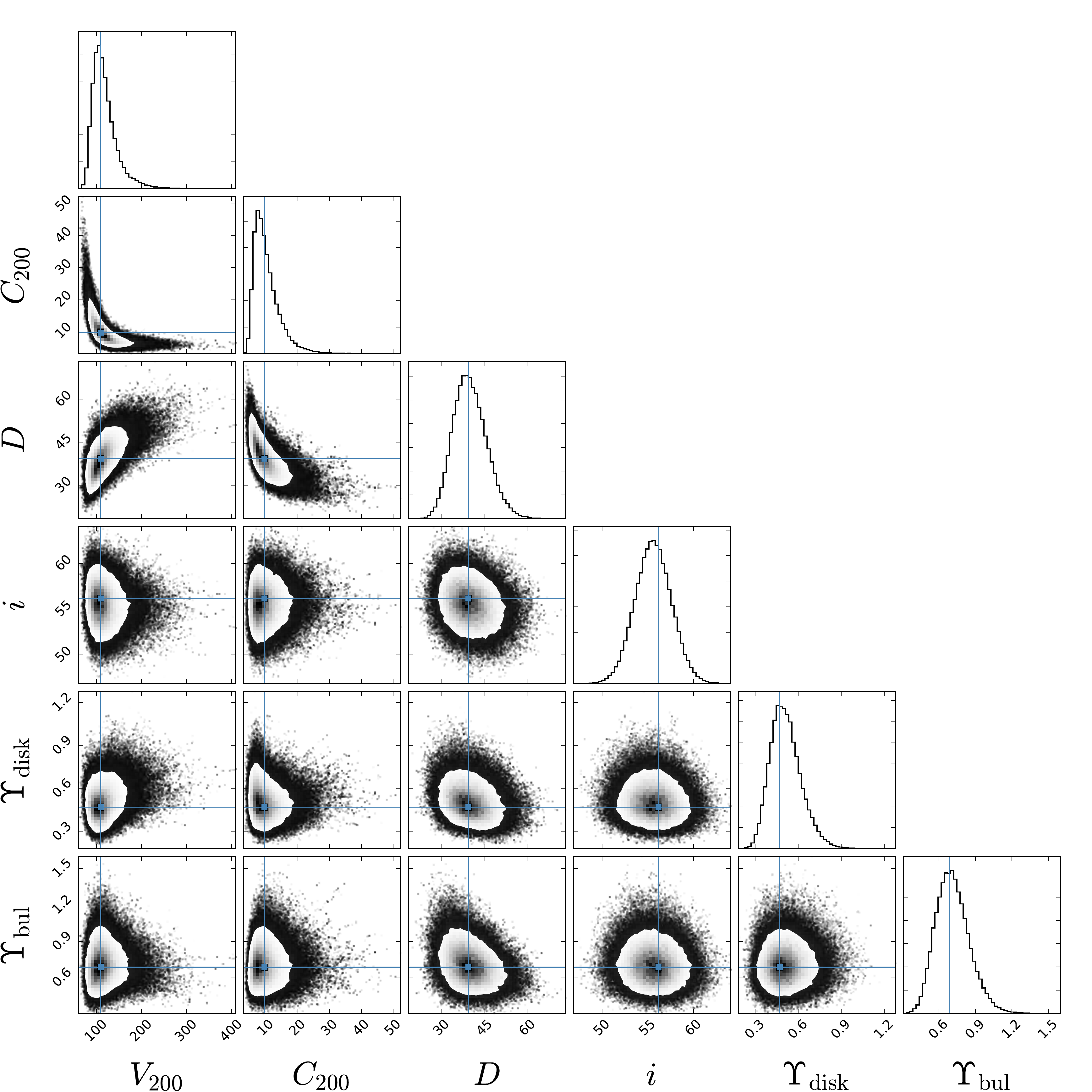}
	}
	\hspace{1cm}
	\subfloat[DC14 (no priors)]{%
		\includegraphics[width=0.38\textwidth]{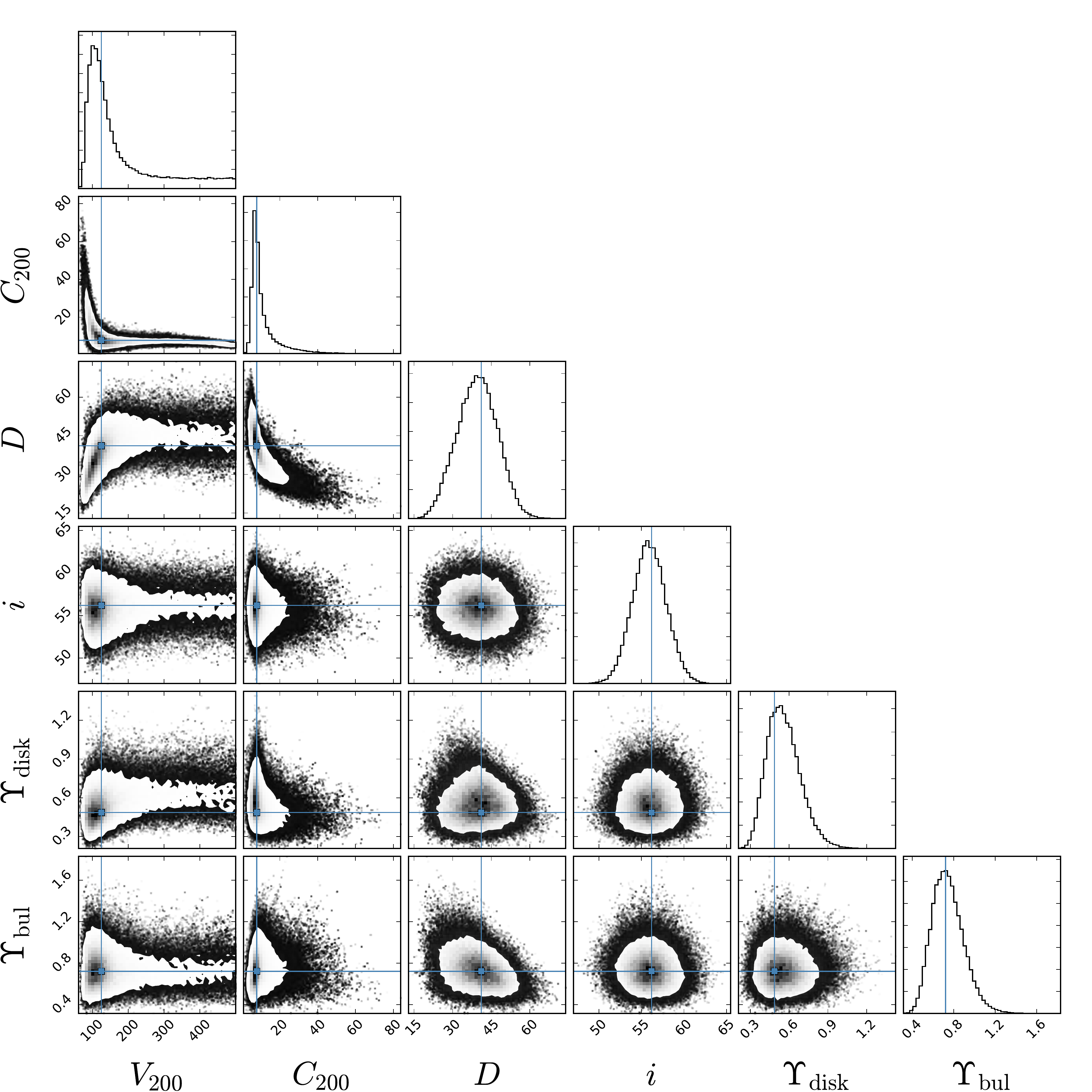}
	}
	\caption{Same as Fig. \ref{Fig: N2 corner plots}, but for UGC\,6176.}
	\label{Fig: U corner plots}
\end{figure*}

\end{appendix}

\end{document}